\documentclass[nofootinbib,preprint,showpacs,showkeys,amsmath,superscriptaddress,amssymb]{revtex4}

\usepackage{titlesec}
\usepackage{graphicx,epsfig,epstopdf}
\usepackage{bm}
\usepackage{epstopdf}
\usepackage{amssymb,amsmath,amsfonts,latexsym}
\usepackage{dcolumn}
\usepackage{bm}
\usepackage{hyperref}
\usepackage[italic]{hepnames}
\usepackage[utf8]{inputenc}
\newcommand{\nn}{\nonumber\\}
\usepackage{breqn}             
\makeatletter                  
\let\cat@comma@active\@empty   
\makeatother                   

\newcommand{\bea}{\begin{eqnarray}}
\newcommand{\ena}{\end{eqnarray}}
\newcommand{\be}{\begin{equation}}
\newcommand{\en}{\end{equation}}
\newcommand{\Tr}{\mbox{\rm{tr}}}

\begin{document}
\title{Exclusive semileptonic decays of $D$ and $D_s$ mesons \\in the covariant confining quark model}

\author{M.~A.~Ivanov}
\email{ivanovm@theor.jinr.ru}
\affiliation{Bogoliubov Laboratory of Theoretical Physics, \\
Joint Institute for Nuclear Research, 141980 Dubna, Russia}

\author{J.~G.~K\"{o}rner}
\email{jukoerne@uni-mainz.de}
\affiliation{PRISMA Cluster of Excellence, Institut f\"{u}r Physik, \\
Johannes Gutenberg-Universit\"{a}t, 
D-55099 Mainz, Germany}

\author{J.~N.~Pandya}
\email{jnpandya-apphy@msubaroda.ac.in}
\affiliation{Applied Physics Department, Faculty of Technology and Engineering, \\ The Maharaja Sayajirao University of Baroda, Vadodara 390001, Gujarat, India}

\author{P.~Santorelli}
\email{Pietro.Santorelli@na.infn.it}
\affiliation{Dipartimento di Fisica ``E.~Pancini'', Universit\`{a} di Napoli Federico II,\\ Complesso Universitario di Monte S. Angelo, Via Cintia, Edificio 6, 80126 Napoli, Italy}
\affiliation{Istituto Nazionale di Fisica Nucleare, Sezione di Napoli, 80126 Napoli, Italy}

\author{N.~R.~Soni}
\email{nrsoni-apphy@msubaroda.ac.in}
\affiliation{Applied Physics Department, Faculty of Technology and Engineering, \\ The Maharaja Sayajirao University of Baroda, Vadodara 390001, Gujarat, India}

\author{C.~T.~Tran}
\email{tranchienthang1347@gmail.com}
\thanks{corresponding author}
\affiliation{Dipartimento di Fisica ``E.~Pancini'', Universit\`{a} di Napoli Federico II,\\ Complesso Universitario di Monte S. Angelo, Via Cintia, Edificio 6, 80126 Napoli, Italy}
\affiliation{Istituto Nazionale di Fisica Nucleare, Sezione di Napoli, 80126 Napoli, Italy}


\begin{abstract}
Recently, the BESIII collaboration has reported numerous measurements of various $D_{(s)}$ meson semileptonic decays with significantly improved precision. Together with similar studies carried out at {\it BABAR}, Belle, and CLEO, new windows to a better understanding of weak and strong interactions in the charm sector have been opened. In light of new experimental data, we review the theoretical description and predictions for the semileptonic decays of $D_{(s)}$ to a pseudoscalar or a vector meson. This review is essentially an extended discussion of our recently published results obtained in the framework of the covariant confining quark model. 

 \end{abstract}

\maketitle

\section{Introduction}
\label{sec:introduction}

One of the fundamental ingredients of the Standard Model (SM) of particle physics is the Cabibbo-Kobayashi-Maskawa (CKM) matrix~\cite{Cabibbo:1963yz, Kobayashi:1973fv} which describes the quark mixing and holds the key to $CP$-violating phenomena. Precise determination of the CKM matrix elements is therefore crucially important. In this respect, semileptonic weak decays of mesons play an important role in our understanding of the SM since they provide the most direct way to extract the CKM matrix elements from experiments. Purely leptonic decays of mesons can also be used for the same purpose, but in many cases, they are not as experimentally accessible as semileptonic decays due to helicity suppression. Moreover, in semileptonic decays, the appearance of one (and only one) hadron in the final state gives rise to a richer phenomenology in comparison with purely leptonic decays, and, at the same time, keeps semileptonic decays theoretically cleaner compared to nonleptonic ones (for reviews, see e.g.,~\cite{Richman:1995wm, Asner:2008nq}).

Recently, the study of semileptonic decays of charm mesons has gained a great deal of attention, thank to the development of experimental facilities and progress in theoretical studies. Many collaborations have provided more and more precise measurements of these decays, which allow the extraction of the CKM matrix element $|V_{cd}|$ and $|V_{cs}|$ to an increasingly better accuracy. For example, the Particle Data Group (PDG) recently reported the values~\cite{Tanabashi:2018oca} $|V_{cd}|=0.2140\pm 0.0029\pm 0.0093$ and $|V_{cs}|=0.967\pm 0.025$ based on the measurements of the decays $D\to\pi(K)\ell\nu$ at {\it BABAR}~\cite{Lees:2014ihu, Aubert:2007wg}, Belle~\cite{Widhalm:2006wz}, BESIII~\cite{Ablikim:2015ixa}, and CLEO~\cite{Besson:2009uv}. Note that such extraction of $|V_{cd(s)}|$ requires theoretical calculation of the form factors characterizing the hadronic transitions $D\to\pi(K)$. Here, the form factors $f_+^{D\pi}(0)=0.666\pm 0.029$ and $f_+^{DK}(0)=0.747\pm 0.019$ obtained from a recent lattice QCD calculation~\cite{Aoki:2016frl} were used.

Semileptonic $D_{(s)}$ decays also offer stringent tests of the SM in the charm sector including the CKM matrix unitarity, isospin symmetry, $CP$-violation, and lepton flavor universality (LFU). Recently, many semimuonic charm decays have been measured for the first time at BESIII~\cite{Ablikim:2016sqt, Ablikim:2017omq, Ablikim:2018frk}. This sheds more light on the search for possible LFU violations and new physics beyond the SM at the precision frontier. Last but not least, semileptonic $D_{(s)}$ decays allow one to probe into the strong interaction effects happening in the transition between the initial and final mesons. These effects are parametrized by the hadronic invariant form factors which are functions of the momentum transfer squared ($q^2$) between the mesons. As a result, measurements of the form factors can be used to test different theoretical models and improve the inputs of theoretical calculations. 

Weak decays of hadrons are characterized by the interplay between weak and strong interactions. While the structure of the weak interaction is rather simple in the SM, the dynamics of the strong interaction is theoretically challenging. The reason is simple: transitions between hadrons are related to the bound state effects and hadronization, which are characterized by nonperturbative dynamics. Therefore, one needs nonperturbative methods to take into account the strong interaction in these decays. These methods include lattice QCD, QCD sum rules, and quark models. Very recently~\cite{Soni:2017eug, Soni:2018adu}, we studied a large set of $D_{(s)}$ semileptonic decays where
the hadron in the final state is one of $\pi$, $\rho$, $\omega$, $K$, $K^\ast(892)$, $\eta$, $\eta^\prime$ in the case of $D$ decays, and $\phi$, $K$, $K^\ast(892)$, $\eta$, $\eta^\prime$ in the case of $D_s$ decays. We also considered the decays $D^+_{(s)}\to D^0\ell^+\nu_\ell$. In these studies, the form factors were calculated in the whole physical range of momentum transfer by using our covariant confining quark model (CCQM).

In this review, we provide a detailed theoretical description of the $D _{(s)}$ semileptonic decays and summarize our theoretical predictions obtained in the CCQM, which were published in the recent papers~\cite{Soni:2017eug, Soni:2018adu}. We also discuss recent experimental data and other theoretical results. The rest of the review is organized as follows. In Sec.~\ref{sec:ff}, we introduce the semileptonic matrix element and define the corresponding hadronic form factors. We also discuss several form factor parametrizations that are commonly used in the literature. In Sec.~\ref{sec:hel} we describe the helicity technique and use it to obtain the twofold decay distribution. Section~\ref{sec:4fold} is devoted to the derivation of the full angular fourfold distribution and the construction of physical observables that can be studied experimentally. In Sec.~\ref{sec:CCQM} we briefly describe the CCQM and its application in the calculation of the hadronic form factors. Our predictions for the form factors, the decay branching fractions, and other physical observables are presented in Sec.~\ref{sec:results}. A detailed comparison of our results with other theoretical approaches and experimental data is provided. Finally, we briefly conclude in Sec.~\ref{sec:conclusion}.
\section{Matrix element and form factors}
\label{sec:ff}
\begin{figure}[htbp]
\includegraphics[scale=0.6]{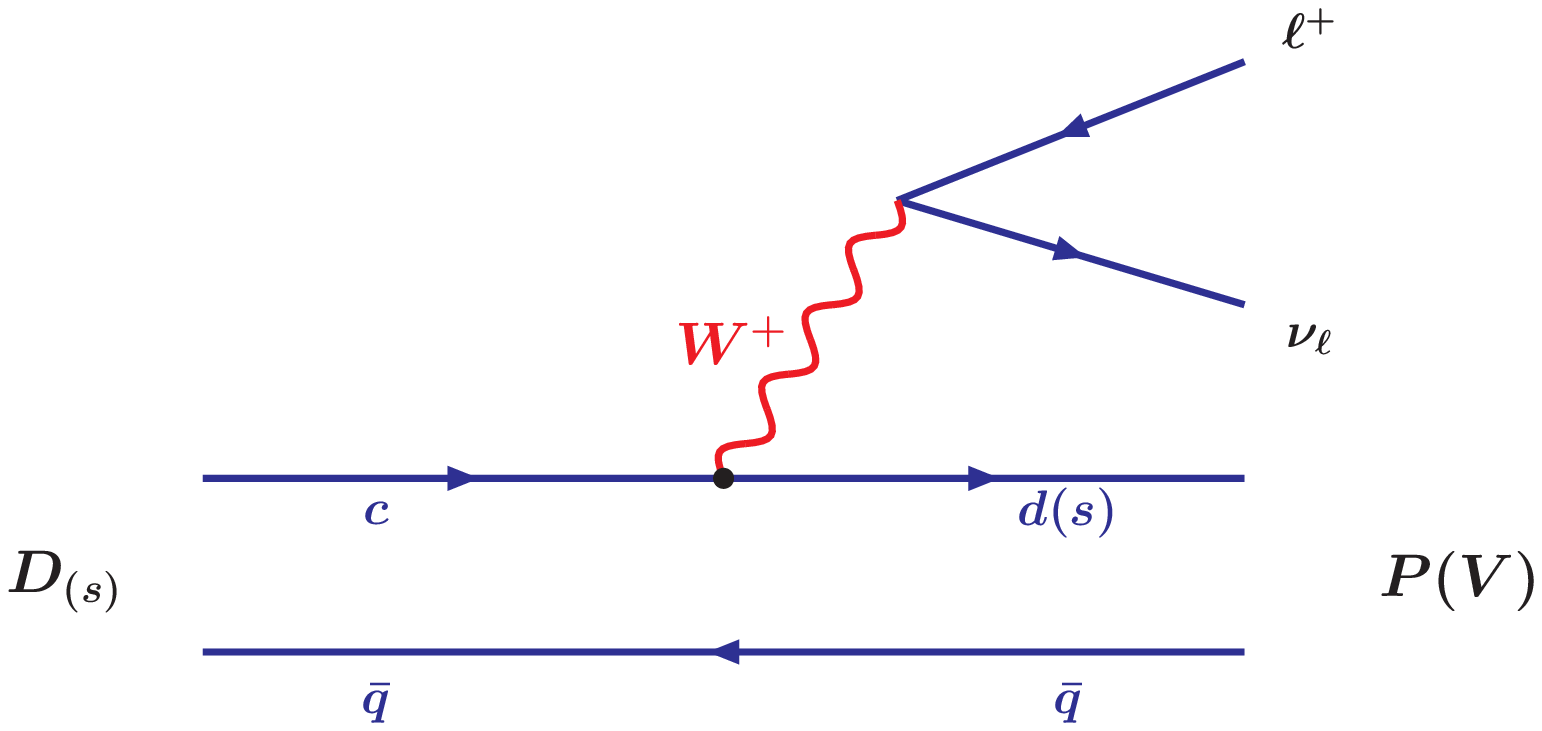}
\caption{Feynman diagram for semileptonic $D_{(s)}$ decays. The spectator quark $\bar q$ can be $\bar u$ ($D^0$), $\bar d$ ($D^+$), or $\bar s$ ($D_s$). $P$ ($V$) stands for pseudoscalar (vector) meson.}
\label{fig:feyn}
\end{figure}

In the SM, semileptonic decays of $D_{(s)}$ meson proceed via two sub-processes $c\to d(s)W^{\ast +}$ and $W^{\ast +}\to\ell\nu_\ell$ as depicted in Fig.~\ref{fig:feyn}. The theoretical description of the weak interactions in these decays is straighforward. However, the difficult part lies in the prediction of the strong interactions bounding quarks inside hadrons, which are large at the typical decay energies. Due to the fact that the $W^{\ast +}$ boson decays into a lepton pair which is invisible to the strong force, the weak and strong interactions can be well separated. This makes semileptonic decays theoretically cleaner than nonleptonic decays.

In the SM, the matrix element for semileptonic decays of the $D_{(s)}$ meson to a pseudoscalar ($P$) or a vector ($V$) meson in the final state is written as
\be
\mathcal M(D_{(s)} \to (P,V) \ell^+ \nu_{\ell}) = \frac{G_F}{\sqrt{2}} V^\ast_{cq} H^\mu L_\mu,
\en
where $G_F$ is the Fermi constant. The leptonic and hadronic currents are given by
\bea
L_\mu &=& \bar{\nu}_\ell \gamma_\mu(1-\gamma_5) \ell,
\label{eq:lept-current}\\
H^\mu &=& \left\langle (P,V) |V^\mu-A^\mu | D_{(s)}\right\rangle,
\label{eq:hadr-current}
\ena
with $V^\mu=\bar{q}\gamma^\mu c$ and $A^\mu=\bar{q}\gamma^\mu\gamma_5c$ being the flavor-changing vector and axial-vector currents, respectively. All strong effects have been absorbed into the hadronic current $H^\mu$, which is also referred to as the hadronic matrix element.

The hadronic matrix element is constructed from four-vectors appearing in the transition, namely, the four momenta and polarization vectors of the mesons. In the case of $D_{(s)}(p_1)\to P(p_2)$, there are two independent four-vectors, which can be taken as $P = p_1 + p_2$ and $q=p_1-p_2$. One can then parametrize $H^\mu$ in terms of two invariant form factors depending on the momentum transfer squared ($q^2$) between the initial and final mesons as follows:
\be 
\label{eq:ff-PP}
\langle P(p_2)| V^\mu  | D_{(s)}(p_1) \rangle
= F_+(q^2) P^\mu + F_-(q^2) q^\mu.
\en
Note that for $P\to P^\prime$ transitions, the axial-vector current $A^\mu$ does not contribute, and therefore, has been omitted in Eq.~(\ref{eq:ff-PP}). In the case of $D_{(s)}(p_1)\to V(p_2,\epsilon_2)$, the additional polarization vector $\epsilon_2$ of the meson $V$ gives rise to four form factors which can be defined as
\bea
\label{eq:ff-PV}
\langle V(p_2,\epsilon_2)
|V^\mu-A^\mu | D_{(s)}(p_1) \rangle
&=& \frac{\epsilon^{\ast}_{2\alpha}}{M_1+M_2}
\Big[-g^{\mu\alpha}P\cdot qA_0(q^2) + P^{\mu}P^{\alpha}A_+(q^2)\nn
&&+ q^{\mu}P^\alpha A_-(q^2) 
+ i\varepsilon^{\mu\alpha Pq} V(q^2)\Big],
\ena
where we have used the short notation $\varepsilon^{\mu\alpha Pq}\equiv \varepsilon^{\mu\alpha\nu\beta }P_\nu q_\beta$. It is worth noting that  $\epsilon_2^\ast\cdot p_2=0$, and the mesons are on shell: $p_1^2=m^2_{D_{(s)}}\equiv M_1^2$, $p_2^2=m_{P,V}^2 \equiv M_2^2$. The vector current $V^\mu$ contributes to the form factor $V(q^2)$, while the axial-vector current $A^\mu$ contributes to $A_{\pm,0}(q^2)$.

There exists another way to define the form factors, which is more common in the literature, proposed by Bauer, Stech, and Wirbel (BSW)~\cite{Wirbel:1985ji}:
\bea
\label{eq:ff-BSW}
\langle P(p_2)|V^\mu| D_{(s)}(p_1) \rangle
&=& F_1(q^2) \left[P^\mu-\frac{M_1^2-M_2^2}{q^2}q^\mu\right]+F_0(q^2) \frac{M_1^2-M_2^2}{q^2} q^\mu,\\
\langle V(p_2,\epsilon_2) |V^\mu-A^\mu| D_{(s)}(p_1) \rangle
&=& 
-(M_1+M_2)\epsilon_2^{\ast\mu}A_1(q^2)
+\frac{\epsilon_2^\ast\cdot q}{M_1+M_2}P^\mu A_2(q^2)\nn
&&+2M_2\frac{\epsilon_2^\ast\cdot q}{q^2}q^\mu\left[A_3(q^2) -A_0(q^2)\right]
+\frac{2i\varepsilon^{\mu\epsilon_2^\ast p_2 p_1}}{M_1+M_2}V(q^2),
\ena
where $A_3(q^2)$ is the abbreviation for
\be 
A_3(q^2)=\frac{M_1+M_2}{2M_2}A_1(q^2)-\frac{M_1-M_2}{2M_2}A_2(q^2).
\en
The BSW form factors satisfy the constraints $F_0(0)=F_1(0)$ and $A_0(0)=A_3(0)$, so there is no singularity at $q^2=0$. The form factors $F_0(q^2)$ and $F_1(q^2)$ can be associated with the exchange of particles with quantum numbers $J^P=0^+$ and $J^P=1^-$. The form factors $A_{1,2}(q^2)$ and  $V(q^2)$ are associated with $J^P=1^+$ and $J^P=1^-$, respectively~\cite{Wirbel:1985ji, Richman:1995wm}.
The relations between our form factors defined in Eqs.~(\ref{eq:ff-PP},\ref{eq:ff-PV}) and the BSW form factors read
\bea
A_0^{\rm BSW}(q^2) &=& \frac{M_1-M_2}{2M_2}\Big[A_0^{\rm our}(q^2)-A_+(q^2) - \frac{q^2}{M_1^2-M_2^2} A_-(q^2) \Big],\nn
A_1(q^2) &=& \frac{M_1-M_2}{M_1+M_2} A_0^{\rm our}(q^2),\qquad
F_0(q^2) = F_+ + \frac{q^2}{M_1^2-M_2^2} F_-(q^2), \\
A_2(q^2) &=& A_+(q^2), \qquad V^{\rm BSW}(q^2) = V^{\rm our}(q^2),\qquad F_1(q^2) = F_+(q^2).\nonumber
\label{eq:ff-rel}
\ena

In semileptonic decays of $D$ and $D_s$ mesons, the tau mode is kinematically forbidden. For the light lepton modes, the limit of zero lepton mass is often used. In this limit, the terms in the hadronic matrix elements that are proportional to $q^\mu$ do not contribute to the decay rates since $q^\mu L_\mu=0$ as $m_\ell\to 0$. Therefore, in the literature, more focus is on the form factors $F_+(q^2)$, $A_1(q^2)$, $A_2(q^2)$, and $V(q^2)$. For the $D_{(s)}\to P$ transitions, the normalization of the form factor $F_+(q^2)$ at maximum recoil $(q^2=0)$ has been  studied extensively. Regarding the $D_{(s)}\to V$ transitions, since the form factor $A_1(q^2)$ appears in all helicity amplitudes (see Sec.~\ref{sec:hel}), it is usually factored out by defining the following ratios at maximum recoil
\be
r_2 = \frac{A_2(q^2=0)}{A_1(q^2=0)},\qquad r_V = \frac{V(q^2=0)}{A_1(q^2=0)}.
\en
From the experimental point of view, these ratios can be obtained without any assumption about the total decay rates or the CKM matrix elements.

Before moving to the next section, it is worth discussing several parametrizations of the form factors that are commonly used in the literature. There are two essential features one wishes to know about a form factor $F_i(q^2)$: its normalization, usually at $q^2=0$, and its shape. The second becomes more important when the available kinematical range is large, since most of the theoretical calculations are best applicable only for a typically limited $q^2$ region. For example, LQCD works best at large $q^2$, while QCDSR {\textemdash} at small $q^2$. The knowledge of the form factor's shape is then used to extrapolate the results to the rest $q^2$ range where direct calculation is less reliable. The lack of this knowledge leads to the main source of uncertainties when extracting the CKM matrix elements from experimental data on exclusive semileptonic decays. There is no QCD-based theory yet that can fully describe the complexity of strong interaction dynamics encoded in the hadronic form factors. However, several constraints on the behavior of the form factors can be obtained by using kinematic and dispersion relations, and by assuming some limits, such as the heavy quark limit and the large energy effective theory limit. For a detailed review on this subject, we refer the reader to~\cite{Hill:2006ub} and references therein. For a more recent review, see Appendix~A.5 of~\cite{Aoki:2019cca}.

In early studies of beauty and charm semileptonic decays, the functional form of form factors were usually assumed to obey the simple pole model (also known as the nearest pole dominance)
\be 
\label{eq:simple-pole}
F_i(q^2)=\frac{F_i(0)}{1-q^2/m_{\rm pole}^2},
\en
which is based on the vector-dominance physical picture of the decays. The pole mass $m_{\rm pole}$ is the mass of the lowest-lying vector meson that has the appropriate quantum numbers dictated by the corresponding hadronic current. For example, $m_{\rm pole}=m_{D_s^{\ast +}}$ for $D^0\to K^-\ell^+\nu$, and $m_{\rm pole}=m_{D^{\ast +}}$ for $D^0\to \pi^-\ell^+\nu$. It soon became clear that this model was an oversimplification of the real dynamics: fitting to experimental data yields ``non-physical" values of the pole mass. Also, when the kinematical range is large, two free parameters ($F_i(0)$ and $m_{\rm pole}$) are not enough to accommodate experimental data well.

In order to introduce a parametrization that fits data and, at the same time, has some physical meaning, Becirevic and Kaidalov~\cite{Becirevic:1999kt} came up with the modified pole model
\be 
\label{eq:BK-para}
F_i(q^2)=\frac{F_i(0)}{(1-q^2/m_{\rm pole}^2)(1-\alpha\, q^2/m_{\rm pole}^2)},
\en
where $m_{\rm pole}$ is usually fixed to the physical value explained above, and $\alpha$ is a free parameter that takes into account contributions from higher states in the form of an additional effective pole. This ansatz has been widely used in lattice calculations and experimental studies of semileptonic form factors due to its elasticity for data fitting and the ability to satisfy several constraints on the form factors. The modified pole model, however, faced the same difficulty as the single pole model did, when applied to a large variety of decays. While it still can be used to fit experimental data, the motivation for such simplification, which was originally proposed for beauty decays, turned out not quite valid for charm decays~\cite{Hill:2006ub, Dobbs:2007aa}. 

A more systematic and model-independent parametrization of semileptonic form factors has been developed by several groups based on rather general properties of the form factors including QCD dispersion relations and analyticity (see e.g.,~\cite{Bourrely:1980gp, Boyd:1994tt, Lellouch:1995yv, Boyd:1997qw, Boyd:1997kz, Arnesen:2005ez, Becher:2005bg}). In particular, this parametrization provides better control of theoretical uncertainties in lattice calculations. The basic idea is to perform an analytic continuation of the form factors into the complex $t\equiv q^2$ plane. The physical semileptonic region is then given by $m_\ell^2\leq t \leq t_-$, with $t_-= (M_1-M_2)^2$. A generic form factor contains poles and a branch cut $[t_+,\infty)$ along the real axis, where $t_+=(M_1+M_2)^2$ is the pair-production threshold. One then (effectively) ``factors out" the poles, and expands the rest of the form factor into a series around some kinetic point $t_0$. In order for the series to converge efficiently, the expansion is done with respect to a new variable $z(t,t_0)$, which  is a kinematic function of $t$ defined by
\be 
\label{eq:z-def}
z(t,t_0)=\frac{\sqrt{t_+-t}-\sqrt{t_+-t_0}}{\sqrt{t_+-t}+\sqrt{t_+-t_0}}.
\en
This conformal transformation maps the branching cut onto the unit $z$ circle, and the rest of the complex $t$ plane onto the open unit $z$ disk. The kinematic point $t_0$ is abitrary ($t_0<t_+$) and corresponds to $z=0$. The ``default" choice is a rather intermediate value, $t_0=t_+(1-\sqrt{1-t_- /t_+})$, which helps minimize the expansion parameter $|z|_{\rm max}$.

The parametrization is often referred to as ``$z$-expansion" or ``$z$-parametrization" and has the form
\be 
\label{eq:z-para}
F_i(t)=\frac{1}{P(t)\phi(t,t_0)}\sum_{k=0}^K a_k(t_0)[z(t,t_0)]^k,
\en
where the so-called Blaschke factor $P(t)$ accommodates the resonances below the pair-production threshold $t_+$, the so-called outer function $\phi(t,t_0)$ is an abitrary function analytic outside of the cut, which is perturbatively calculable and does not affect physical observables, and $a_k(t_0)$ are the expansion coefficients to be determined. The Blaschke factor is used to remove sub-threshold poles, for instance, $P(t)=1$ for $D\to\pi$, and $P(t)=z(t,m^2_{D^\ast_s})$ for $D\to K$. The outer function is chosen in such a way to provide a bound on the expansion coefficients. The typical bound is the unitarity condition $\sum^K_{k=0}a_k^2\leq 1$ for any $K$, which implies that the exclusive production rate of $M_1M_2$ states induced by the given current must not exceed the inclusive one. This bound corresponds to the following standard choice of the outer function for the $F_+$ form factor
\be
\label{eq:phi}
\phi(t,t_0)=\beta(\sqrt{t_+-t}+\sqrt{t_+-t_0})\frac{(\sqrt{t_+-t}+\sqrt{t_+-t_-})^{3/2}}{(\sqrt{t_+-t}+\sqrt{t_+})^5}\frac{t_+-t}{(t_+-t_0)^{1/4}},
\en
where the parameter $\beta$ is calculated by using operator product expansion techniques, which at leading order yields $\beta=\sqrt{\pi m_c^2/3}$ for charm decays. The outer function for other form factors can be found in~\cite{Boyd:1997kz}. 

The $z$-expansion using the outer function in Eq.~(\ref{eq:phi}) is often referred to as the Boyd-Grinstein-Lebed parametrization. An alternative choice of the outer function was proposed by Bourrely, Caprini, and Lellouch in~\cite{Bourrely:2008za}, which results in a simple parametrization that combines the pole factorization and the $z$-expansion as follows:
\be
\label{eq:BCL1}
F_i(t)=\frac{1}{1-t/m^2_{\rm pole}}\sum_{k=0}^Kb_k(t_0)[z(t,t_0)]^k.
\en
One of the advantages of this parametrization is the simplicity to translate the near-threshold behavior of the form factors into a useful constraint on the expansion coefficients. For the form factor $F_+$, the asymptotic behavior ${\rm Im}F_+(t)\sim (t-t_+)^{3/2}$ near $t_+$ is imposed to obtain the final constrained form for the parametrization
\be
\label{eq:BCL2}
F_+(t)=\frac{1}{1-t/m^2_{\rm pole}}\sum_{k=0}^{K-1}b_k\left[z^k-(-1)^{k-K}\frac{k}{K}z^K\right].
\en

\section{Helicity amplitudes and decay distribution}
\label{sec:hel}
\begin{figure}[htbp]
\includegraphics[scale=0.5]{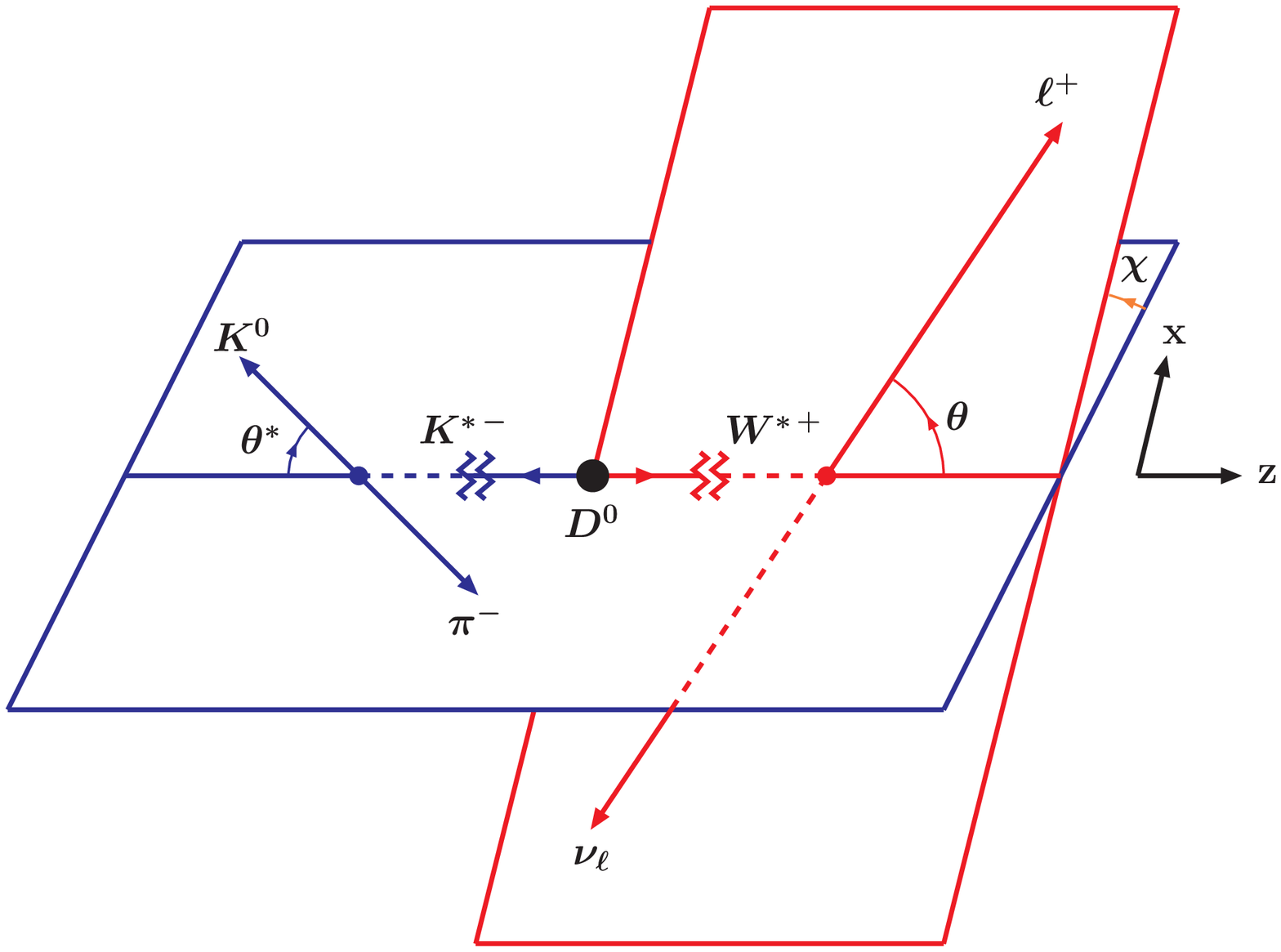}
\caption{Definition of the angles $\theta$, $\theta^\ast$, and $\chi$ in
the cascade decay $D^0\to K^{\ast-}(\to K^0\pi^-)\ell^+\nu_\ell$. The polar angles $\theta$ and $\theta^\ast$ are defined in the rest frames of the $W^\ast$ and the $K^\ast$, respectively. $\chi$ is the azimuthal angle between the two decay planes.}
\label{fig:4fold}
\end{figure}

Let us first consider the twofold differential decay distribution in terms of $q^2$ and the polar angle $\theta$. The polar angle $\theta$ is defined in the rest frame of the $W^\ast$ as the angle between the momentum of the final charged lepton and the direction opposite to the daughter meson's momentum (see Fig.~\ref{fig:4fold}). One has 
\be
\frac{d^2\Gamma}{dq^2 d\cos\theta} = 
\frac{|{\bf p_2}|}{(2\pi)^3 32M_1^2}\left(1-\frac{m_\ell^2}{q^2}\right)
\cdot\sum\limits_{\rm pol}|\mathcal{M}|^2
=\frac{G^2_F |V_{cq}|^2 }{(2\pi)^3}
\frac{|{\bf p_2}|}{64 M_1^2}\left(1-\frac{m_\ell^2}{q^2}\right)
H^{\mu\nu} L_{\mu\nu},
\label{eq:2-fold-dis}
\en
where  $|{\bf p_2}|=\lambda^{1/2}(m_1^2,m_2^2,q^2)/2m_1 $
is the momentum of the daughter meson in the $D_{(s)}$ rest frame, with $\lambda(x,y,z) \equiv x^2+y^2+z^2-2(xy+yz+zx)$ being the K{\"a}ll{\'e}n function. 

The hadronic tensor is given by
\bea
H^{\mu\nu} &=& \sum\limits_{\rm pol}\,
\langle X|V^\mu-A^\mu | D_{(s)}\rangle\cdot 
\langle X|V^\nu-A^\nu | D_{(s)}\rangle^\dagger
\nn
&=&
\left\{
\begin{array}{lr}
H_\mu H_\nu^\dagger
& \qquad\text{for}\qquad  D_{(s)}\to P
\\
T_{\mu\alpha} T_{\nu\beta}^\dagger \left( -g^{\alpha\beta}+\frac{p_2^\alpha p_2^\beta}{M_2^2}\right)
& \qquad\text{for}\qquad  D_{(s)}\to V
\end{array}
\right. ,
\label{eq:hadr_tensor}
\ena
where the tensor $T_{\mu\alpha}$ is defined by the relation $H_\mu=\epsilon_2^{\ast \alpha} T_{\mu\alpha}$. One can easily obtain the explicit expression for $T_{\mu\alpha}$ by comparing this relation with Eqs.~(\ref{eq:hadr-current}) and~(\ref{eq:ff-PV}). 

The unpolarized leptonic tensor for the process $W^{\ast+}\to \ell^+ \nu_\ell$ $\left(  W^{\ast-}\to \ell^-\bar \nu_\ell\right)$ is given  by 
\bea
L_{\mu\nu} &=& \left\{\begin{array}{lr}
\Tr\left[(\not\! k_1 + m_\ell) O_\mu \not\! k_2 O_\nu\right]
& \qquad\text{for}\qquad  W^{\ast -}\to \ell^-\bar\nu_\ell 
\\
\Tr\left[ (\not\! k_1 - m_\ell) O_\nu \not\! k_2 O_\mu\right]
& \qquad\text{for}\qquad  W^{\ast +}\to \ell^+ \nu_\ell 
                   \end{array}\right.
\nn
&=&
8 \big( 
  k_1^\mu k_2^\nu  + k_1^\nu k_2^\mu 
- k_1\cdot k_2g^{\mu\nu}
  \pm  i \varepsilon^{\mu \nu \alpha \beta} k_{1\alpha} k_{2\beta}
\big),
\label{eq:lept_tensor}
\ena
where $O_\mu=\gamma_\mu(1-\gamma_5)$, and $k_1$ ($k_2$) is the momentum of the charged lepton (neutrino). The upper/lower sign 
refers to the two $(\ell^-\bar\nu_\ell)/(\ell^+\nu_\ell)$ configurations.
The sign change results from the parity violating part of 
the leptonic tensor. In our case we have to use the lower sign
in Eq.~(\ref{eq:lept_tensor}). We use the following convention for the $\gamma_5$ matrix and the Levi-Civita tensor in Minkowski space: 
\bea
&&
\gamma^0 = \left(\begin{array}{lr}
                           I & 0 \\[-1ex]
                           0 & -I
                    \end{array}\right), \qquad
\gamma^k = \left(\begin{array}{lr}
                               0     & \sigma_k \\[-1ex]
                           -\sigma_k & 0
                    \end{array}\right), \quad
\gamma^5 = \gamma_5 = i\gamma^0\gamma^1\gamma^2\gamma^3
=  \left(\begin{array}{lr}
                           0 & I \\[-1ex]
                           I & 0
                    \end{array}\right), 
\label{eq:Levy-Civita}\\
&&
\Tr\left(\gamma_5\gamma^\mu\gamma^\nu\gamma^\alpha\gamma^\beta \right)
= 4i\varepsilon^{\mu\nu\alpha\beta},
\qquad
\Tr\left(\gamma_5\gamma_\mu\gamma_\nu\gamma_\alpha\gamma_\beta \right)
= 4i \varepsilon_{\mu\nu\alpha\beta},
\qquad
\varepsilon_{0123} = -\varepsilon^{0123} = +1.
\nonumber
\ena

The Lorentz contraction in Eq.~(\ref{eq:2-fold-dis}) can be  evaluated in terms of the so-called helicity amplitudes as described  in~\cite{Korner:1987kd, Korner:1989qb, Korner:1989ve, Faessler:2002ut, Gratrex:2015hna}. First, one defines an orthonormal and complete helicity basis
$\epsilon^\mu(\lambda_W)$ with three spin-1 components  orthogonal to
the momentum transfer $q^\mu$, i.e., $\epsilon^\mu(\lambda_W) q_\mu=0$, for 
$\lambda_W=\pm,0$, and one spin-0 (time) component $\lambda_W=t$ with
$\epsilon^\mu(t)= q^\mu/\sqrt{q^2}$. The orthonormality and completeness relations read 
\bea
\epsilon^\ast_\mu(m)\epsilon^\mu(n) &=& g_{mn} 
\quad (\text{orthonormality}),
\nn
\epsilon_\mu(m)\epsilon^{\ast}_{\nu}(n)g_{mn} &=& g_{\mu\nu} \quad\,\, (\text{completeness}),
\label{eq:orth-compl}
\ena
with $m,n=t,\pm,0$ and $g_{mn}={\rm diag}(+,-,-,-)$.

With the help of the completeness relation one then rewrites the contraction
of the leptonic and hadronic tensors in Eq.~(\ref{eq:2-fold-dis}) as follows:
\bea
L^{\mu\nu}H_{\mu\nu} &=& 
L_{\mu'\nu'}\epsilon^{\mu'}(m)\epsilon^{\ast\mu}(m')g_{mm'}
\epsilon^{\ast \nu'}(n)\epsilon^{\nu}(n')g_{nn'}H_{\mu\nu}
\nn
&=& L(m,n) g_{mm'} g_{nn'} H(m',n'),
\label{eq:contraction}
\ena
where $L(m,n)$ and $H(m,n)$ are the leptonic and hadronic tensors in the helicity-component space. One has
\be
L(m,n) = \epsilon^\mu(m)\epsilon^{\ast \nu}(n)
L_{\mu\nu},
\qquad
H(m,n) = \epsilon^{\ast\mu}(m)\epsilon^\nu(n)H_{\mu\nu}.
\label{eq:hel_tensors}
\en
The point is that the two tensors can now be evaluated in two different
Lorentz frames. The leptonic tensor $L(m,n)$ will be evaluated
in the $W^\ast$ rest frame while the hadronic tensor $H(m,n)$ {\textemdash} in the $D_{(s)}$ rest frame.

In order to express the helicity-components of the hadronic tensor $H(m,n)$ through the invariant form factors given in Eqs.~(\ref{eq:ff-PP}) and~(\ref{eq:ff-PV}), one has to define the polarization vector $\epsilon(\lambda_W)$ explicitly. In the $D_{(s)}$ rest frame, the momenta and polarization vectors can be written as
\be
\begin{array}{ll}
p^\mu_1 = (M_1,0,0,0),\quad & \quad
\epsilon^\mu(t) =
\frac{1}{\sqrt{q^2}}(q_0,0,0,|{\bf p_2}|),
\\
p^\mu_2 = (E_2,0,0,-|{\bf p_2}|), \quad &\quad
\epsilon^\mu(\pm) = 
\frac{1}{\sqrt{2}}(0,\mp 1,-i,0),
\\
q^\mu   = (q_0,0,0,+|{\bf p_2}|), \quad &\quad
\epsilon^\mu(0) =
\frac{1}{\sqrt{q^2}}(|{\bf p_2}|,0,0,q_0),\\
\end{array}
\label{eq:Dframe}
\en
where $E_2 = (M_1^2+M_2^2-q^2)/2 M_1$ and $q_0=(M_1^2-M_2^2+q^2)/2 M_1$.

For the $D_{(s)}\to P$ transition one has
\be
H(m,n) = \big[\epsilon^{\ast \mu}(m)H_\mu\big]\cdot 
         \big[\epsilon^{\ast \nu}(n)H_\nu\big]^\dagger
\equiv
H_m H^{\dagger}_n.
\label{eq:hel_pp_def}
\en
The helicity amplitudes $H_m$ are written in terms of the
invariant form factors as follows:
\be
H_t   = \frac{1}{\sqrt{q^2}}(Pq F_+ + q^2 F_-),
\qquad
H_\pm = 0,
\qquad
H_0   = \frac{2M_1|{\bf p_2}|}{\sqrt{q^2}} F_+ .
\label{eq:hel_pp}
\en
It should be noted that $H_{t}=H_{0}$ at maximum recoil ($q^{2}=0$), and $H_{0}=0$ at zero recoil ($q^{2}=q^2_{\rm max})$.

For the $D_{(s)}\to V$ transition, in addition to the $W^\ast$ polarization vector $\epsilon(\lambda_W)$, one needs the explicit representation of the polarization vector $\epsilon_2(\lambda_V)$ of the daughter vector meson. In the $D_{(s)}$ rest frame, the helicity components of the vector $\epsilon_2(\lambda_V)$ read
\be
\epsilon^\mu_2(\pm) = 
\frac{1}{\sqrt{2}}(0,\pm 1,-i,0),
\qquad
\epsilon^\mu_2(0) = 
\frac{1}{M_2}(|{\bf p_2}|,0,0,-E_2).
\label{eq:vect_pol}
\en
The hadronic tensor for the $D_{(s)}\to V$ transition is then rewritten in terms of the corresponding helicity amplitudes as follows:
\bea 
H(m,n) &=&  
\epsilon^{\ast \mu}(m) \epsilon^{ \nu}(n)H_{\mu\nu}
=
\epsilon^{\ast \mu}(m) \epsilon^{ \nu}(n) 
T_{\mu\alpha}
\epsilon_2^{\ast\alpha}(r)\epsilon_2^{\beta}(s)\delta_{rs}
T^{\dagger}_{\beta\nu}
\nn
&=&
\epsilon^{\ast \mu}(m)\epsilon_2^{\ast\alpha}(r)
T_{\mu\alpha} \cdot
\Big[\epsilon^{\ast \nu}(n)\epsilon_2^{\ast\beta}(s)T_{\nu\beta}
\Big]^\dagger\delta_{rs}
\equiv 
H_{mr} H^{\dagger}_{nr}.
\label{eq:hel_vv_def}
\ena 

Angular momentum conservation dictates that 
$r=m$ and $s=n$ for $m,n=\pm,0$, and $r,s=0$ for $m,n=t$. One then obtains the following non-zero helicity amplitudes 
\bea
H_{t} &\equiv & H_{t0}=
\epsilon^{\ast \mu}(t)\epsilon_2^{\ast \alpha}(0)T_{\mu\alpha}=
\frac{M_1|{\bf p_2}|}{M_2(M_1+M_2)\sqrt{q^2}}
\left[P\cdot q(-A_0+A_+)+q^2 A_-\right],\nn
H_{\pm} &\equiv & H_{\pm\pm}=
\epsilon^{\ast \mu}(\pm)\epsilon_2^{\ast \alpha}(\pm)T_{\mu\alpha}=
\frac{-P\cdot q A_0\pm 2M_1|{\bf p_2}| V}{M_1+M_2},\\
H_{0} &\equiv &  H_{00}=
\epsilon^{\ast \mu}(0)\epsilon_2^{\ast \alpha}(0)T_{\mu\alpha}= 
\frac{-P\cdot q(M_1^2 - M_2^2 - q^2) A_0 + 4M_1^2|{\bf p_2}|^2 A_+}{2M_2(M_1+M_2)\sqrt{q^2}}.\nonumber
\label{eq:hel_vv}
\ena
Note that the helicity amplitudes satisfy the zero-recoil relations $H_{t}=0$ and $H_{\pm}=H_{0}$. Also, at maximum recoil ($q^{2}=0$), the dominating helicity amplitudes are
$H_{t}$ and $H_{0}$, and $H_{t}(0)=H_{0}(0)$. 

We are done with the hadronic tensor $H(m,n)$. Let us know turn to the leptonic tensor $L(m,n)$, which is evaluated in the $W^\ast$ rest frame, where the charged lepton and the neutrino are back-to-back, i.e., ${\bf k_1}+{\bf k_2}=0$. Again, one needs explicit expressions for the momenta and polarization vectors in this frame. One has
\bea
q^\mu   &=& (\sqrt{q^2},0,0,0),
\nn
k^\mu_1 &=& 
(E_1,|{\bf k_1}|\sin\theta\cos\chi, 
|{\bf k_1}| \sin\theta\sin\chi,|{\bf k_1}| \cos\theta),
\nn
k^\mu_2 &=& (|{\bf k_1}|,-|{\bf k_1}|\sin\theta\cos\chi,
-|{\bf k_1}|\sin\theta\sin\chi,-|{\bf k_1}|\cos\theta),
\label{eq:lept_basis}
\ena
with $E_1=(q^2+m^2_\ell)/2\sqrt{q^2}$ and 
$|{\bf k_1}|=(q^2-m_\ell^2)/2\sqrt{q^2}$ being the energy and three-momentum of the charged lepton in the $W^\ast$ rest frame. The azimuthal angle $\chi$ is defined in Fig.~\ref{fig:4fold}. Here we have chosen a right-handed $(x,y,z)$ coordinate system such that the $z$ axis is the direction of the $W^\ast$ in the parent rest frame, and the momentum of $\ell^+$ lies in the $(xz)$ plane.
The longitudinal, transverse, and time helicity-component of the polarization vector $\epsilon(\lambda_W)$ in the $W^\ast$ rest frame are given by
\be
\epsilon^\mu(0)=(0,0,0,1), \quad
\epsilon^\mu(\pm) = 
\frac{1}{\sqrt{2}}(0,\mp 1,-i,0), \quad
\epsilon(t)=(1,0,0,0).
\en
One obtains [the matrix columns and rows are ordered in the sequence 
$(t,+,0,-)$]
\bea 
\label{eq:lt1}
\lefteqn{(2q^2v)^{-1} L(m,n)(\theta,\chi)=}\nn
&=&\left( \begin{array}{cccc} 
0 & 0 & 0 & 0 \\[-1ex]
0 & (1\mp\cos\theta)^2 & \mp \frac{2}{\sqrt{2}} (1\mp\cos\theta) \sin\theta 
e^{i\chi} & \sin^2\theta e^{2i\chi} \\[-1ex]
0 & \mp \frac{2}{\sqrt{2}} (1\mp\cos\theta) \sin\theta 
e^{-i\chi} & 2\sin^2\theta 
& \mp \frac{2}{\sqrt{2}} (1\pm\cos\theta) \sin\theta 
e^{i\chi} \\[-1ex]
0 & \sin^2\theta e^{-2i\chi} 
& \mp \frac{2}{\sqrt{2}} (1\pm\cos\theta) \sin\theta 
e^{-i\chi} & (1\pm\cos\theta)^2 \\[-1ex]
\end{array} \right)\\
&&+\delta_\ell 
\left( \begin{array}{cccc} 
4 & - \frac{4}{\sqrt{2}} \sin\theta e^{i\chi} & 4 \cos\theta &
 \frac{4}{\sqrt{2}} \sin\theta e^{i\chi} \\[-1ex]
- \frac{4}{\sqrt{2}} \sin\theta e^{i\chi} & 
2 \sin^2\theta & 
- \frac{2}{\sqrt{2}} \sin 2\theta e^{i\chi} & 
- 2 \sin^2\theta e^{2i\chi} \\[-1ex]
4\cos\theta & 
- \frac{2}{\sqrt{2}} \sin 2\theta e^{-i\chi} & 
4\cos^2\theta & 
\frac{2}{\sqrt{2}} \sin 2\theta e^{i\chi} \\[-1ex]
\frac{4}{\sqrt{2}} \sin\theta e^{-i\chi} & 
- 2 \sin^2\theta e^{-2i\chi} & 
\frac{2}{\sqrt{2}} \sin 2\theta e^{-i\chi} & 
2\sin^2\theta\nn
\end{array} \right),
\ena 
where we have introduced the velocity-type parameter $v\equiv 1- m_\ell^2/q^2$ and the helicity-flip factor $\delta_\ell\equiv m_\ell^2/2q^2$. The upper/lower signs in the nonflip part of Eq.~(\ref{eq:lt1}) stand for the two configurations $(\ell^-\bar\nu_\ell)/(\ell^+\nu_\ell)$.

In order to obtain the polar angle distribution, one integrates Eq.~(\ref{eq:lt1}) over the azimuthal angle 
$\chi$ as follows: $L(m,n)(\theta)= \int d(\chi/2\pi) L(m,n)(\theta,\chi)$. The integration yields
\bea
\label{lt2}
\lefteqn{(2q^2v)^{-1} L(m,n)(\theta)=}\\
&=&
\left( \begin{array}{cccc} 
0 & 0 & 0 & 0 \\[-1ex]
0 & (1\mp\cos\theta)^2 &0 & 0 \\[-1ex]
0 & 0 & 2\sin^2\theta 
& 0 \\[-1ex]
0 & 0 &0 & (1\pm\cos\theta)^2 \\[-1ex]
\end{array} \right)
+\delta_\ell 
\left( \begin{array}{cccc} 
4 & 0 & 4 \cos\theta &
 0 \\[-1ex]
0 & 
2 \sin^2\theta & 
 & 0 \\[-1ex]
4\cos\theta & 
0 & 
4\cos^2\theta & 
0 \\[-1ex]
0 & 0 & 
0 & 
2\sin^2\theta \nn
\end{array} \right).
\ena

Finally, the differential decay distribution over $q^2$ and $\cos\theta$ reads
\bea
\frac{d\Gamma(D_{(s)}\to X \ell^+\nu_\ell)}{dq^2d(\cos\theta)} &=&
\frac{G_F^2 |V_{cq}|^2 |{\bf p_2}| q^2 v^2}{32 (2\pi)^3 M_1^2}
\Big[
(1+\cos^2\theta){\cal H}_U + 2\sin^2\theta{\cal H}_L 
+2\cos\theta{\cal H}_P
\nn
&&+ 2\,\delta_\ell \left( \sin^2\theta{\cal H}_U 
+ 2\cos^2\theta{\cal H}_L
+ 2{\cal H}_S - 4\cos\theta{\cal H}_{SL} \right)
\Big],
\label{eq:distr2}
\ena
where $\mathcal{H}_i$'s are bilinear combinations of the helicity amplitudes whose definitions are given in Table~\ref{tab:bilinears}. Note the zero-recoil relations
$2{\cal H}_U={\cal H}_L={\cal H}_T={\cal H}_I$ and 
${\cal H}_P={\cal H}_A={\cal H}_S={\cal H}_{SA}={\cal H}_{ST}={\cal H}_{S}=0$.
Similar relations hold for the imaginary parts. At maximum recoil, the dominating helicity structure functions are ${\cal H}_L$, ${\cal H}_S$, and ${\cal H}_{SL}$, and ${\cal H}_L(0)={\cal H}_S(0)={\cal H}_{SL}(0)$. 
\begin{table}[htbp] 
\begin{center}
\caption{ Definition of helicity structure functions and their parity 
properties. The indices stand for Unpolarized-transverse ($U$), Parity-odd ($P$), Transverse-interference ($T$), Longitudinal ($L$),  transverse-longitudinal Interference ($I$), Scalar ($S$), Scalar-Transverse interference ($ST$), Scalar-Longitudinal interference ($SL$), parity-Asymmetric ($A$), and Scalar-Asymmetric interference ($SA$).}
\def\arraystretch{0.8}
\begin{ruledtabular}
\begin{tabular}{llll} 
Parity-conserving  &   Parity-violating  \\
\hline
${\cal H}_U   = |H_{+}|^2 + |H_{-}|^2$   &  
${\cal H}_P   = |H_{+}|^2 - |H_{-}|^2$   \\
${\cal H}_L    = |H_{0}|^2 $   & 
${\cal H}_{A} = \tfrac 12   {\rm Re}(  H_{+}  H_{0}^\dagger 
                         - H_{-} H_{0}^\dagger )$  
\\
${\cal H}_{T} ={\rm Re}( H_{+}H_{-}^{\dagger})$
 &  
${\cal H}_{IA} = \tfrac 12   {\rm Im}(  H_{+}  H_{0}^\dagger 
                         - H_{-} H_{0}^\dagger )$
\\ 
${\cal H}_{IT} ={\rm Im}( H_{+}H_{-}^{\dagger})$
&  
${\cal H}_{SA} = \tfrac 12 \, {\rm Re}(  H_{+}  H_{t}^\dagger 
                         - H_{-} H_{t}^\dagger )$
\\ 
${\cal H}_{I}  =  \tfrac 12 \, {\rm Re}(  H_{+}  H_{0}^\dagger 
                         + H_{-} H_{0}^\dagger )$ 
 &  
${\cal H}_{ISA} = \tfrac 12 \, {\rm Im}(  H_{+}  H_{t}^\dagger 
                         - H_{-} H_{t}^\dagger )$
\\
${\cal H}_{II}  =  \tfrac 12 \, {\rm Im}(  H_{+}  H_{0}^\dagger 
                         + H_{-} H_{0}^\dagger )$ 
  &  
\\
${\cal H}_S    = |H_{t}|^2$   &  
\\
 ${\cal H}_{ST} =  \tfrac 12 \,{\rm Re}(  H_{+}  H_{t}^\dagger 
                         + H_{-} H_{t}^\dagger )$
  & 
\\ 
${\cal H}_{IST} =  \tfrac 12 \,{\rm Im} ( H_{+}  H_{t}^\dagger 
                         + H_{-} H_{t}^\dagger) $
 &  
\\ 
${\cal H}_{SL} =  {\rm Re}(  H_{0} H_{t}^\dagger ) $
 &  
 \\
${\cal H}_{ISL} =  {\rm Im}(  H_{0} H_{t}^\dagger ) $
  &  
 \\
${\cal H}_{\rm tot} =  {\cal H}_U+{\cal H}_L+\delta_{\ell}(
{\cal H}_U+{\cal H}_L+3{\cal H}_S ) $
 &  \\
\end{tabular}
\end{ruledtabular}
\label{tab:bilinears}
\end{center}
\end{table}

Integrating Eq.~(\ref{eq:distr2}) over $\cos\theta$, one obtains the differential decay distribution over $q^2$
\be
\frac{d\Gamma(D_{(s)}\to X \ell^+\nu_\ell)}{dq^2} =
\frac{G_F^2 |V_{cq}|^2 |{\bf p_2}| q^2 v^2}{12 (2\pi)^3 M_1^2}
\Big[ {\cal H}_U+{\cal H}_L+\delta_{\ell}
({\cal H}_U+{\cal H}_L+3{\cal H}_S) \Big].
\label{eq:distr1}
\en

\section{Fourfold distribution and physical observables}
\label{sec:4fold}
The tensor contraction $L_{\mu\nu}H^{\mu\nu}$ reveals more fruitful structures and physical properties when one considers the cascade decays 
$D_{(s)}\to V(\to P_1P_2) \ell^+\nu_\ell$. In particular, one can study the polarization 
of the $V$ meson. As an example, we consider the decay $D^0\to K^{\ast\,-}(\to K^0\pi^-) \ell^+\nu_\ell$, the kinematics of which is depicted in Fig.~\ref{fig:4fold}. 

In the narrow-width approximation, the hadronic tensor reads
\be
H_{\mu\nu}=T_{\mu\alpha}(T_{\nu\beta})^\dagger
\frac{3}{2\,|{\bf p_3}|}\mathcal{B}(K^\ast\to K^0\pi)p_{3\alpha'}p_{3\beta'}
S^{\alpha\alpha'}(p_2)S^{\beta\beta'}(p_2)\,,
\label{eq:had-ten-4}
\en
where 
$S^{\alpha\alpha'}(p_2)=-g^{\alpha\alpha'}+p_2^\alpha 
p_2^{\alpha'}/M_2^2$ 
is the standard spin-1 tensor,
and  $p_3$ and $p_4$
are the momenta of the $K^0$ and $\pi$ mesons, respectively. One has $p_2=p_3+p_4$, $p_3^2=M_3^2$, $p_4^2=M_4^2$, and 
$ |{\bf p_3}| = \lambda^{1/2}(M_2^2,M_3^2,M_4^2)/(2M_2)$ is the three-momentum of the $K^0$ meson in the $K^\ast$ rest frame.

One needs explicit representations for the momenta and polarization vectors in the $K^\ast$ rest frame. These vectors are given by
\be
\begin{array}{ll}
p^\mu_2 = (M_2,0,0,0), \qquad & \qquad 
\epsilon^\mu_2(+) = \frac{1}{\sqrt{2}}(0,+1,-i,0),\\
p^\mu_3 = (E_3 , +|{\bf p_3}|\sin\theta^\ast , 0 , -|{\bf p_3}|\cos\theta^\ast),\qquad & \qquad
\epsilon^\mu_2(-) = \frac{1}{\sqrt{2}}(0,-1,-i,0),
\\
p^\mu_4 = (E_4 , -|{\bf p_3}|\sin\theta^\ast , 0 ,+|{\bf p_3}|\cos\theta^\ast), \qquad & \qquad
\epsilon^\mu_2(0) = (0,0,0,-1),
\end{array}
\label{eq:Kframe}
\en
where we have set the azimuthal angle $\chi^\ast$
of the $(K^0,\pi)$-plane to zero without loss of generality. 

The spin 1 tensor $S^{\alpha\alpha'}(p_2)$ is then written as 
\be
S^{\alpha\alpha'}(p_2)=-g^{\alpha\alpha'}
+\frac{p_2^\alpha p_2^{\alpha'}}{M_2^2}
=\sum\limits_{m=\pm,0}\epsilon_2^\alpha(m)
\epsilon_2^{\ast\alpha'}(m) \, . 
\label{eq:S=1-tensor}
\en
Using the same technique as in Eq.~(\ref{eq:contraction}) one can rewrite
the contraction of the leptonic and hadronic tensors in terms of the helicity components. 

Finally, one obtains the fourfold decay distribution as follows:
\be
\frac{d\Gamma(D^0\to K^{\ast-}(\to K^0\pi^-) \ell^+\nu_\ell)}
     {dq^2\,d\cos\theta\,d(\chi/2\pi)\,d\cos\theta^\ast} 
=
\frac{ G_F^2|V_{cs}|^2 |{\bf p_2}| q^2 v^2}{12 (2\pi)^3 M_1^2}\,
{\cal B}(K^\ast\to K^0\pi)\,W(\theta^\ast,\theta,\chi),
\label{eq:four-width}
\en
where
\bea
\label{eq:4angular}
W(\theta^\ast,\theta,\chi) &=&
  \frac{9}{32}(1+\cos^2\theta)\sin^2\theta^\ast{\cal H}_U
+ \frac{9}{8}\sin^2\theta\,\cos^2\theta^\ast{\cal H}_L
\mp \frac{9}{16}\cos\theta\,\sin^2\theta^\ast{\cal H}_P
\nn
&&-\frac{9}{16}\sin^2\theta\sin^2\theta^\ast\cos 2\chi {\cal H}_T
    \mp\frac{9}{8} \sin\theta\sin 2\theta^\ast\cos\chi {\cal H}_A
\nn
&&
    +\frac{9}{16}\sin 2\theta\sin 2\theta^\ast\cos\chi {\cal H}_I
    \pm\frac{9}{8}\sin\theta\sin 2\theta^\ast\sin\chi{\cal H}_{II}
\nn
&&
    -\frac{9}{16}\sin 2\theta\sin 2\theta^\ast\sin\chi {\cal H}_{IA}
    +\frac{9}{16}\sin^2\theta\sin^2\theta^\ast\sin 2\chi
{\cal H}_{IT}
\nn
&+&\delta_\ell\Big[
 \frac{9}{4}\cos^2\theta^\ast{\cal H}_S
    -\frac{9}{2}\cos\theta\cos^2\theta^\ast{\cal H}_{SL} 
    +\frac{9}{4}\cos^2\theta\cos^2\theta^\ast{\cal H}_{L}
\\
&&
\qquad    +\frac{9}{16}\sin^2\theta\sin^2\theta^\ast{\cal H}_{U} 
    +\,\frac{9}{8}\sin^2\theta\sin^2\theta^\ast\cos 2\chi {\cal H}_{T}
\nn
&&\qquad 
    +\frac{9}{4}\sin\theta\sin 2\theta^\ast\cos\chi {\cal H}_{ST}
    -\frac{9}{8}\sin 2\theta\sin 2\theta^\ast\cos\chi {\cal H}_{I}
\nn
&&\qquad 
    -\frac{9}{4}\sin\theta\sin 2\theta^\ast\sin\chi{\cal H}_{ISA}
    +\frac{9}{8}\sin 2\theta\sin 2\theta^\ast\sin\chi {\cal H}_{IA}
\nn
&&\qquad 
    -\frac{9}{8}\sin^2\theta\sin^2\theta^\ast\sin 2\chi {\cal H}_{IT}
\Big].\nonumber
\ena
The upper/lower signs in the nonflip part of Eq.~(\ref{eq:4angular}) correspond to the two configurations $(\ell^-\bar\nu_\ell)/(\ell^+\nu_\ell)$. In this case, one uses the lower signs.
In these decay, $CP$ symmetry is conserved and there are no final-state strong interaction effects. As a result, one may assume that all helicity amplitudes are real, which implies the vanishing of all terms proportional to $\sin\chi$ and $\sin2\chi$. The derivation of the fourfold distribution can also be done by using the Wigner $d$-function (see e.g.,~\cite{Gutsche:2015mxa}).

The fourfold distribution allows one to define a number of physical
observables which can be measured experimentally. We first define a normalized angular decay distribution 
$\widetilde W(\theta^\ast,\theta,\chi)$ through
\be
\widetilde W(\theta^\ast,\theta,\chi)=\frac{W(\theta^\ast,\theta,\chi)}
{ {\cal H}_{\rm tot}}.
\label{eq:normdis}
\en
\noindent 
Integration of $\widetilde W(\theta^\ast,\theta,\chi)$ over
$\cos\theta^\ast,\,\cos\theta$, and $\chi/2\pi$ gives $1$.

By integrating Eq.~(\ref{eq:4angular}) over $\cos\theta^\ast$ and 
$\chi$ one obtains the twofold ($q^2,\cos\theta$) distribution given in~Eq.~(\ref{eq:distr2}). The normalized $\theta$ distribution is described by a 
tilted parabola
\be
\widetilde W(\theta)=\frac{a+b\cos\theta+c\cos^{2}\theta}{2(a+c/3)}.
\label{eq:theta-distr}
\en
The linear coefficient $b/2(a+c/3)$ in Eq.~(\ref{eq:theta-distr}) can be extracted by defining a 
forward-backward asymmetry as follows:
\bea
\mathcal{A}_{FB}(q^2) = 
\frac{d\Gamma(F)-d\Gamma(B)}{d\Gamma(F)+d\Gamma(B)}
&=&
\frac{ \int_{0}^{1} d\!\cos\theta\, d\Gamma/d\!\cos\theta
      -\int_{-1}^{0} d\!\cos\theta\, d\Gamma/d\!\cos\theta }
     { \int_{0}^{1} d\!\cos\theta\, d\Gamma/d\!\cos\theta
      +\int_{-1}^{0} d\!\cos\theta\, d\Gamma/d\!\cos\theta} 
\nn
&=& \frac{b}{2(a+c/3)}=\frac34 \frac{{\cal H}_P
-4\delta_\ell{\cal H}_{SL}}{{\cal H}_{\rm tot}}.
\label{eq:fbAsym}
\ena
The quadratic coefficient $c/2(a+c/3)$ in Eq.~(\ref{eq:theta-distr}) is extracted by taking the second derivative of $\widetilde W(\theta)$. We therefore define a lepton-side convexity parameter by
\be
C_F^\ell(q^2) = \frac{d^{2}\widetilde W(\theta)}{d(\cos\theta)^{2}}
= \frac{c}{a+c/3} 
= \frac34 (1-2\delta_\ell)
\frac{ {\cal H}_U - 2 {\cal H}_L }{ {\cal H}_{\rm tot} }.
\label{eq:convex_lep}
\en 

It is worth noting that in the $D_{(s)}\to V$ decays, the forward-backward asymmetry receives contributions from a purely parity-violating source originated by the $VA$ interaction
(results in ${\cal H}_P$), and a parity-conserving source by
the $VV$ and $AA$ interactions (results in ${\cal H}_{SL}$).
The parity-conserving parity-odd contribution ${\cal H}_{SL}$ arises from the 
interference of the $(0^{+};1^{-})$ and $(0^{-};1^{+})$ components of the 
$VV$ and $AA$ current products, respectively. In the case of $D_{(s)}\to P$,
 the forward-backward asymmetry arises solely from the
$(0^{+};1^{-})$ interference term of the $VV$ current product.

By integrating the fourfold distribution over $\cos\theta$ and 
$\chi$ one obtains the angular distribution over the angle $\theta^\ast$ whose normalized form is described by an untilted parabola (without a linear term)
\be
\widetilde {W} (\theta^\ast)=\frac{a'+c'\cos^{2}\theta^\ast}{2(a'+c'/3)}.
\en
The $\cos\theta^\ast$ distribution can therefore be characterized by a hadron-side convexity parameter defined by
\be
C_F^h(q^2) = \frac{d^{2}\widetilde W(\theta^\ast)}{d(\cos\theta^{\ast})^{2}}
=\frac{c'}{a'+c'/3}=-\frac32 
\frac{ {\cal H}_U - 2 {\cal H}_L 
      +\delta_\ell( {\cal H}_U - 2 {\cal H}_L -6{\cal H}_S )}{ 
{\cal H}_{\rm tot} }.
\label{eq:convex_had}
\en

There is another way to extract information from the $\cos\theta^\ast$ distribution. To do this one writes the distribution as
\be
\widetilde W(\theta^\ast)=\frac34\left(2F_L(q^2)\cos^2\theta^\ast+F_T(q^2)\sin^2\theta^\ast\right),
\en
where $F_L(q^2)$ and $F_T(q^2)$ are the longitudinal and transverse polarization fractions of the $K^\ast$ meson, and are given by
\be
F_L(q^2)=\frac{\mathcal{H}_L+\delta_\ell(\mathcal{H}_L+3\mathcal{H}_S)}{\mathcal{H}_{\rm tot}},\qquad F_T(q^2)=\frac{(1+\delta_\ell)\mathcal{H}_U}{\mathcal{H}_{\rm tot}},\qquad F_L(q^2)+F_T(q^2)=1.
\en
\noindent
The hadron-side convexity parameter and the polarization fractions of the $K^\ast$ meson are related by
\be 
C_F^h(q^2)=\frac32 \left( 2F_L(q^2)-F_T(q^2) \right)=\frac32 \left(3F_L(q^2)-1\right).
\en

The remaining coefficient functions ${\cal H}_{T}(1 - 2\delta_\ell)$,
${\cal H}_{I}(1 - 2\delta_\ell)$, and 
$ ({\cal H}_{A} + 2 \delta_\ell{\cal H}_{ST})$ in Eq.(\ref{eq:4angular}) can be 
projected out by calculating the
appropriate trigonometric moments of the normalized decay distribution
$\widetilde W(\theta^\ast,\theta,\chi)$. The trigonometric moments are
written as
\be
W_{i} = \int d\!\cos\theta\,d\!\cos\theta^\ast d(\chi/2\pi)
M_{i}(\theta^\ast,\theta,\chi)\widetilde W(\theta^\ast,\theta,\chi) 
\equiv  \left\langle M_{i}(\theta^\ast,\theta,\chi)\right\rangle,
\en
where $M_{i}(\theta^\ast,\theta,\chi)$ defines the trigonometric moment that 
is being taken.
One finds 
\bea
W_T(q^2) &\equiv& \left\langle  \cos 2\chi \right\rangle
= -\frac{1 - 2\delta_\ell}{2} \frac{{\cal H}_{T}}{{\cal H}_{\rm tot}},
\nn
W_I(q^2) &\equiv& \left\langle
\cos\theta\cos\theta^{\ast}\cos \chi \right\rangle
= \frac{9\pi^2(1 - 2\delta_\ell)}{512}
 \frac{{\cal H}_{I}}{{\cal H}_{\rm tot}},
\\
W_A(q^2) &\equiv&\left\langle\sin\theta\cos\theta^{\ast}\cos \chi \right\rangle  
=  \frac{3\pi}{16}
\frac{ {\cal H}_{A} + 2 \delta_\ell{\cal H}_{ST} }{ {\cal H}_{\rm tot} }.
\nonumber
\label{eq:W}
\ena
The coefficient functions ${\cal H}_{T}(1 - 2\delta_\ell)$,
${\cal H}_{I}(1 - 2\delta_\ell)$, and 
$ ({\cal H}_{A} + 2 \delta_\ell{\cal H}_{ST})$ can also be projected out by
taking piecewise sums and differences of different sectors of the angular 
phase space \cite{Korner:1989qb}.

Finally, we consider the longitudinal and transverse polarizations of the final charged
lepton. One obtains the polarizations of the lepton by leaving the helicities $\lambda_\ell$ of the lepton unsummed. The longitudinal polarization can be obtained directly from the difference between the helicity flip (hf) and nonflip (nf) structures as follows:
\be
P^\ell_L(q^2) =\frac{{\cal H}_{\rm nf}-\delta_{\ell}{\cal H}_{\rm hf}}
{{\cal H}_{\rm nf}+\delta_{\ell}{\cal H}_{\rm hf}}= 
\frac{ {\cal H}_U + {\cal H}_L 
      - \delta_\ell( {\cal H}_U + {\cal H}_L + 3 {\cal H}_S )}
{ {\cal H}_{\rm tot} }.
\label{eq:lpolarization}
\en
The transverse polarization can be calculated using the representation of the
polarized lepton tensor given in the Appendix 
of~\cite{Gutsche:2015mxa}. One obtains
\be
P^\ell_T(q^2) = -\frac{3\pi\sqrt{\delta_\ell}}{4\sqrt{2}}
\frac{ {\cal H}_P + 2 {\cal H}_{SL}} { {\cal H}_{\rm tot} }.
\label{eq:tpolarization}
\en
For the decays $D_{(s)} \to P\ell^{+}\nu_{\ell}$ 
there exists a simple relation between $P^\ell_T(q^2)$ and
$A_{FB}(q^{2})$ which reads 
\be
P^{\ell}_T(q^2) = \frac{\pi\sqrt{q^2}}{2m_\ell} A_{FB}(q^2). 
\en
It should be noted that the the lepton polarization depends on the frame in which it is defined. The longitudinal and transverse polarizations in
(\ref{eq:lpolarization}) and (\ref{eq:tpolarization}) are calculated in the
$W^{\ast +}$ rest frame.

\section{Form factors in the covariant confining quark model}
\label{sec:CCQM}
The theoretical description of the $D_{(s)}$ semileptonic decays that we have provided so far is model independent. Let us now turn to the calculation of the form factors, which depends on the nonperturbative method being applied. From the theoretical point of view, the calculation of hadronic form factors is the most difficult part in understanding semileptonic decays. Among various approaches, lattice QCD (LQCD) provides predictions with high accuracy and well controlled uncertainty. Therefore, LQCD results are often used for the extraction of the CKM matrix elements from experiments. However, since this approach requires extremely large computational sources, most of LQCD studies have been focusing on the key channels $D\to\pi(K)$ (for reviews, see e.g.,~\cite{Aoki:2016frl, Aoki:2019cca}).

Several $D_{(s)}$ semileptonic decay form factors have been calculated using QCD sum rules (QCDSR)~\cite{Ball:1991bs, Ball:1993tp, Colangelo:2001cv, Du:2003ja} and QCD light-cone sum rules (LCSR)~\cite{Khodjamirian:2000ds, Ball:2006yd, Azizi:2010zj, Offen:2013nma, Duplancic:2015zna, Fu:2018yin}. Most of these studies focused on the channels $D\to\pi(K)$ and $D_{(s)}\to\eta^{(\prime)}$. In~\cite{Wu:2006rd}, the authors used LCSR in the context of heavy quark effective theory to calculate the form factors of $D_{(s)}$ semileptonic transitions  to $\pi$, $K^{(\ast)}$, $\eta$, $\rho$, $\omega$, and $\phi$. In the sum rules approach, the form factors are calculated for a limited kinematic region at small $q^2$. The values of the form factors at high $q^2$ are obtained by doing extrapolation, assuming a certain form factor behavior. This makes the predictions for the high-$q^2$ region less reliable, especially when the available kinematical range is large.

Semileptonic $D_{(s)}$ form factors have been studied extensively in the framework of various phenomenological quark models. We mention here the Isgur-Scora-Grinstein-Wise (ISGW) model~\cite{Isgur:1988gb} and its updated version ISGW2~\cite{Scora:1995ty}, the relativistic quark model using a quasipotential approach~\cite{Faustov:1996xe}, the chiral quark model~\cite{Palmer:2013yia}, the constituent quark model (CQM)~\cite{Melikhov:2000yu}, the model combining heavy meson and chiral symmetries (HM$\chi$T)~\cite{Fajfer:2004mv, Fajfer:2005ug}, and the light-front quark model (LFQM)~\cite{Wei:2009nc, Verma:2011yw, Cheng:2017pcq}. Several semileptonic $D_{(s)}$ decays were also studied in the large energy effective theory~\cite{Charles:1998dr}, chiral perturbation theory~\cite{Bijnens:2010jg}, the so-called chiral unitary approach ($\chi$UA)~\cite{Sekihara:2015iha}, and a new approach assuming pure heavy quark symmetry~\cite{Dai:2018vzz}. Recently, a simple expression for $D\to K$ semileptonic form factors was studied in~\cite{Pham:2018bxs}. We also mention here early attempts to account for flavor symmetry breaking in pseudoscalar meson decay constants by the authors of~\cite{Gershtein:1976aq, Khlopov:1978id}. It is worth noting that each method has only a limited range of applicability, and their combination will give a better picture of the underlined physics~\cite{Melikhov:2000yu}. 

An alternative quark-model approach to the study of semileptonic $D_{(s)}$ form factors has been carried out recently by us in the framework of the covariant confining quark model~\cite{Soni:2017eug, Soni:2018adu}. The CCQM is an effective quantum field approach to the calculation of hadronic transitions~\cite{Efimov:1988yd, Efimov:1993zg, Branz:2009cd}. The key assumption is that hadrons interact via the exchange of constituent quarks. This is realized by using a relativistic invariant Lagrangian describing the coupling of a hadron to its constituent quarks, the coupling strength of which is determined by the so-called compositeness condition $Z_H=0$~\cite{Salam:1962ap, Weinberg:1962hj},
where $Z_H$ is the wave function renormalization constant of the hadron $H$. To better understand the physical meaning of this condition, one should note that $Z_H^{1/2}$ is the matrix element between a physical bound state and the corresponding bare state. The  $Z_H=0$ condition ensures the absence of any bare state in the physical state and, therefore, provides an effective description of a bound state. It also helps avoid double counting of hadronic degrees of freedom.

Starting with the effective Lagrangian written in terms of constituent quark and hadron quantum fields, one uses Feynman rules to evaluate quark diagrams describing the matrix elements of hadronic transitions. This approach is self-consistent and the calculation of form factors is straightforward. The model requires a small number of free parameters including the constituent quark masses, the effective size parameters of hadrons, and a universal infrared cutoff parameter. Notably, in the CCQM, all form factors are obtained in the whole kinematical range without using any extrapolation. Also, the model allows the study of multiquark states in a consistent manner. The CCQM has been successfully applied to a large number of studies involving mesons~\cite{Ivanov:1999ic, Ivanov:2006ni, Ivanov:2015woa, Dubnicka:2018gqg, Ivanov:2011aa}, baryons~\cite{Gutsche:2013pp, Gutsche:2017hux, Gutsche:2018nks, Gutsche:2012ze}, and multiquark states~\cite{Dubnicka:2011mm, Goerke:2016hxf, Goerke:2017svb, Gutsche:2017twh}. More detailed descriptions of the CCQM can be found in these references. In what follows we just mention some main features of the model for completeness.

For a meson $M$, the effective Lagrangian describing the quark-hadron interaction is written as
\begin{dmath}\label{eq:int_lagrange}
\mathcal{L}_{\rm int}  =  g_M M(x) \!\! \int \!\! dx_1 dx_2 F_M(x;x_1,x_2) \bar{q}_2(x_2) \Gamma_M q_1(x_1) +  {\rm H.c.},
\end{dmath}
where $g_M$ is the coupling strength which is determined by the compositeness condition mentioned above, and $\Gamma_M$ is the Dirac matrix containing the meson's quantum numbers:  $\Gamma_M = \gamma_5$ for pseudoscalar mesons, and $\Gamma_M = \gamma_\mu$ for vector mesons.  Here, $F_M(x,x_1,x_2)$ is the quark-hadron vertex function which effectively describes the quark distribution inside the meson and is given by
\be
\label{eq:vertex_function}
F_M(x,x_1,x_2) = \delta^{(4)} \Big(x - \sum_{i=1}^2 w_i x_i \Big)\cdot\Phi_M \big((x_1 - x_2)^2\big),
\en
where $w_{q_i} = m_{q_i}/ (m_{q_1} + m_{q_2})$ such that $w_1 + w_2 = 1$. The function $\Phi_M$ 
must have an appropriate falloff behavior in the Euclidean region to guarantee the absense of ultraviolet divergences in the quark loop integrals. In  our previous studies, we have pointed out that the predictions for physical observables are insensitive to the specific form of $\Phi_M$. We therefore assume the following Gaussian form for $\Phi_M$ in the momentum space for simplicity:
\begin{equation} \label{eq:gaussian}
\widetilde{\Phi}_M(-p^2) = \!\! \int \!\! dx e^{ipx}\Phi_M(x^2)=e^{p^2/\Lambda_M^2},
\end{equation}
where $\Lambda_M$ is a model parameter characterizing the finite size of the meson. 

The form factors are calculated by evaluating the one-loop Feynman diagram shown in Fig.~\ref{fig:mass}. In the CCQM, the matrix element of the hadronic transition is written as a convolution of quark propagators and vertex functions as follows:
\begin{figure}[t]
\includegraphics[width=0.65\textwidth]{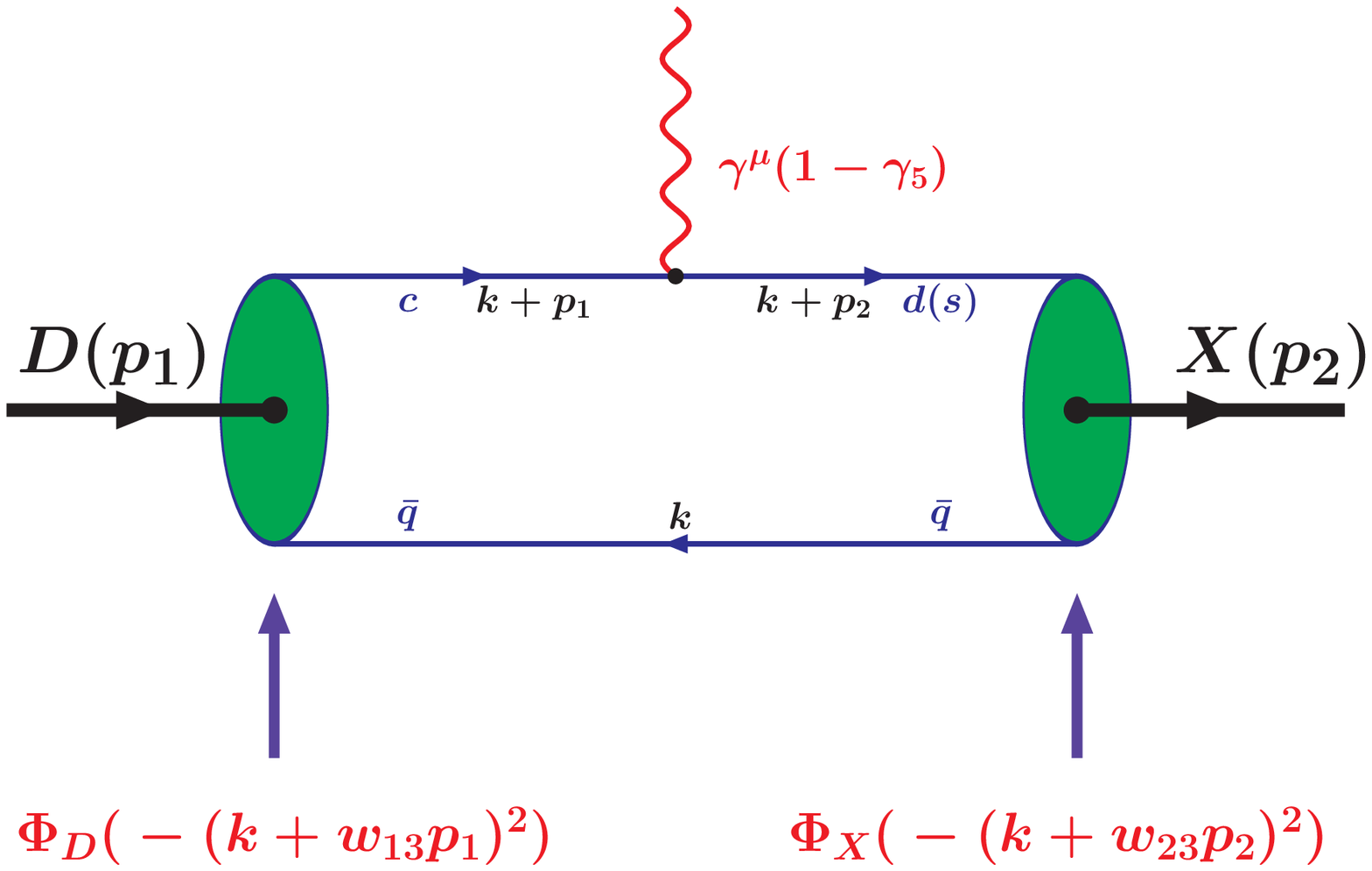}
\vspace*{-5mm}
\caption{Quark model diagram for the $D_{(s)}$-meson semileptonic decay.}
\label{fig:mass}
\end{figure}
\bea
\langle P(p_2)
|\bar{q} O^\mu c
| D_{(s)}(p_1) \rangle
&=&
N_c g_{D_{(s)}} g_P \!\! \int \!\!\frac{d^4k}{ (2\pi)^4 i} 
\widetilde\Phi_{D_{(s)}}\!\! \left(-(k+w_{13} p_1)^2\right)
\widetilde\Phi_P\left(-(k+w_{23} p_2)^2\right)
\nn
&&\times
\Tr \big[
O^{\mu} S_1(k+p_1) \gamma^5 S_3(k) \gamma^5 S_2(k+p_2) 
\big],
\\
\label{eq:PP'}
\langle V(p_2,\epsilon_2)
|\bar{q} O^\mu c
| D_{(s)}(p_1) \rangle
&=&
N_c g_{D_{(s)}} g_V \!\! \int\!\! \frac{d^4k}{ (2\pi)^4 i}
\widetilde\Phi_{D_{(s)}} \!\! \left(-(k+w_{13} p_1)^2\right)
\widetilde\Phi_V\left(-(k+w_{23}p_2)^2\right)
\nn
&&\times
\Tr \big[ 
O^{\mu} S_1(k+p_1)\gamma^5 S_3(k) \not\!\epsilon_2^{\ast} S_2(k+p_2) \big],
\label{eq:PV}
\ena
where $N_c=3$ is the number of colors and $O^\mu$ is a short notation for $\gamma^\mu(1-\gamma_5)$. Here, there are three quarks taking part in the process. We therefore use the double-subscript notation $w_{ij}=m_{q_j}/(m_{q_i}+m_{q_j})$ ($i,j=1,2,3$) such that $w_{ij}+w_{ji}=1$. In Fig.~\ref{fig:mass}, one has $q_1=c$, $q_2=d(s)$, and $q_3=q$. We use the Fock-Schwinger representation for the quark propagators $S_i$, which brings in the integrations over the Schwinger parameters $\alpha_i$:
\be
S_i(k) = (m_{q_i}+\not\! k)
\int\limits_0^\infty\! d\alpha_i \exp[-\alpha_i(m_{q_i}^2-k^2)].
\en
It is worth mentioning that all loop integrations are performed in Euclidean space. One uses the Wick rotation to transform the Minkowski space to the Euclidean space.

Details about the techniques we used for the loop-integration evaluation can be found in our previous papers (see e.g., Refs.~\cite{Branz:2009cd, Ivanov:2015tru}). By applying these techniques, one finally obtains the expression for a form factor $F$ in the form of a threefold integral
\be
F   =N_c\, g_{D_{(s)}} g_{(P,V)} \!\! \int\limits_0^{1/\lambda^2}\!\! dt t
\!\! \int\limits_0^1\!\! d\alpha_1
\!\! \int\limits_0^1\!\! d\alpha_2  
\delta\Big(1 -  \alpha_1-\alpha_2 \Big) 
f(t\alpha_1,t\alpha_2),
\label{eq:3fold}
\en
where $f(t\alpha_1,t\alpha_2)$ is the resulting integrand for the form factor $F$, and $\lambda$ is the infrared cutoff parameter, which is introduced to rule out possible thresholds related to the creation of free quarks. This means the parameter $\lambda$ effectively assures the confinement of constituent quarks inside hadrons. This method is general and can be applied for diagrams with an arbitrary number of loops and propagators. In the CCQM, we take $\lambda$ to be universal for all physical processes.

The CCQM contains several free parameters which cannot be obtained from first-principle calculations. These parameters include the hadron size parameters $\Lambda_M$, the constituent quark masses $m_q$, and the universal infrared cutoff parameter $\lambda$. In order to determine their values, we fit our results for a set of radiative and leptonic decay constants to experimental data or LQCD~\cite{Ivanov:2006ni, Ivanov:2011aa}. The results of the fit for those parameters needed in this paper are listed in Tables~\ref{tab:size_parameter} and~\ref{tab:quark_mass}. We list also the fit result for several leptonic decay constants in Table~\ref{tab:decay-const}. As a byproduct, we include our predictions for the $D_{(s)}$ purely leptonic decays in Table~\ref{tab:leptonic}.
\begin{table*}
\caption{Meson size parameters (in GeV).}\label{tab:size_parameter}
\renewcommand{\arraystretch}{0.7}
\begin{ruledtabular}
\begin{tabular}{ccccccccccc}
$\Lambda_{D}$ & $\Lambda_{D_s}$ & $\Lambda_{K}$ & $\Lambda_{K^{*}}$ & $\Lambda_\phi$ & $\Lambda_\rho$ & $\Lambda_\omega$ & $\Lambda_{\eta}^{q\bar{q}}$ & $\Lambda_{\eta}^{s\bar{s}}$  & $\Lambda_{\eta'}^{q\bar{q}}$ & $\Lambda_{\eta'}^{s\bar{s}}$\\
\hline
1.600 & 1.750 & 1.014 & 0.805 & 0.880 & 0.610 & 0.488 & 0.881 & 1.973 & 0.257 & 2.797
\end{tabular}
\end{ruledtabular}
\end{table*}
\begin{table}
\caption{Quark masses and infrared cutoff parameter (in GeV).}\label{tab:quark_mass}
\renewcommand{\arraystretch}{0.7}
\begin{ruledtabular}
\begin{tabular}{ccccc}
$m_{u/d}$ & $m_s$ &  $m_c$ &  $m_b$ & $\lambda$ \\
\hline
0.241 & 0.428 & 1.672 & 5.05 & 0.181
\end{tabular}
\end{ruledtabular}
\end{table}
\begin{table}
\caption{Leptonic decay constants $f_H$ (in MeV)}
\renewcommand{\arraystretch}{0.7}
\begin{ruledtabular}
\begin{tabular}{ccll|ccll}
$f_H$ & CCQM & Other & Ref. & $f_H$ & CCQM & Other & Ref. \\
\hline
$f_{D_s}$		& 257.7	&	$249.0(1.2)	$ & PDG~\cite{Tanabashi:2018oca} & $f_K$			& 157.0 	& $155.6(0.4)	$		& PDG~\cite{Tanabashi:2018oca} \\				
$f_D$			& 206.1	&	$211.9(1.1)$ & PDG~\cite{Tanabashi:2018oca} 	
& $f_\pi$		& 130.3	& $130.2(1.7)$ 	& PDG~\cite{Tanabashi:2018oca}\\
$f_{D_s}/f_D$& 	1.25		& $1.173(3)$	& PDG~\cite{Tanabashi:2018oca} & $f_K/f_\pi$		& 	1.20		& $1.1928(26)$	& PDG~\cite{Tanabashi:2018oca}\\				
$f_{D^*}$		& 244.3	& $223.5(8.4)$	& LQCD~\cite{Lubicz:2017asp}	& $f_{D_s^*}$ 	& 272.1 & $268.8(6.6)$ 	& LQCD~\cite{Lubicz:2017asp}\\				
$f_{K^*}$		& 226.8 	& $222 \pm 8$	& QCDSR~\cite{Ball:2006eu}	&$f_\rho$		& 218.3 	& $208.5 \pm 5.5 \pm 0.9$	& LQCD~\cite{Sun:2018cdr} \\ 
$f_\phi$			& 226.6	& $238 \pm 3$	& LQCD~\cite{Chakraborty:2017hry} & $f_\omega$	& 198.4 	& $194.60 \pm	3.24$ & LFQM~\cite{Verma:2011yw}
\end{tabular}
\end{ruledtabular}\label{tab:decay-const}
\end{table}
\begin{table}[!htb]
\caption{Leptonic $D_{(s)}^+$ branching fractions}
\renewcommand{\arraystretch}{0.7}
\begin{ruledtabular}
\begin{tabular}{lll|lll}
Channel & CCQM & PDG Data~\cite{Tanabashi:2018oca} & Channel & CCQM & PDG Data~\cite{Tanabashi:2018oca}\\
\hline
$D^+ \to e^+\nu_e$				& $8.42 \times 10^{-9}$ & $< 8.8 \times 10^{-6}$					& $D_s^+ \to e^+\nu_e$			& $1.40 \times 10^{-7}$ 	& $< 8.3 \times 10^{-5}$\\
$D^+ \to \mu^+\nu_\mu$		& $3.57 \times 10^{-4}$ & $(3.74 \pm 0.17) \times 10^{-4}$	& $D_s^+ \to \mu^+\nu_\mu$		& $5.97 \times 10^{-3}$ 	& $(5.50 \pm 0.23) \times 10^{-3}$\\
$D^+ \to \tau^+\nu_\tau$		& $0.95 \times 10^{-3}$ & $< 1.2 \times 10^{-3}$					& $D_s^+ \to \tau^+\nu_\tau$		& $5.82 \ \%$ 				& $(5.48 \pm 0.23) \%$
\end{tabular}
\end{ruledtabular}\label{tab:leptonic}
\end{table}

Once the model parameters are fixed, the calculation of the form factors is straightforward. One can use \textsc{mathematica} or \textsc{fortran} code to evaluate the threefold integral in Eq.~(\ref{eq:3fold}). It is important to note that in the CCQM, the form factors are calculable in the entire range of momentum transfer without using any extrapolation. For convenience in the calculation of other physical observables such as the branching fractions, as well as for easy representation, we choose a double-pole parametrization to interpolate the calculated values of the form factors as follows:
\begin{equation}
\label{eq:double_pole}
F(q^2) = \frac{F(0)}{1-a \hat{s} + b \hat{s}^2},\qquad \hat{s}=\frac{q^2}{m_{D_{(s)}}^2}.
\end{equation}
This chosen parametrization works perfectly for all form factors. The difference between the exact calculated values and the interpolation curves is less than 1\% for any value of $q^2$. 

In Table~\ref{tab:form_factors} we present our results for the parameters $F(0)$, $a$, and $b$ appearing in the parametrization Eq.~(\ref{eq:double_pole}). In what follows we will compare our form factors with other theoretical approaches and available experimental measurements. For easy comparison, we relate all form factors to the BSW definition given in Eq.~(\ref{eq:ff-BSW}). 

\begin{table}[!htb]
\caption{Parameters of the double-pole parametrization 
Eq.~(\ref{eq:double_pole}) for the form factors.} \label{tab:form_factors}
\renewcommand{\arraystretch}{0.7}
\begin{ruledtabular}
\begin{tabular}{lccrlrcc}
$F$ & $F(0)$ & $a$ & $b$ & $F$ & $F(0)$ & $a$ & $b$\\
\hline
$A_+^{D\to\rho}$ 		& 0.57 	& 0.96 	& 0.15		& $A_-^{D\to\rho}$ 		& $-0.74$ 	& 1.11 	& 0.22\\
$A_0^{D\to\rho}$ 		& 1.47 	& 0.47 	& $-0.10$ 	& $V^{D\to\rho}$ 			& 0.76 		& 1.13 	& 0.23\\
$A_+^{D\to\omega}$ 	& 0.55 	& 1.01 	& 0.17 		& $A_-^{D\to\omega}$	& $-0.69$ 	& 1.17 	& 0.26\\
$A_0^{D\to\omega}$ 	& 1.41 	& 0.53 	& $-0.10$ 	& $V^{D\to\omega}$ 		& 0.72 		&1.19 	& 0.27\\
$A_+^{D\to K^*}$	& 0.68	& 0.86 	& 0.09		& $A_-^{D\to K^*}$		& $-0.90$	& 0.96	& 0.14\\
$A_0^{D\to K^*}$	& 2.08	& 0.40	& $-0.10$	& $V^{D\to K^*}$		& 0.90		& 0.97	& 0.13\\
$A_+^{D_s\to\phi}$	& 0.67	& 1.06	& 0.17		& $A_-^{D_s\to\phi}$		& $-0.95$	& 1.20	& 0.26\\
$A_0^{D_s\to\phi}$	& 2.13	& 0.59	& $-0.12$	& $V^{D_s\to\phi}$			& 0.91		& 1.20	& 0.25\\
$A_+^{D_s\to K^*}$	& 0.57	& 1.13 	& 0.21		& $A_-^{D_s\to K^*}$		& $-0.82$	& 1.32	& 0.34\\
$A_0^{D_s\to K^*}$	& 1.53	& 0.61	& $-0.11$	& $V^{D_s\to K^*}$		& 0.80		& 1.32	& 0.33\\
$F_+^{D \to \pi}$		& 0.63 	& 0.86	& 0.09		& $F_-^{D \to \pi}$			& $-0.41$	& 0.93	& 0.13\\
$F_+^{D \to K}$			& 0.77 	& 0.73	& 0.05		& $F_-^{D \to K}$			& $-0.39$	& 0.78	& 0.07\\
$F_+^{D \to \eta}$		& 0.36 	& 0.93	& 0.12		& $F_-^{D \to \eta}$		& $-0.20$	& 1.02	& 0.18\\
$F_+^{D \to \eta'}$		& 0.36	& 1.23	& 0.23		& $F_-^{D \to \eta'}$		& $-0.03$& 2.29	& 1.71\\
$F_+^{D \to D^0}$		& 0.91	& 5.88	& 4.40		& $F_-^{D \to D^0}$		& $-0.026$& 6.32	& 8.37\\
$|F_+^{D_s \to \eta}|$	& 0.49	& 0.69	& 0.002		& $F_-^{D_s \to \eta}$	& $+0.26$	& 0.74	& 0.008\\
$F_+^{D_s \to \eta'}$	& 0.59	& 0.88	& 0.018		& $F_-^{D_s \to \eta'}$	& $-0.23$	& 0.92	& 0.009\\
$F_+^{D_s\to K}$		& 0.60 	& 1.05	& 0.18		& $F_-^{D_s\to K}$ 		& $-0.38$	& 1.14	& 0.24\\
$F_+^{D_s \to D^0}$	& 0.92	& 5.08	& 2.25		& $F_-^{D_s \to D^0}$ 	& $-0.34$	& 6.79	& 8.91
\end{tabular}
\end{ruledtabular}
\end{table}

As a common drawback of phenomenological quark models, it is difficult to rigorously quantify the theoretical uncertainty of predictions in the CCQM. In order to estimate the uncertainties of our form factors, we assume that the main source of uncertainty is the errors of the model parameters obtained when fitting. These parameters are determined from a least-squares fit to experimental data and some lattice results.
The errors of the fitted parameters are allowed to be within 10$\%$. We then calculate the propagation of these errors to the form factors and found the uncertainties on the form factors to be of order 10\%--15\% at $q^2=0$, and 20\%--30\% at $q^2_{\rm max}$. It should be kept in mind, however, that this is only a rough estimate.

Finally, it should be mentioned that in this paper, all physical quantities such as the mass and lifetime of mesons and leptons, the CKM matrix elements, and other constants are taken from the recent report of the PDG~\cite{Tanabashi:2018oca}. In particular, the values $|V_{cd}|=0.218$ and $|V_{cs}|=0.997$ were used in our calculation of the branching fractions. 
\section{Results and Discussion}
\label{sec:results}
\subsection{$D^0 \to (\pi^-,K^-) \ell^+ \nu_\ell$ and $D^+ \to (\pi^0,K^0) \ell^+ \nu_\ell$}
\label{subsec:D-pi-K}

The decays $D \to (\pi,K) \ell^+ \nu_\ell$ have been studied extensively, both theoretically and experimentally, as the key channels for the determination of the CKM matrix elements $|V_{cd}|$ and $|V_{cs}|$, as well as for the study of heavy-to-light semileptonic form factors. In particular, the normalization and the shape of the form factors $F_+^{D\to \pi}$ and $F_+^{D\to K}$ have been studied experimentally by FOCUS~\cite{Link:2004dh}, Belle~\cite{Widhalm:2006wz}, CLEO~\cite{Besson:2009uv}, {\it BABAR}~\cite{Aubert:2007wg, Lees:2014ihu}, and BESIII~\cite{Ablikim:2015ixa, Ablikim:2017lks}. 

\begin{figure*}[!htbp]
\renewcommand{\arraystretch}{0.3}
\begin{tabular}{cc}
\includegraphics[width=0.45\textwidth]{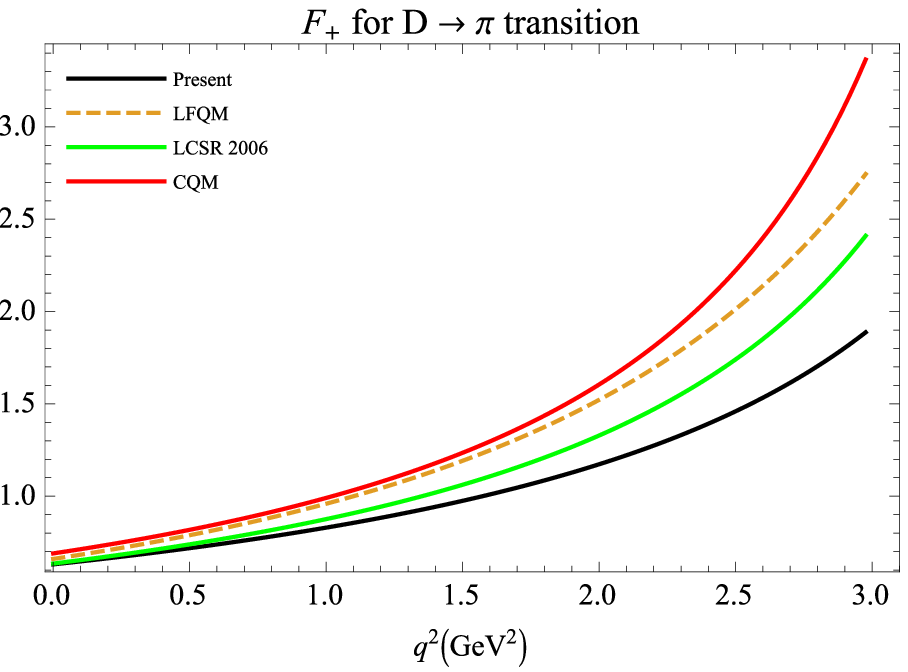}&
\hfill\includegraphics[width=0.45\textwidth]{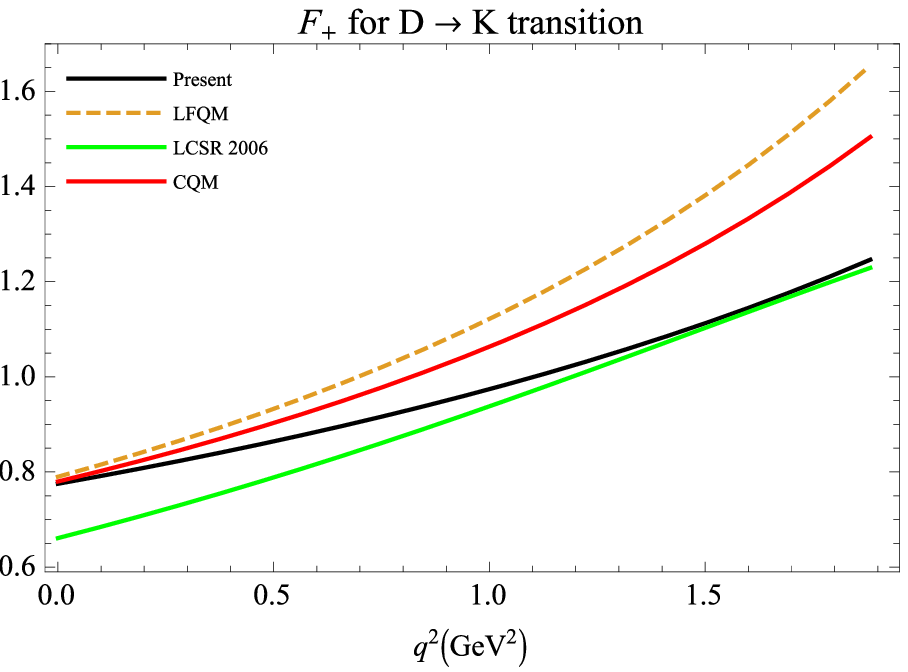}
\end{tabular}
\vspace*{-3mm}
\caption{Form factor $F_+(q^2)$ for $D \to (\pi, K)$ in our model, LCSR~\cite{Wu:2006rd}, LFQM~\cite{Verma:2011yw}, and CQM~\cite{Melikhov:2000yu}.}
\label{fig:D-pi-K}
\end{figure*}

In Table~\ref{tab:ff-D-pi-K} we compare the form factors $F_+^{D\to \pi}(0)$ and  $F_+^{D\to K}(0)$ with those obtained in the LCSR, LFQM, and CQM, and with published LQCD results. One sees that the predictions are very consistent. In order to have a better picture of the form factors in the whole $q^2$ range, we plot their $q^2$ dependence in Fig.~\ref{fig:D-pi-K}, which shows very good agreement between different approaches in the small-$q^2$ region.

\begin{table}[!htbp]
\caption{Comparison of $F_+(0)$ for $D \to (\pi,K)$ transitions.}
\renewcommand{\arraystretch}{0.7}
\begin{ruledtabular}
\begin{tabular}{lccllll}
& CCQM & Other & Ref. & LQCD & Experiment & Ref.\\
\hline
$D\to\pi$ & $0.63\pm 0.09$ & $0.635^{+0.060}_{-0.057}$ & LCSR~\cite{Wu:2006rd} & $0.612(35)$~\cite{Lubicz:2017syv} & $0.637\pm 0.009$ & BESIII~\cite{Ablikim:2015ixa}\\
{} & {} & 0.66 & LFQM~\cite{Verma:2011yw} & $0.666(29)$~\cite{Na:2011mc}\\
{} & {} & 0.69 & CQM~\cite{Melikhov:2000yu} & $0.64(3)(6)$~\cite{Aubin:2004ej}\\
$D\to K$ & $0.77\pm 0.11$ & $0.661^{+0.067}_{-0.066}$ & LCSR~\cite{Wu:2006rd} & $0.765(31)$~\cite{Lubicz:2017syv} & $0.737\pm 0.004$ & BESIII~\cite{Ablikim:2015ixa}\\
{} & {} & 0.79 & LFQM~\cite{Verma:2011yw} & $0.747(19)$~\cite{Na:2010uf}& $0.727 \pm 0.011$ & {\it BABAR}~\cite{Aubert:2007wg}\\
{} & {} & 0.78 & CQM~\cite{Melikhov:2000yu} & $0.73(3)(7)$~\cite{Aubin:2004ej}
\end{tabular}
\end{ruledtabular}
\label{tab:ff-D-pi-K}
\end{table}

A recent lattice calculation of the $D\to \pi (K)\ell\nu$ form factors was done by the ETM collaboration~\cite{Lubicz:2017syv, Lubicz:2018rfs}. In this work, they also considered an additional tensor form factor associated with the tensor four-fermion operator which may appear beyond the SM. The tensor form factor is defined by~\cite{Ivanov:2017hun, Tran:2018kuv}
\be
\langle P(p_2)|\bar{q}\sigma^{\mu\nu}(1-\gamma^5)c|D(p_1)\rangle 
=\frac{iF^T(q^2)}{M_1+M_2}\left(P^\mu q^\nu - P^\nu q^\mu 
+i \varepsilon^{\mu\nu Pq}\right).
\en
\begin{figure*}[htbp]
\renewcommand{\arraystretch}{0.3}
\begin{tabular}{cc}
\includegraphics[width=0.48\textwidth]{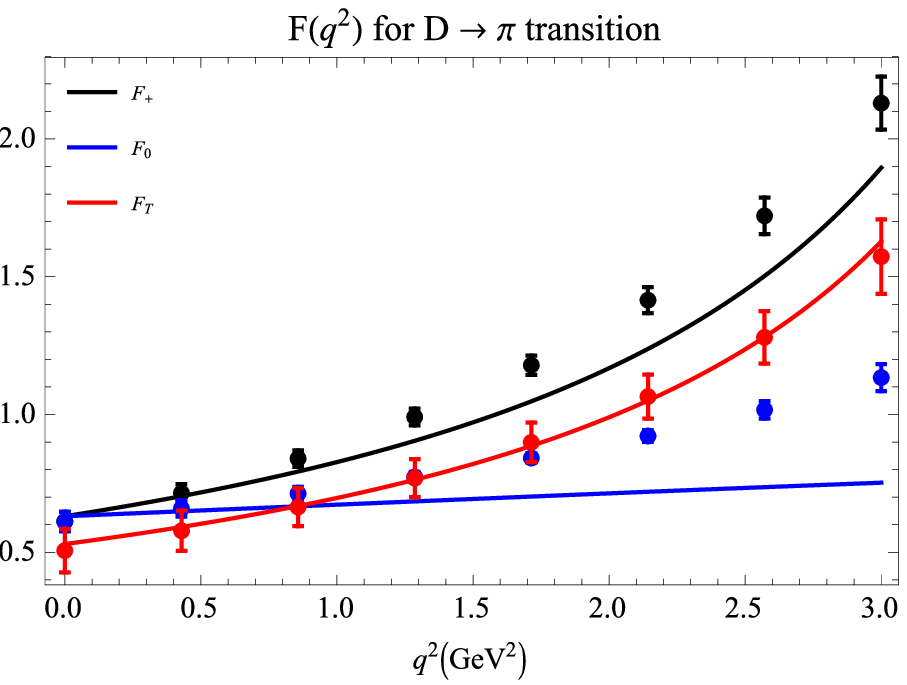}&
\includegraphics[width=0.48\textwidth]{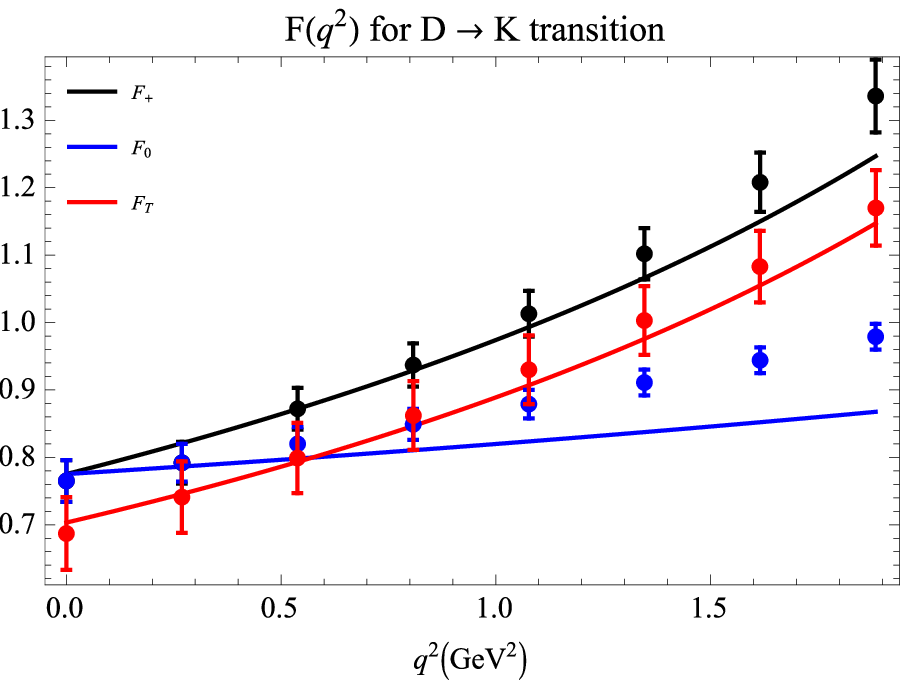}
\end{tabular}
\vspace*{-3mm}
\caption{$D\to \pi(K)\ell\nu$ form factors obtained in our model (solid lines) and in lattice calculation (dots with error bars) by the ETM collaboration~\cite{Lubicz:2017syv, Lubicz:2018rfs}.}
\label{fig:LQCD}
\end{figure*}
In Fig.~\ref{fig:LQCD} we compare the form factors $F_0(q^2)$, $F_+(q^2)$, and $F_T(q^2)$ of the $D\to \pi (K)\ell\nu$ transitions with the ETM results. One sees that our $F_0(q^2)$ agrees with the ETM only in the low $q^2$ region. However, our $F_+(q^2)$ is very close to that of the ETM. Note that we obtained $F_0(q^2)$ by evaluating $\langle P(p_2)|\bar{q}\gamma^\mu c| D_{(s)}(p_1) \rangle$. Meanwhile, the ETM collaboration directly calculated the scalar matrix element $\langle P(p_2)|\bar{q}c|D(p_1)\rangle$ and then determined $F_0(q^2)$ using the equation of motion. The tensor form factor, which is determined directly from the corresponding matrix element without any additional assumptions, shows full agreement between the two studies. The values of the form factors and their ratios at maximum recoil are also presented in Table~\ref{tab:scalar} for comparison. 
\begin{table*}[!htb]
\caption{$D\to \pi(K)\ell\nu$ form factors and their ratios at $q^2=0$.}
\label{tab:scalar}
\renewcommand{\arraystretch}{0.7}
\begin{ruledtabular}
\begin{tabular}{lcccccc}
&$f_+^{D\pi}(0)$ & $f_+^{DK}(0)$ & $f_T^{D\pi}(0)$ & $f_T^{DK}(0)$ &$f_T^{D\pi}(0)/f_+^{D\pi}(0)$ & $f_T^{DK}(0)/f_+^{DK}(0)$\\
\hline
CCQM & $0.63\pm 0.09$ & $0.78\pm 0.12$ & $0.53\pm 0.08$ & $0.70\pm 0.11$ & $0.84\pm 0.17$ & $0.90\pm 0.18$\\
ETM~\cite{Lubicz:2017syv, Lubicz:2018rfs} & $0.612(35)$ & $0.765(31)$ & $0.506(79)$ & $0.687(54)$ & $0.827(114)$ & $0.898(50)$
\end{tabular}
\end{ruledtabular}
\end{table*}

In Table~\ref{tab:Br-D-pi-K} we present theoretical predictions for the branching fractions of the decays $D \to (\pi,K) \ell^+\nu_\ell$ together with latest experimental results. Our predictions for the decays $D \to K \ell^+\nu_\ell$ agree well with experimental data. For $D \to \pi \ell^+\nu_\ell$, our results are smaller by about $20\%$.
\begin{table}[!htbp]
\caption{Branching fractions for $D \to (\pi,K) \ell^+\nu_\ell$ (in \%). We used the lifetime values $\tau_{D^+}=1.04$~$ps$ and $\tau_{D^0}=0.41$~$ps$ from PDG~\cite{Tanabashi:2018oca} in our calculation.}
\renewcommand{\arraystretch}{0.7}
\begin{ruledtabular}
\begin{tabular}{lccccll}
Channel & CCQM & HM$\chi$T~\cite{Fajfer:2004mv} & LFQM~\cite{Cheng:2017pcq} & LCSR~\cite{Wu:2006rd}  & Experiment & Ref.\\
\hline
$D^0 \to \pi^-e^+\nu_e$	& 0.22	& 0.27 & & $0.278^{+0.035}_{-0.030}$ & $0.293 \pm 0.004$	& PDG~\cite{Tanabashi:2018oca}\\
$D^0 \to \pi^-\mu^+\nu_\mu$	& 0.22	& & & $0.275^{+0.035}_{-0.030}$ &  $0.272 \pm 0.010$	& BESIII~\cite{Ablikim:2018frk}\\
& & & & & $0.231\pm 0.032$ & Belle~\cite{Widhalm:2006wz}\\
$D^+ \to \pi^0e^+\nu_e$	 & 0.29 & 0.33 & $0.41 \pm 0.03$ & $0.352^{+0.045}_{-0.038}$ & $0.372\pm 0.017$	& PDG~\cite{Tanabashi:2018oca}\\
& & & & & $0.350 \pm 0.015$ & BESIII~\cite{Ablikim:2018frk}\\					
$D^+ \to \pi^0\mu^+\nu_\mu$	 & 0.28 & & $0.41 \pm 0.03$ & $0.349^{+0.045}_{-0.038}$\\  \\
$D^0 \to K^-e^+\nu_e$	& 3.63 & 3.4 & & $3.20^{+0.47}_{-0.43}$ & $3.503 \pm 0.029$ 	&  PDG~\cite{Tanabashi:2018oca}\\
$D^0 \to K^-\mu^+\nu_\mu$ & 3.53	& & & $3.15^{+0.46}_{-0.42}$ & $3.413 \pm 0.040$ & BESIII~\cite{Ablikim:2018evp}\\
& & & & & $3.45\pm 0.23$ & Belle~\cite{Widhalm:2006wz}\\
$D^+ \to \bar{K}^0 e^+\nu_e$ & 9.28 & 8.4 & $10.32 \pm 0.93$ & $8.12^{+1.19}_{-1.08}$ & $8.73 \pm 0.10$	 & PDG~\cite{Tanabashi:2018oca}\\
$D^+ \to \bar{K}^0\mu^+\nu_\mu$	& 9.02	& & $10.07 \pm 0.91$ & $7.98^{+1.16}_{-1.06}$ & $8.72 \pm 0.19$ & BESIII~\cite{Ablikim:2016sqt}
\end{tabular}
\end{ruledtabular}\label{tab:Br-D-pi-K}
\end{table}

Several precise tests of isospin invariance and LFU were carried out by CLEO and BESIII. For this purpose, one defines the following rate ratios
\be
I^\ell_{\pi(K)}=\frac{\Gamma(D^0\to \pi^-(K^-)\ell^+\nu_\ell)}{\Gamma(D^+\to \pi^0(K^0)\ell^+\nu_\ell)},\qquad
R_{\pi(K)}=\frac{\Gamma(D\to \pi(K)\mu^+\nu_\mu)}{\Gamma(D\to \pi(K)e^+\nu_e)}.
\en
Isospin invariance predicts $I^\ell_\pi=2$ and $I^\ell_K=1$, while LFU expects $R_{\pi(K)}\approx 1$. Recently, BESIII~\cite{Ablikim:2018frk} provided new result for $\mathcal{B}(D^0 \to \pi^- \mu^+ \nu_\mu)$ and the first result for $\mathcal{B}(D^+ \to \pi^0 \mu^+ \nu_\mu)$. Combining with their previous measurements of the corresponding electron modes, they obtained the ratios $R_{\pi^-}=0.922\pm 0.030\pm 0.022$ and $R_{\pi^0}=0.964\pm 0.037\pm 0.026$,
which agree with LFU expectation within $1.7\sigma$ and $0.5\sigma$, respectively.
Similar test of LFU was done for the decays $D^0\to K^-\ell^+\nu_\ell$~\cite{Ablikim:2018evp} with significantly improved precision. In Table~\ref{tab:isospin} we summarize the result of these tests and also provide the values from the CCQM. The results show no significant violations of isospin symmetry or LFU.
\begin{table}[!htbp]
\caption{Test of isospin invariance and LFU.}
\renewcommand{\arraystretch}{0.7}
\begin{ruledtabular}
\begin{tabular}{ccllccll}
Ratio & CCQM & Experiment & Ref. & Ratio & CCQM & Experiment & Ref.\\
\hline
$I^e_\pi$		& 1.97	& $2.03 \pm 0.14 \pm 0.08$ & CLEO~\cite{Dobbs:2007aa} & $R_{\pi^-}$		 & 0.98	& $0.922 \pm 0.030 \pm 0.022$	& BESIII~\cite{Ablikim:2018frk} \\
$I^e_K$		& 0.99	& $1.06 \pm 0.02 \pm 0.03$ & CLEO~\cite{Dobbs:2007aa}  & $R_{\pi^0}$		 & 0.98	& $0.964 \pm 0.037 \pm 0.026$	& BESIII~\cite{Ablikim:2018frk} \\	 &			& $1.08 \pm 0.22 \pm 0.07$ & BES~\cite{Ablikim:2004ku} & $R_{K^-}$		& 0.97	& $0.974 \pm 0.007 \pm 0.012$ & BESIII~\cite{Ablikim:2018evp}				
\end{tabular}
\end{ruledtabular}\label{tab:isospin}
\end{table}

\subsection{$D^0 \to \rho^- \ell^+ \nu_\ell$ and $D^+ \to \rho^0 \ell^+ \nu_\ell$ }
\label{subsec:D-rho}
The decays $D^{+(0)} \to \rho^{0(-)} e^+ \nu_e$ were measured first by CLEO~\cite{Coan:2005iu, CLEO:2011ab}, and then by BESIII~\cite{Ablikim:2018qzz} with improved precision. The two collaborations also provided the form factor ratios $r_2$ and $r_V$, which are presented in Table~\ref{tab:ff-D-rho}. Our prediction for $r_2$ is compatible with both experiments. Our value for $r_V$ agrees with CLEO data, but is somewhat lower than BESIII result.
\begin{table}[!htbp]
\caption{Ratios of the $D\to \rho$ transition form factors at maximum recoil.}
\renewcommand{\arraystretch}{0.7}
\begin{ruledtabular}
\begin{tabular}{ccccccll}
 Ratio & CCQM & CQM~\cite{Melikhov:2000yu} & LFQM~\cite{Verma:2011yw} & LCSR~\cite{Wu:2006rd} & HM$\chi$T~\cite{Fajfer:2005ug}& Experiment & Ref.\\
\hline
 $r_2$		& $0.93\pm 0.19$ 			&  0.83 & 0.78 & 0.62 & 0.51& $0.83 \pm 0.12$ &CLEO~\cite{CLEO:2011ab}\\
 & & & & & & $0.845 \pm 0.068$ &BESIII~\cite{Ablikim:2018qzz}	\\
 							 $r_V$ 	& $1.26\pm 0.25$ 		 	&  1.53 & 1.47 & 1.34 & 1.72& $1.48 \pm 0.16$ &CLEO~\cite{CLEO:2011ab}\\
& & & & & & $1.695 \pm 0.097$ &BESIII~\cite{Ablikim:2018qzz}	 			
\end{tabular}
\end{ruledtabular}
\label{tab:ff-D-rho}
\end{table}

\begin{figure*}[htbp]
\renewcommand{\arraystretch}{0.3}
\begin{tabular}{cc}
\includegraphics[width=0.45\textwidth]{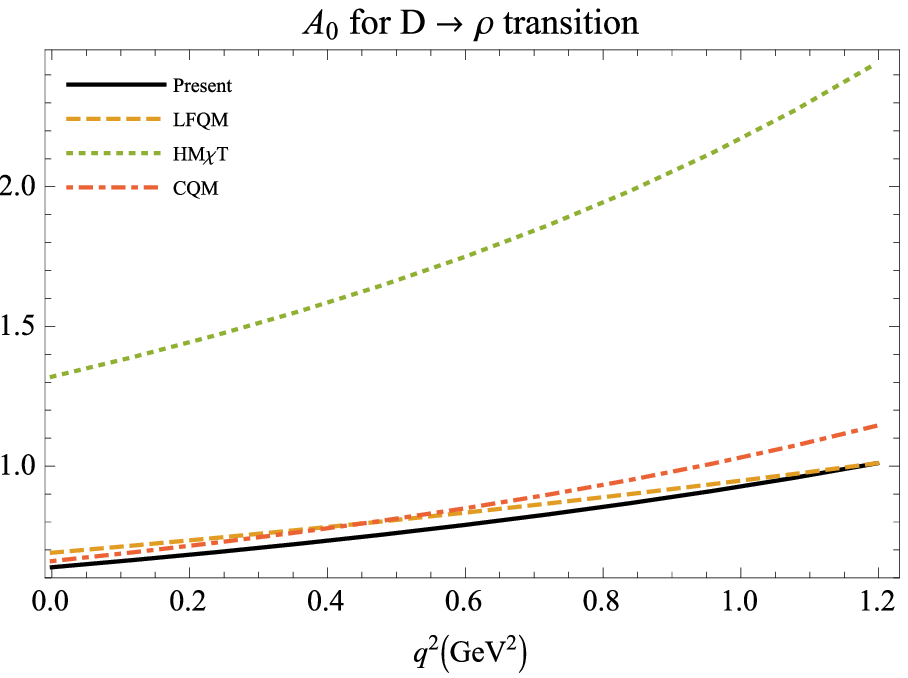}&
\includegraphics[width=0.45\textwidth]{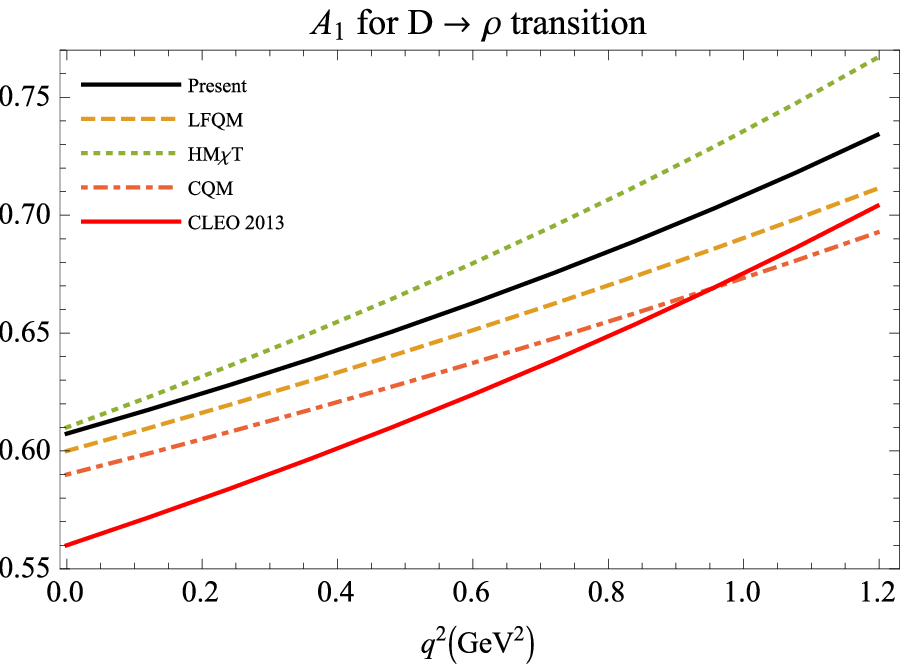}\\
\includegraphics[width=0.45\textwidth]{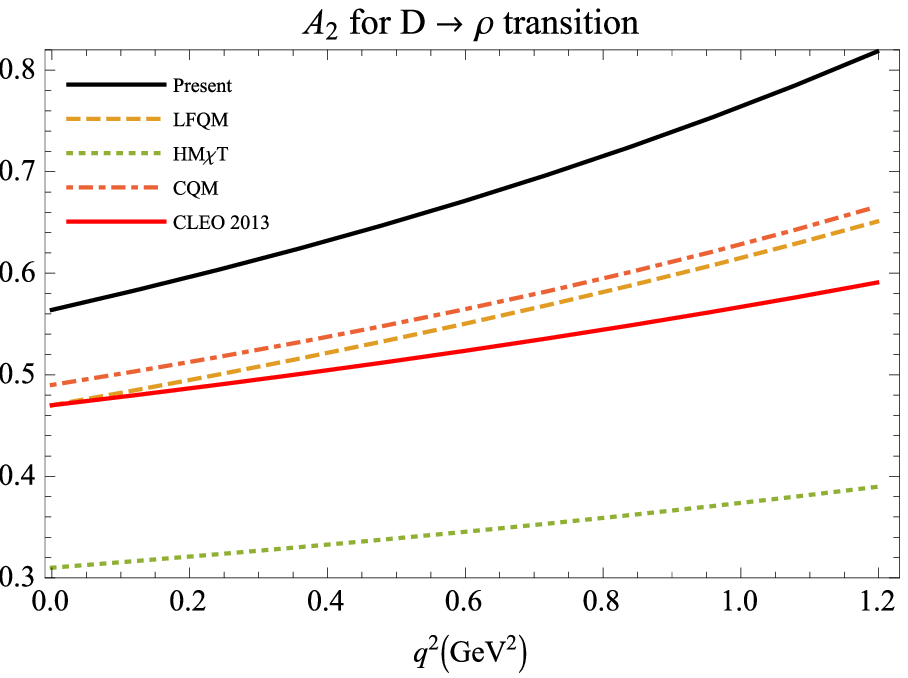}&
\includegraphics[width=0.45\textwidth]{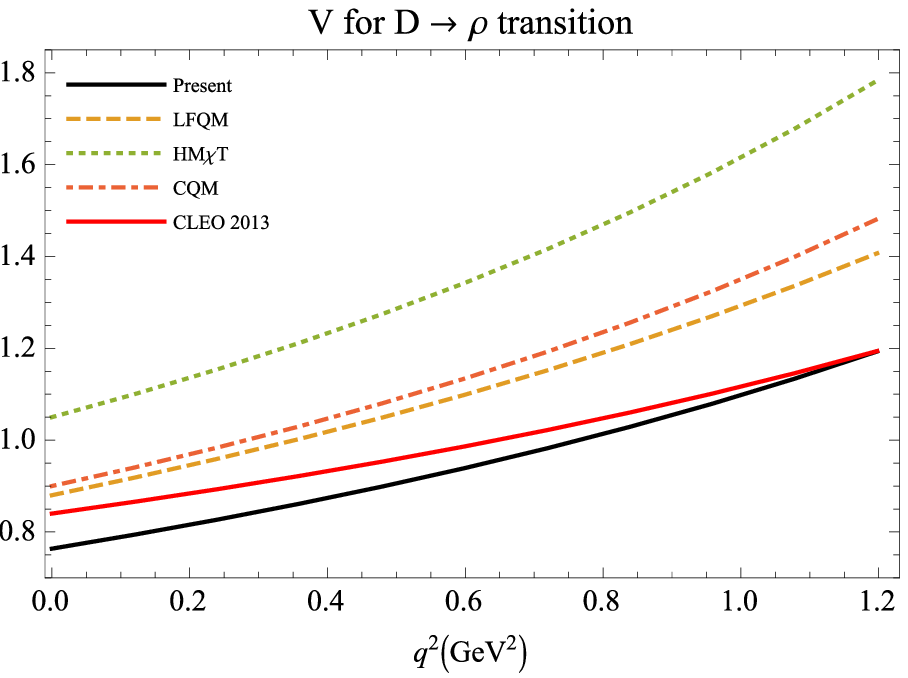}
\end{tabular}
\vspace*{-3mm}
\caption{Form factors for $D \to \rho$ in our model, LFQM~\cite{Verma:2011yw}, HM$\chi$T~\cite{Fajfer:2005ug}, CQM~\cite{Melikhov:2000yu}, and CLEO data~\cite{CLEO:2011ab}.}
\label{fig:D-rho}
\end{figure*}

Using the 2010~PDG values of $|V_{cd}|$ and the $D^{+(0)}$ lifetimes~\cite{Nakamura:2010zzi}, CLEO obtained the form factor normalizations $A_1(0)=0.56\pm 0.01^{+0.02}_{-0.03}$, $A_2(0)=0.47\pm 0.06\pm 0.04$, and $V(0)=0.84\pm 0.09^{+0.05}_{-0.06}$. These values are in agreement with our results $A_1(0)=0.61\pm 0.09$, $A_2(0)=0.57\pm 0.08$, and $V(0)=0.77\pm 0.11$. In Fig.~\ref{fig:D-rho} we plot the $q^2$ dependence for the $D\to \rho$ form factors from several models for comparison. We also show here the form factors obtained from CLEO data~\cite{CLEO:2011ab} using the single pole model they assumed for data fitting.

\begin{table*}[!htb]
\caption{Branching fractions for $D \to \rho \ell^+ \nu_e$ (in unit of $10^{-3}$).}
\label{tab:Br-D-rho}
\renewcommand{\arraystretch}{0.7}
\begin{ruledtabular}
\begin{tabular}{lccccll}
Channel 	& CCQM &  $\chi$UA~\cite{Sekihara:2015iha} & HM$\chi$T~\cite{Fajfer:2005ug} & LCSR~\cite{Wu:2006rd} & Experiment & Ref.\\
\hline
$D^0 \to \rho^- e^+ \nu_e$ & 1.62	& 1.97 & 2.0 & $1.81^{+0.18}_{-0.13}$ & $1.445 \pm 0.070$ 	& BESIII~\cite{Ablikim:2018qzz}\\
{} & {} & {} & {} & {} & $1.77 \pm 0.16$ & CLEO~\cite{CLEO:2011ab}\\
$D^0 \to \rho^- \mu^+ \nu_\mu$ & 1.55 & 1.84 & {} & $1.73^{+0.17}_{-0.13}$\\	\\
$D^+ \to \rho^0 e^+ \nu_e$ & 2.09 & 2.54 & 2.5 & $2.29^{+0.23}_{-0.16}$ & $1.860 \pm 0.093$ & BESIII~\cite{Ablikim:2018qzz}\\
{} & {} & {} & {} & {} & $2.17\pm 0.12^{+0.12}_{-0.22}$ & CLEO~\cite{CLEO:2011ab}\\
$D^+ \to \rho^0 \mu^+ \nu_\mu$ &  2.01 & 2.37 & {} & $2.20^{+0.21}_{-0.16}$ &  $2.4 \pm 0.4$ & PDG~\cite{Tanabashi:2018oca}
\end{tabular}
\end{ruledtabular}
\end{table*}

Finally, we summarize theoretical predictions for the branching fractions $\mathcal{B}(D \to \rho \ell^+ \nu_e)$ in Table~\ref{tab:Br-D-rho}. It is worth considering that isospin invariance demands the ratio $\Gamma(D^0 \to \rho^- e^+ \nu_e)/2 \Gamma(D^+ \to \rho^0 e^+ \nu_e)$ to be equal to unity, and our calculation yields $0.98$, in agreement with CLEO's result of $1.03 \pm 0.09^{+0.08}_{-0.02}$~\cite{CLEO:2011ab} and the value $0.985 \pm 0.054\pm 0.043$ by  BESIII~\cite{Ablikim:2018qzz}.

\subsection{$D^+ \to \omega \ell^+ \nu_\ell$}
\label{subsec:D-om}
\begin{figure*}[htbp]
\renewcommand{\arraystretch}{0.3}
\begin{tabular}{cc}
\includegraphics[width=0.45\textwidth]{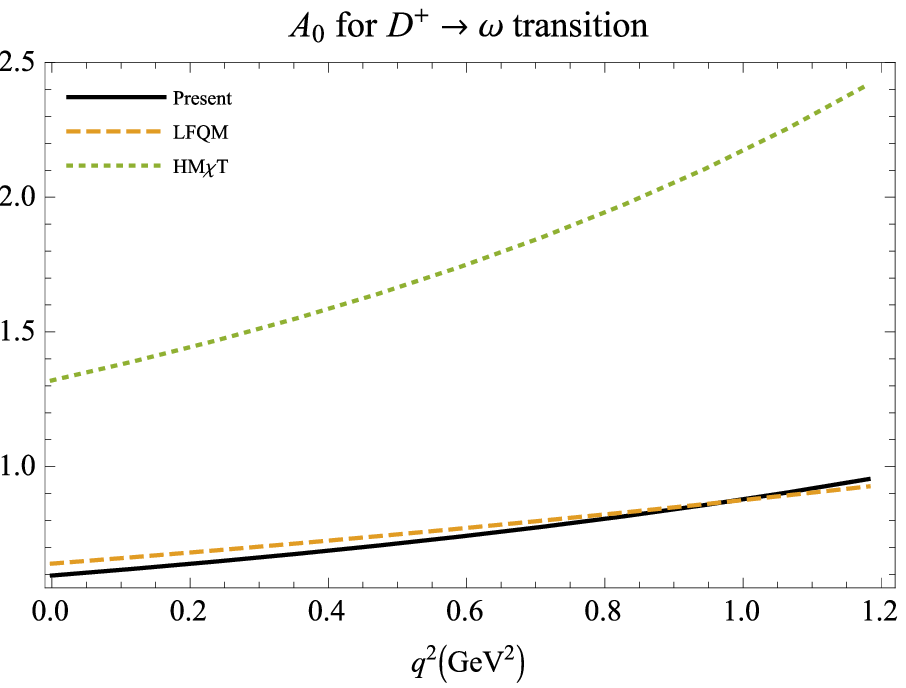}&
\includegraphics[width=0.45\textwidth]{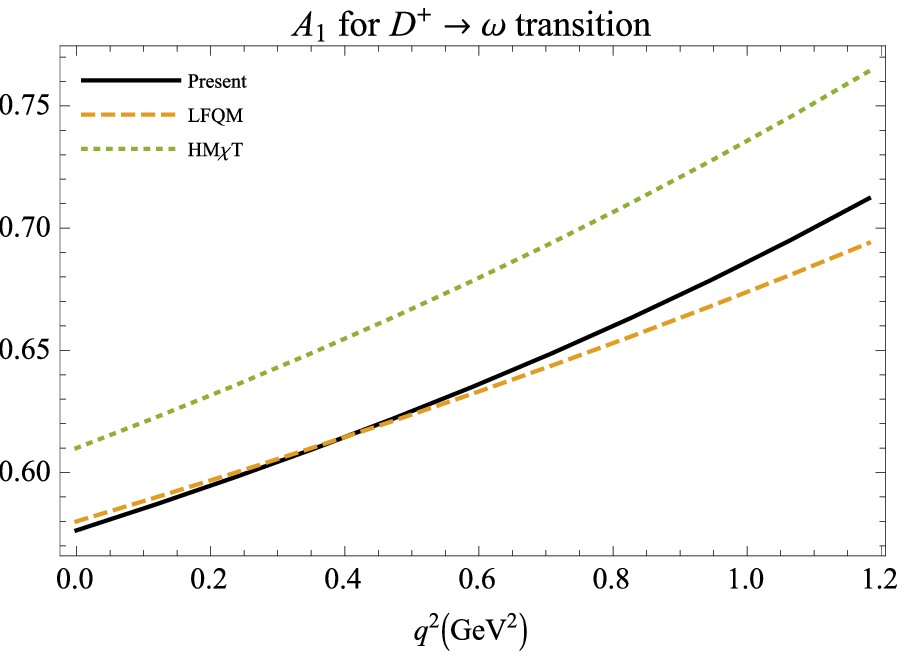}\\
\includegraphics[width=0.45\textwidth]{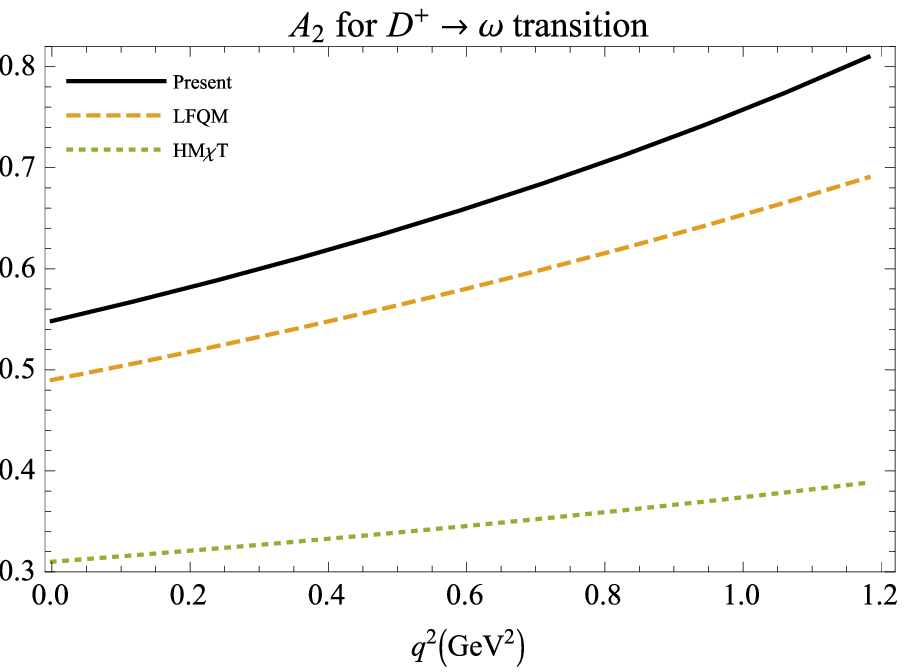}&\includegraphics[width=0.45\textwidth]{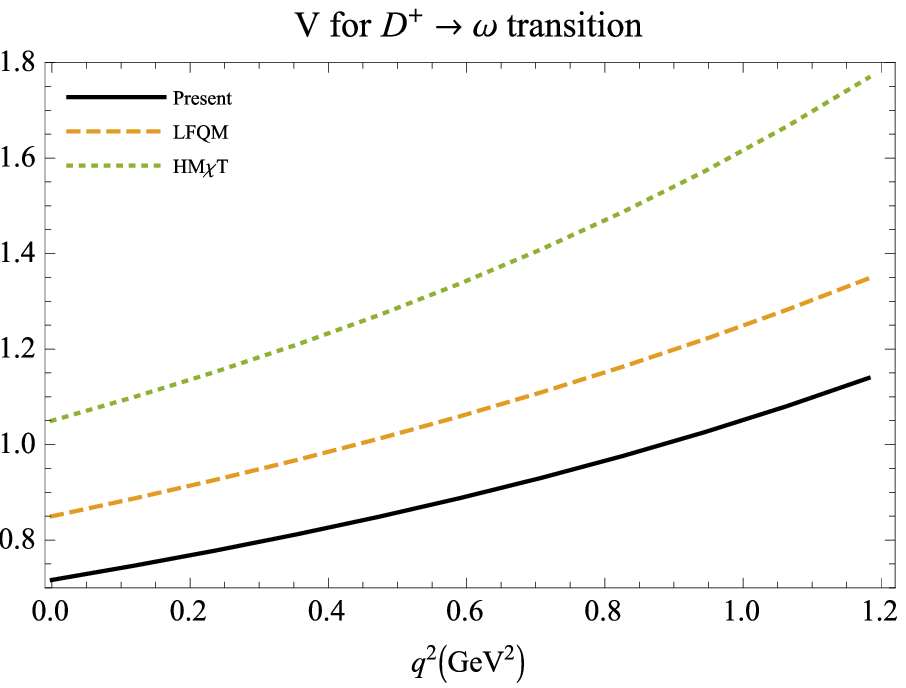}
\end{tabular}
\vspace*{-3mm}
\caption{Form factors for $D^+ \to \omega$ in our model, LFQM~\cite{Verma:2011yw}, and HM$\chi$T~\cite{Fajfer:2005ug}.}
\label{fig:D-om}
\end{figure*}
The decay $D^+ \to \omega e^+ \nu_e$ was observed first by CLEO~\cite{CLEO:2011ab}, then by BESIII  with improved precision~\cite{Ablikim:2015gyp}. BESIII also obtained the form factor ratios $r_2$ and $r_V$ for the first time, which perfectly agree with our prediction (see Table~\ref{tab:ff-D-om}). In Fig.~\ref{fig:D-om} we compare the $q^2$ dependence of the form factors with the LFQM~\cite{Verma:2011yw} and HM$\chi$T~\cite{Fajfer:2005ug}. It is seen that our form factors are very close to those from the LFQM, especially $A_0(q^2)$ and $A_1(q^2)$. In Table~\ref{tab:br-D-om} we summarize theoretical predictions for the branching fractions $\mathcal{B}(D^+ \to \omega \ell^+ \nu_\ell)$ and experimental data for $\mathcal{B}(D^+ \to \omega e^+ \nu_e)$.

\begin{table}[!htbp]
\caption{Ratios of the $D\to \omega$ transition form factors at maximum recoil.}
\renewcommand{\arraystretch}{0.7}
\begin{ruledtabular}
\begin{tabular}{cccccc}
 Ratio & CCQM   & LCSR~\cite{Wu:2006rd} & LFQM~\cite{Verma:2011yw} & HM$\chi$T~\cite{Fajfer:2005ug} & BESIII~\cite{Ablikim:2015gyp}\\
\hline
 $r_2$ 	& $0.95\pm 0.19$ 		 	&  0.60 & 0.84 & 0.51 & $1.06 \pm 0.16$\\
 							 $r_V$ 	& $1.24\pm 0.25$ 		 	&  1.33 & 1.47 & 1.72	& $1.24 \pm 0.11$			
\end{tabular}
\end{ruledtabular}
\label{tab:ff-D-om}
\end{table}
\begin{table*}[!htb]
\caption{Branching fractions for $D^+ \to \omega \ell^+ \nu_\ell$ (in unit of $10^{-3}$).}
\label{tab:br-D-om}
\renewcommand{\arraystretch}{0.7}
\begin{ruledtabular}
\begin{tabular}{lccccccc}
Channel 	& CCQM & LFQM~\cite{Cheng:2017pcq} & HM$\chi$T~\cite{Fajfer:2005ug}  	& LCSR~\cite{Wu:2006rd} & $\chi$UA~\cite{Sekihara:2015iha} & Experiment & Ref.\\
\hline
$D^+ \to \omega e^+ \nu_e$ & 1.85 & $2.1 \pm 0.2$	& 2.5 & $1.93^{+0.20}_{-0.14}$ & 2.46  &  $1.63 \pm 0.14$ 	& BESIII~\cite{Ablikim:2015gyp}\\
& 	& & & & & $1.82 \pm 0.19$ 	& CLEO~\cite{CLEO:2011ab}\\
$D^+ \to \omega \mu^+ \nu_\mu$	& 1.78 & $2.0 \pm 0.2$ & & $1.85^{+0.19}_{-0.13}$ & 2.29
\end{tabular}
\end{ruledtabular}
\end{table*}

\subsection{$D^+ \to \bar{K}^\ast(892)^0 \ell^+ \nu_\ell$ and $D^0 \to K^{\ast}(892)^- \ell^+ \nu_\ell$}
\label{subsec:D-Kv}
The decays $D \to K^{\ast} \ell^+ \nu_\ell$ are Cabibbo favored. However, the modes $D^0 \to K^{\ast -} \ell^+ \nu_\ell$ are experimentally much more challenging than their isospin-symmetric modes $D^+ \to \bar{K}^{\ast 0} \ell^+ \nu_\ell$. The reason lies in the reconstruction of the $K^\ast$ meson, which decays mainly into a $K\pi$ pair. For the $D^+ \to \bar{K}^{\ast 0}(\to K^-\pi^+) \ell^+ \nu_\ell$ channel, the final-state charged mesons can be reconstructed better than the neutral kaon or pion appearing in the case of $D^0 \to K^{\ast -} \ell^+ \nu_\ell$~\cite{Richman:1995wm}.
As a result, the $D^+ \to \bar{K}^{\ast 0} \ell^+ \nu_\ell$ modes have been extensively studied in various experiments. In particular, the ratios of the form factors have been measured mostly in these modes, with the average values of $r_2=0.802\pm 0.021$ and $r_V=1.49 \pm 0.05$~\cite{Tanabashi:2018oca} (see Table~\ref{tab:ff-D-Kv}). These ratios were also measured in the decay $D^0 \to K^{\ast -} \mu^+ \nu_\mu$ by FOCUS~\cite{Link:2004uk} and $D^0 \to K^{\ast -} e^+ \nu_e$ by BESIII~\cite{Ablikim:2018lmn}. The precise measurement of $r_2$ by BESIII~\cite{Ablikim:2018lmn} is, however, lower than the average value obtained in the $D^+ \to \bar{K}^{\ast 0} \ell^+ \nu_\ell$ channel by about $2\sigma$. More experimental study should be dedicated to address this discrepancy.
\begin{table}[!htbp]
\caption{Ratios of the $D \to K^\ast$ transition form factors at maximum recoil.}\label{tab:ff-D-Kv}
\renewcommand{\arraystretch}{0.7}
\begin{ruledtabular}
\begin{tabular}{ccccccccl}
Ratio & CCQM    & CQM & LFQM& LCSR &  HM$\chi$T & PDG~\cite{Tanabashi:2018oca} & Other & Ref.\\
& & \cite{Melikhov:2000yu} & \cite{Verma:2011yw} & \cite{Wu:2006rd} &  \cite{Fajfer:2005ug} & ($D^+\to \bar{K}^{\ast 0}$) &  ($D^0\to K^{\ast -}$)\\
\hline
 $r_2$ 	& $0.92\pm 0.18$	& 0.74 & 0.83 & 0.60 & 0.50 & $0.802\pm 0.021$ &  $0.67\pm 0.06$ & BESIII~\cite{Ablikim:2018lmn}\\
 & & & & & & & $0.91\pm 0.38$ & FOCUS~\cite{Link:2004uk}\\
						 $r_V$	& $1.22\pm 0.24$  & 1.56 & 1.36 & 1.39 & 1.60 	& $1.49\pm 0.05$ & $1.46\pm 0.07$ & BESIII~\cite{Ablikim:2018lmn}	
\\	 & & & & & & & $1.71\pm 0.76$ & FOCUS~\cite{Link:2004uk}	
\end{tabular}
\end{ruledtabular}
\end{table}

\begin{figure*}[htbp]
\renewcommand{\arraystretch}{0.3}
\begin{tabular}{cc}
\includegraphics[width=0.45\textwidth]{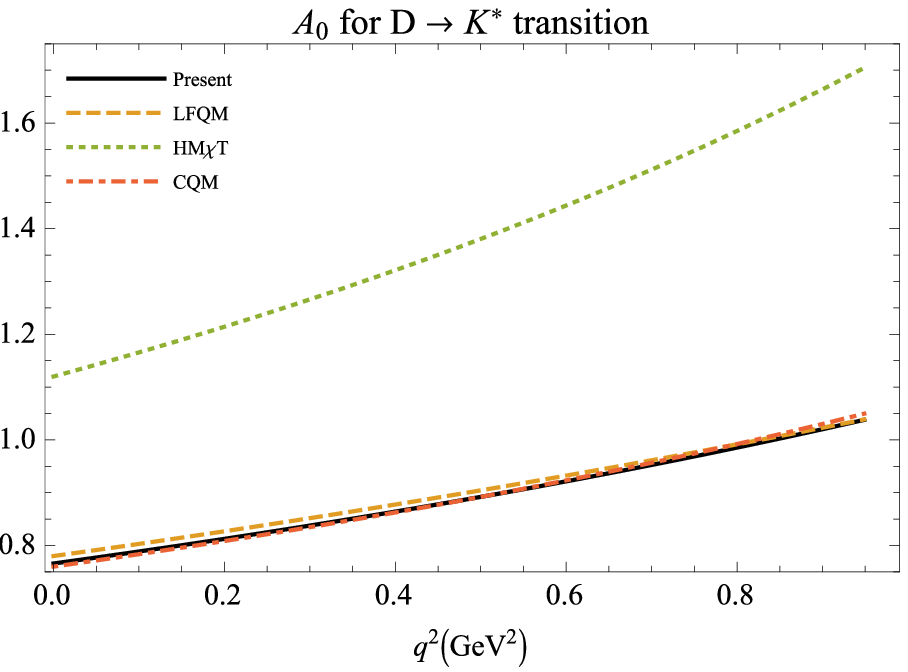}&
\includegraphics[width=0.45\textwidth]{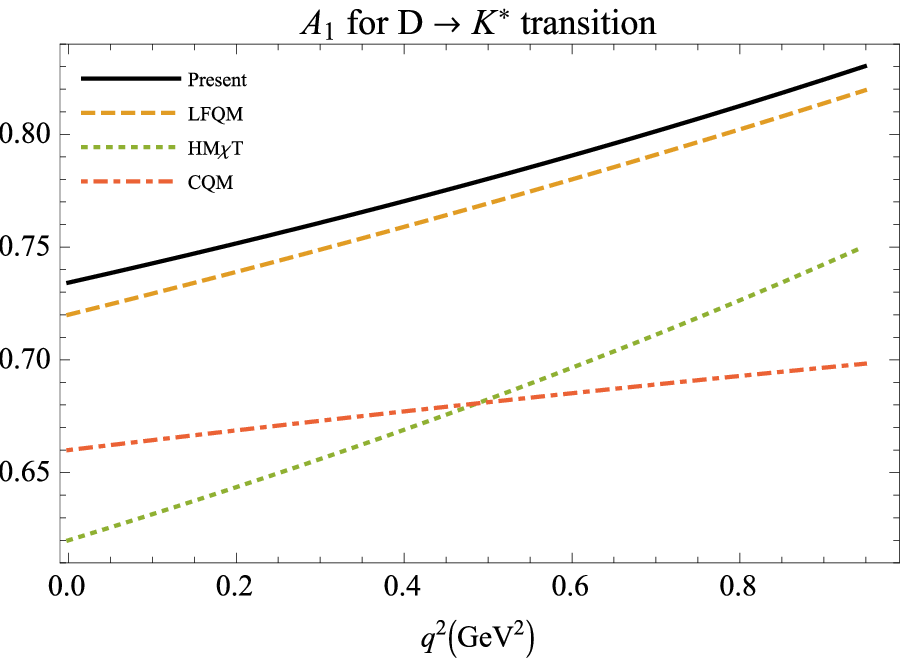}\\
\includegraphics[width=0.45\textwidth]{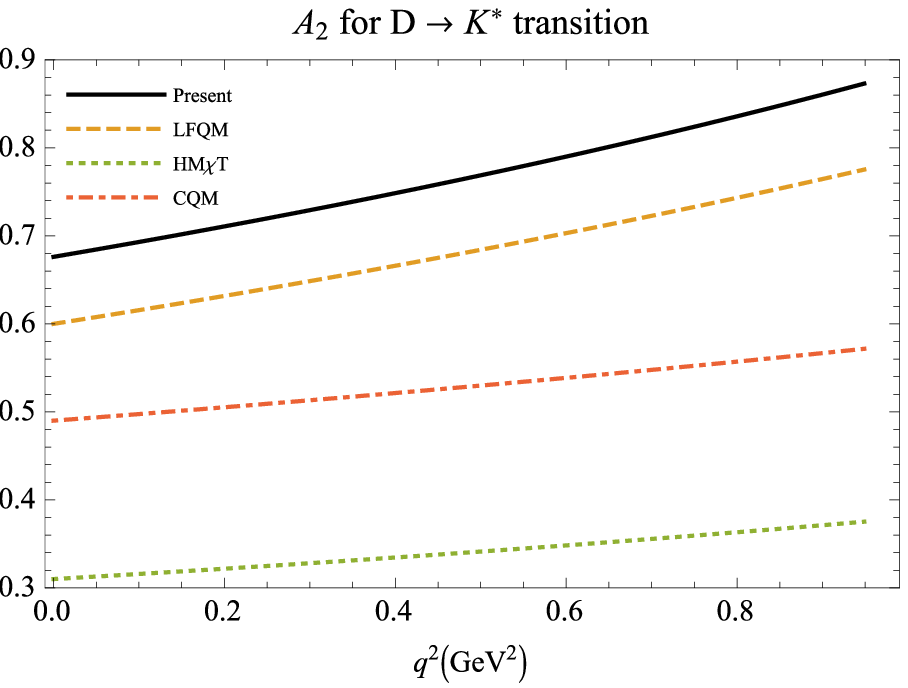}&
\includegraphics[width=0.45\textwidth]{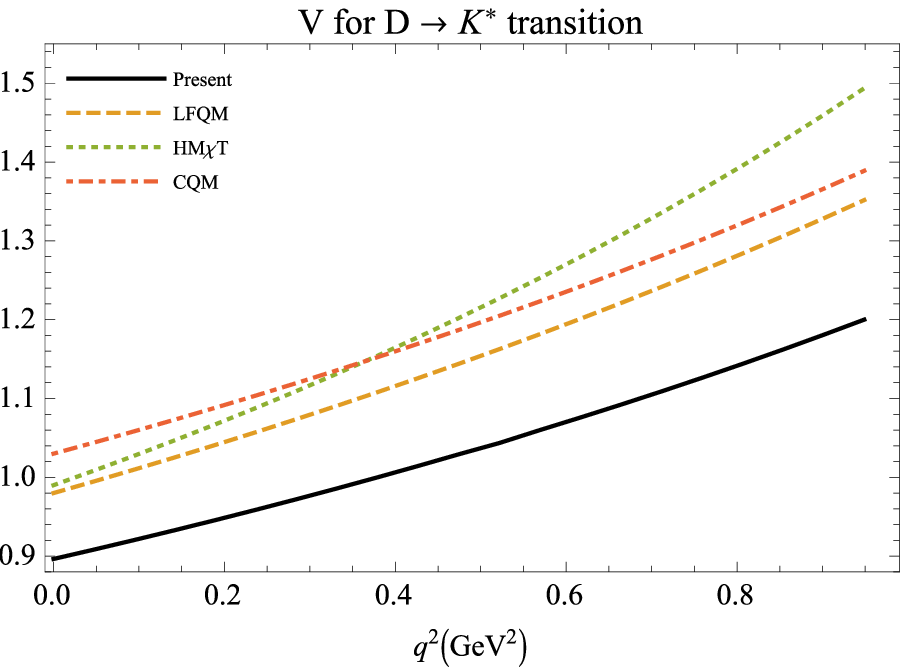}
\end{tabular}
\vspace*{-3mm}
\caption{Form factors for $D \to K^{\ast 0}$ in our model, LFQM~\cite{Verma:2011yw}, HM$\chi$T~\cite{Fajfer:2005ug}, and CQM~\cite{Melikhov:2000yu}.}
\label{fig:D-Kv}
\end{figure*}

In Fig.~\ref{fig:D-Kv} we compare the $q^2$ dependence of the form factors calculated in our model, LFQM~\cite{Verma:2011yw}, HM$\chi$T~\cite{Fajfer:2005ug}, and CQM~\cite{Melikhov:2000yu}. The shape and the normalization of our form factors are close to those from LFQM. It is worth mentioning that the normalization of the form factor $A_1(q^2)$ was obtained in several experiments. BESIII used the values $\tau_{D^+}=(10.40\pm 0.07)\times 10^{-13}$~s and $|V_{cs}|=0.986\pm 0.016$ to obtain $A_1(0)=0.589\pm 0.016$\cite{Ablikim:2015mjo}. {\it BABAR} used the same value for $\tau_{D^+}$ and $|V_{cs}|=0.9729\pm 0.003$ to obtain $A_1(0)=0.620\pm 0.011$~\cite{delAmoSanchez:2010fd}. Our model predicts $A_1(0)=0.74\pm 0.11$, which is slightly larger than both experimental values.
\begin{table}[!htbp]
\caption{Branching fraction for $D \to K^*(892) \ell^+\nu_\ell$ (in \%).}
\renewcommand{\arraystretch}{0.7}
\begin{ruledtabular}
\begin{tabular}{lccccccl}
Channel & CCQM & LFQM~\cite{Cheng:2017pcq} & CQM~\cite{Melikhov:2000yu} & LCSR~\cite{Wu:2006rd} & $\chi$UA~\cite{Sekihara:2015iha} & Experiment & Ref.\\
\hline
$D^0 \to K^{*-}e^+\nu_e$	& 2.96	& & 2.46 & $2.12^{+0.09}_{-0.09}$ & 2.15	& $2.033 \pm 0.066$	& BESIII~\cite{Ablikim:2018lmn}\\
&&&&&&  $2.16 \pm 0.17$ & CLEO~\cite{Coan:2005iu}\\
$D^0 \to K^{*-}\mu^+\nu_\mu$	& 2.80	& & & $2.01^{+0.09}_{-0.08}$ & 1.98 \\
\\
$D^+ \to \bar{K}^{*0}e^+\nu_e$ & 7.61	& $7.5\pm 0.7$	& 6.24  & $5.37^{+0.24}_{-0.23}$ & 5.56 & $5.40\pm 0.10$ & PDG~\cite{Tanabashi:2018oca}\\							
$D^+ \to  \bar{K}^{*0}\mu^+\nu_\mu$	&  7.21	& $7.0\pm 0.7$	&& $5.10^{+0.23}_{-0.21}$ & 5.12 & $5.27\pm 0.16$ & CLEO~\cite{Briere:2010zc}
\end{tabular}
\end{ruledtabular}\label{tab:Br-D-Kv}
\end{table}
In Table~\ref{tab:Br-D-Kv} we present the branching fractions of the decays $D \to K^*(892) \ell^+\nu_\ell$. Our results agree with those obtained in the LFQM, but are larger than other theoretical predictions and experimental results.

\subsection{$D_s^+ \to K^0 \ell^+ \nu_\ell$}
\label{subsec:Ds-K}
In Table~\ref{tab:ff-Ds-K} we present the theoretical predictions for the form factor $F^{D_s K}_+(0)$. Our prediction is comparable to those from LFQM~\cite{Verma:2011yw} and CQM~\cite{Melikhov:2000yu}, however, smaller than the results of LCSR~\cite{Wu:2006rd}  and lattice calculations~\cite{Lubicz:2017syv,Na:2010uf,Aubin:2004ej}. In Fig.~\ref{fig:Ds-K} we plot the $q^2$ dependence for the form factors $F_+(q^2)$ and $F_0(q^2)$. It is seen that the CCQM predicts smaller values for these form factors in the whole $q^2$ range. As a result, our branching fractions are smaller than the predictions of other theoretical approaches, as seen in Table~\ref{tab:Br-Ds-K0}.

\begin{table}[!htbp]
\caption{Comparison of $F_+(0)$ for $D_s \to K^0$ transitions.}
\renewcommand{\arraystretch}{0.7}
\begin{ruledtabular}
\begin{tabular}{cclll}
 CCQM & Other & Ref. & Lattice & Ref.\\
\hline
 $0.60\pm 0.09$ & $0.820^{+0.080}_{-0.071}$ & LCSR~\cite{Wu:2006rd} & $0.765(31)$ & ETM~\cite{Lubicz:2017syv} \\
 {} & 0.66 & LFQM~\cite{Verma:2011yw} & $0.747(19)$ & HPQCD~\cite{Na:2010uf}\\
 {} & 0.72 & CQM~\cite{Melikhov:2000yu} & $0.73(3)(7)$ & Aubin {\it et al.}~\cite{Aubin:2004ej}
\end{tabular}
\end{ruledtabular}
\label{tab:ff-Ds-K}
\end{table}
\begin{figure*}[htbp]
\renewcommand{\arraystretch}{0.3}
\begin{tabular}{cc}
\includegraphics[width=0.45\textwidth]{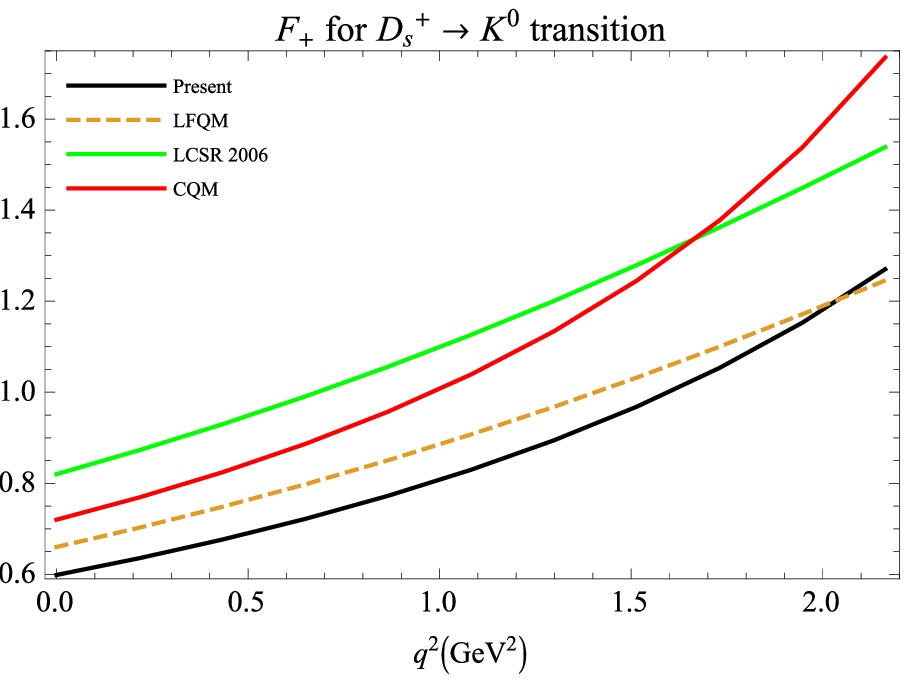} &
\includegraphics[width=0.45\textwidth]{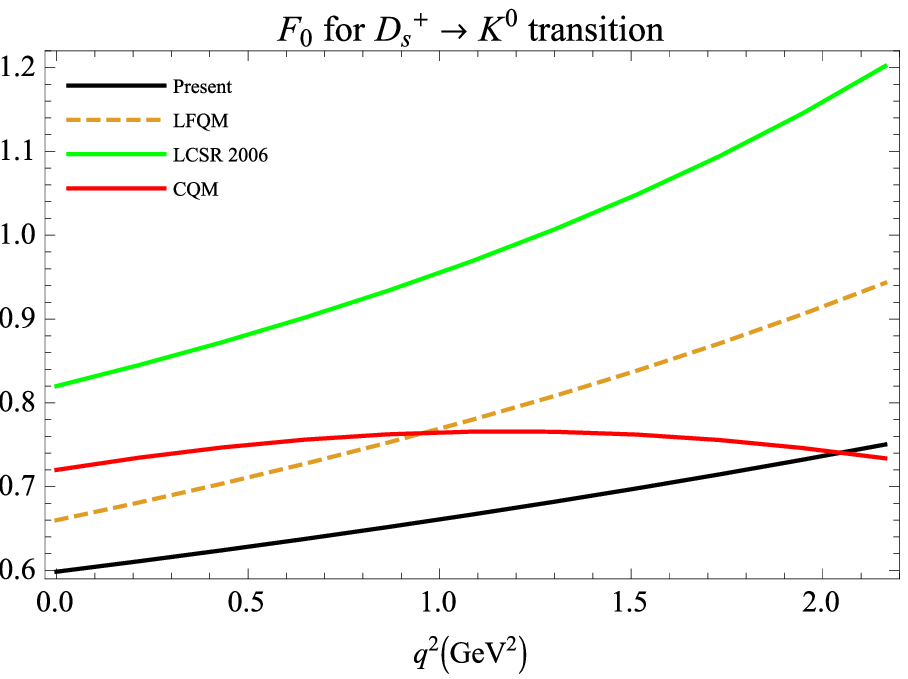}
\end{tabular}
\vspace*{-3mm}
\caption{Form factors $F_{+(0)}(q^2)$ for $D_s^+ \to K^0$ in our model, LFQM~\cite{Verma:2011yw}, LCSR~\cite{Wu:2006rd}, and CQM~\cite{Melikhov:2000yu}.}
\label{fig:Ds-K}
\end{figure*}

Our result for the branching fraction of $D_s^+\to K^0 e^+\nu_e$ is nearly two times smaller than the CLEO central value~\cite{Hietala:2015jqa}. Recently, the BESIII collaboration reported their new measurements of the branching fractions for the decay $D_s^+\to K^0 e^+\nu_e$ with improved precision~\cite{Ablikim:2018upe}. They also obtained for the first time the values of the form factors at maximum recoil. Their result for $\mathcal{B}(D_s^+\to K^0 e^+\nu_e)$ is closer to our prediction, in comparison with the CLEO one~\cite{Hietala:2015jqa}, but is still larger than ours by about $2\sigma$. Regarding the form factor, BESIII obtained the value $f_+^{D_sK}(0)=0.720\pm 0.085$ which marginally agrees with our prediction $0.60\pm 0.09$.

\begin{table*}[!htb]
\caption{Branching fractions for $D_s^+ \to K^0 \ell^+ \nu_\ell$ (in unit of $10^{-3}$).}
\label{tab:Br-Ds-K0}
\renewcommand{\arraystretch}{0.7}
\begin{ruledtabular}
\begin{tabular}{lcccccl}
Channel & CCQM & LFQM~\cite{Cheng:2017pcq} & CQM~\cite{Melikhov:2000yu} & LCSR~\cite{Wu:2006rd} & Experiment & Ref.\\
\hline
$D_s^+ \to K^0 e^+ \nu_e$				& 2.0		& $2.7 \pm 0.2$ &	
3.18	& $3.90^{+0.74}_{-0.57}$ &$3.9 \pm 0.9$ 				
& CLEO~\cite{Hietala:2015jqa}\\
& & & & &  $3.25\pm 0.41$ & BESIII~\cite{Ablikim:2018upe}\\
$D_s^+ \to K^0  \mu^+ \nu_{\mu}$	& 2.0		& $2.6 \pm 0.2$ &	
3.18 & $3.83^{+0.72}_{-0.56}$ 
\end{tabular}
\end{ruledtabular}
\end{table*}

\subsection{$D_s^+ \to \bar{K}^\ast (892)^0\ell^+ \nu_\ell$}
\label{subsec:Ds-Kv}

The decays $D_s^+ \to \bar{K}^{\ast 0}\ell^+ \nu_\ell$ are Cabibbo suppressed and therefore, experimentally challenging. There exist only two measurements of the electron mode $D_s^+ \to \bar{K}^{\ast 0}e^+ \nu_e$ by CLEO~\cite{Yelton:2009aa, Hietala:2015jqa} and BESIII~\cite{Ablikim:2018upe} collaborations. BESIII also measured the form factor ratios for the first time. However, the current uncertainties for these ratios are still large, and cover the wide range of various theoretical predictions (see Table~\ref{tab:ff-Ds-Kv}). 
\begin{table}[!htbp]
\caption{Ratios of the $D_s^+ \to K^{*0}$ transition form factors at maximum recoil.}\label{tab:ff-Ds-Kv}
\renewcommand{\arraystretch}{0.7}
\begin{ruledtabular}
\begin{tabular}{cccccccc}
Ratio & CCQM    & CQM~\cite{Melikhov:2000yu} & LFQM~\cite{Verma:2011yw} & LCSR~\cite{Wu:2006rd} &  HM$\chi$T~\cite{Fajfer:2005ug} & BESIII~\cite{Ablikim:2018upe}\\
\hline
 $r_2$ 	& $0.99\pm 0.20$	& 0.74 & 0.82 & 0.53 & 0.55 & $0.77\pm 0.29$ \\
						 $r_V$	& $1.40\pm 0.28$  & 1.82 & 1.55 & 1.31 & 1.93 	& $1.67\pm 0.38$					
\end{tabular}
\end{ruledtabular}
\end{table}
\begin{table*}[!htb]
\caption{Branching fractions for $D_s^+ \to K^{*0} \ell^+ \nu_\ell$ (in $\%$) in our model, CQM~\cite{Melikhov:2000yu}, $\chi$UA~\cite{Sekihara:2015iha}, HM$\chi$T~\cite{Fajfer:2005ug}, LFQM~\cite{Cheng:2017pcq}, LCSR~\cite{Wu:2006rd}, and experimental data.}
\label{tab:Br-Ds-Kv}
\renewcommand{\arraystretch}{0.7}
\begin{ruledtabular}
\begin{tabular}{cccccccl}
Channel & CCQM & CQM & $\chi$UA & HM$\chi$T & LFQM & LCSR & Experiment \\
\hline						
$D_s^+ \to K^{*0} e^+ \nu_e$	& 0.18	& 0.19 & 0.202 & 0.22 & $0.19\pm 0.02$ & $0.233^{+0.029}_{-0.030}$ &  $0.18 \pm 0.04$~\cite{Hietala:2015jqa}\\
&&&&&&& $0.237 \pm 0.033$~\cite{Ablikim:2018upe}	\\
$D_s^+ \to K^{*0} \mu^+ \nu_\mu$	& 0.17	& 0.19 & 0.189 & 0.22 & $0.19\pm 0.02$ & $0.224^{+0.027}_{-0.029}$
\end{tabular}
\end{ruledtabular}
\end{table*}
More experimental data would be needed to test these predictions, which differ largely between models. In Fig.~\ref{fig:Ds-Kv} we compare the shape of the $D_s\to K^{\ast 0}$ form factors in the entire $q^2$ range between our model, LFQM, HM$\chi$T, and CQM. The results for the branching fractions are presented in Table~\ref{tab:Br-Ds-Kv}. Despite the difference in the form factors, the branching fractions from different models are quite consistent, and agree well with the available experimental data.
\begin{figure*}[htbp]
\renewcommand{\arraystretch}{0.3}
\begin{tabular}{cc}
\includegraphics[width=0.45\textwidth]{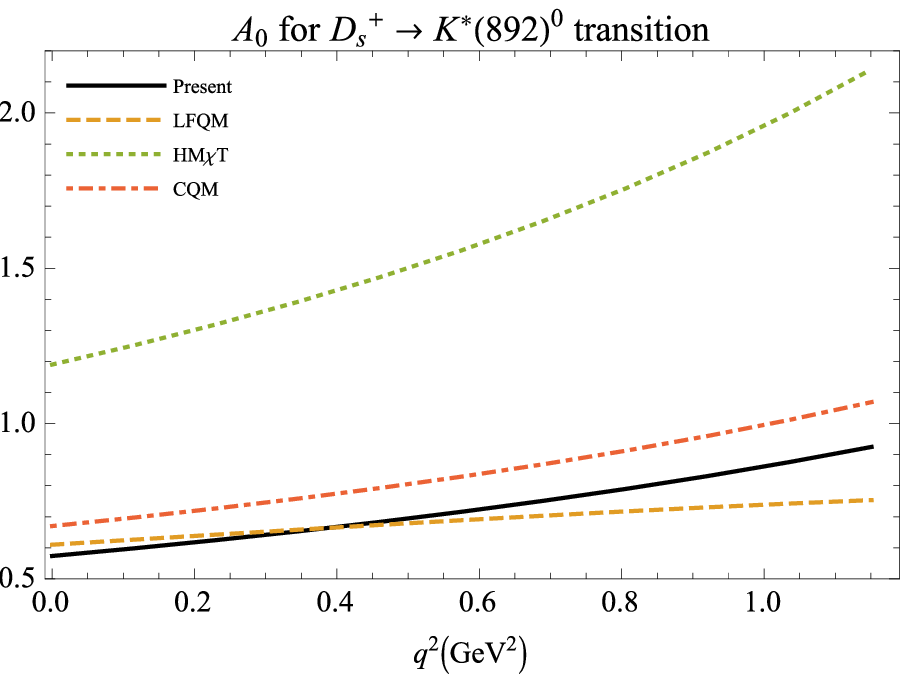} &\includegraphics[width=0.45\textwidth]{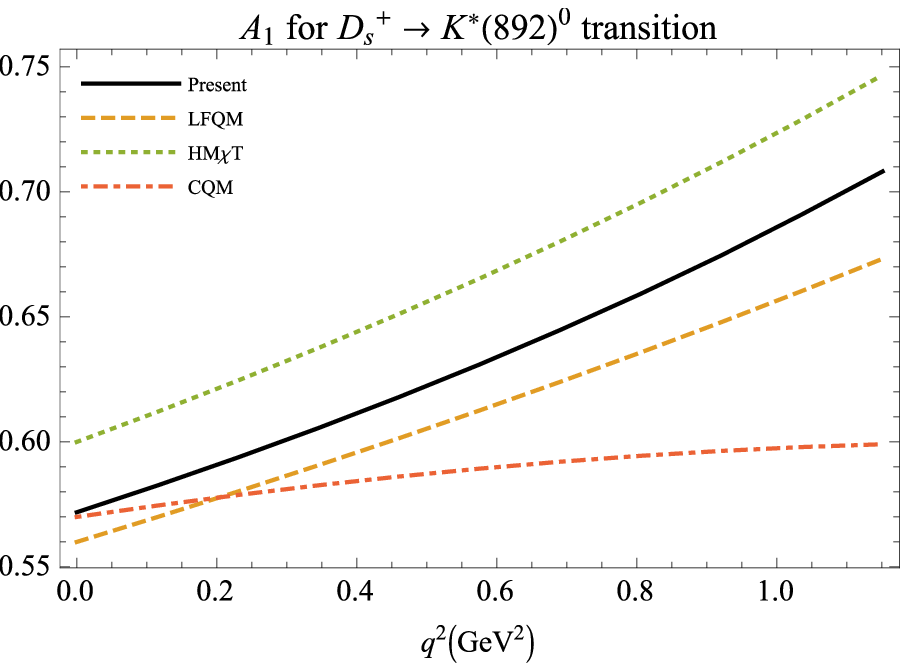} \\
\includegraphics[width=0.45\textwidth]{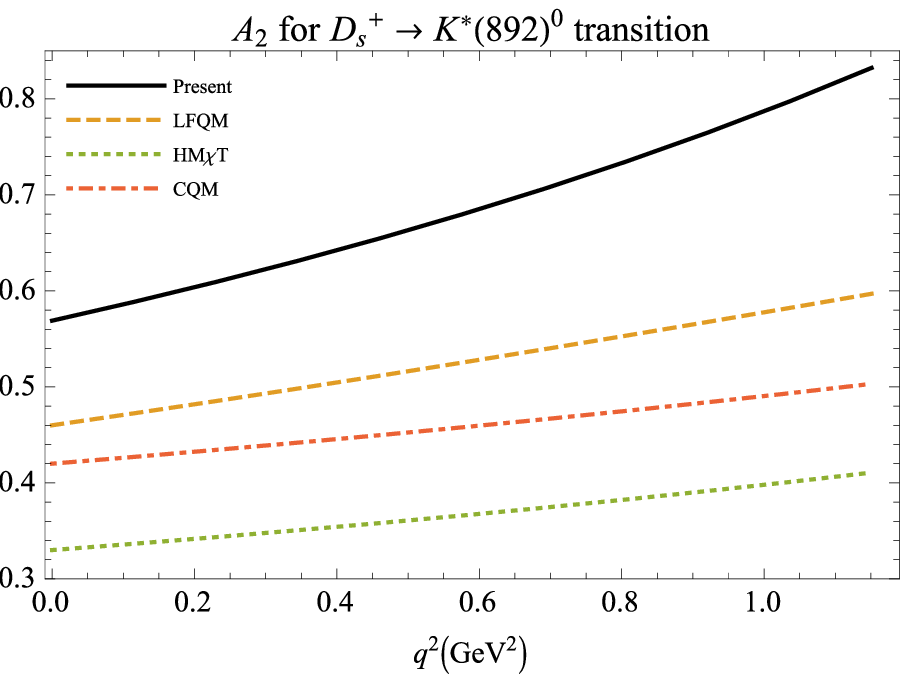} &
\includegraphics[width=0.45\textwidth]{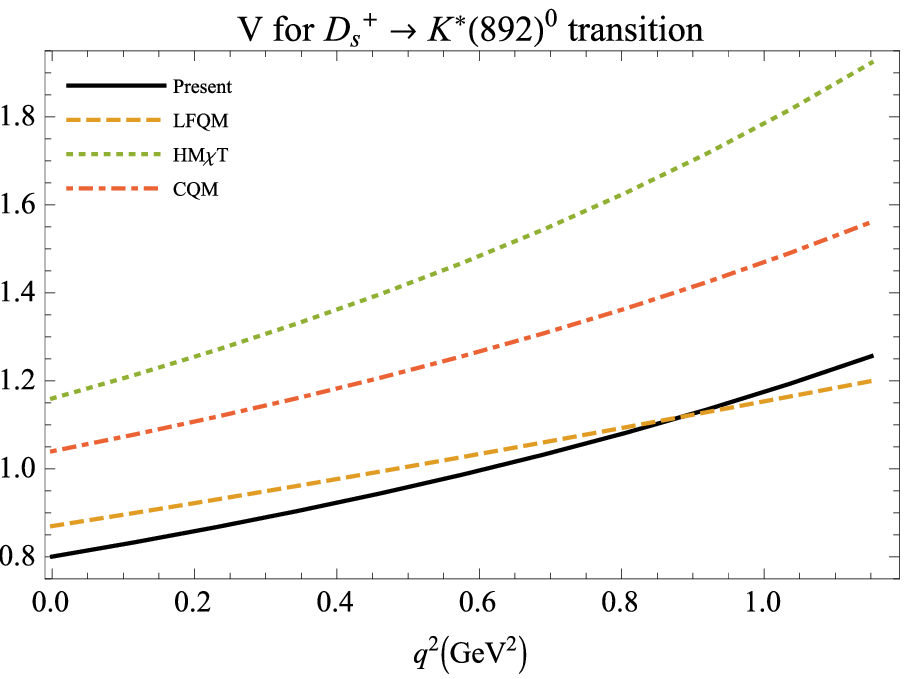}
\end{tabular}
\vspace*{-3mm}
\caption{Form factors for $D_s^+ \to K^*(892)^0$ in our model, LFQM~\cite{Verma:2011yw}, HM$\chi$T~\cite{Fajfer:2005ug}, and CQM~\cite{Melikhov:2000yu}.}
\label{fig:Ds-Kv}
\end{figure*}

\subsection{$D_s^+ \to \phi \ell^+ \nu_\ell$}
\label{subsec:Ds-phi}
The form factor ratios for the $D_s^+ \to \phi$ transition are shown in Table~\ref{tab:ff-Ds-phi}. In general, our results for the ratios $r_2$ and $r_V$ agree well with the PDG data within uncertainty. However, in the case of $r_V(D_s^+ \to \phi)$, our prediction is much lower than that from PDG. Note that our value $r_V(D_s^+ \to \phi)=1.34$ is close to the values $1.42$ from the LFQM~\cite{Verma:2011yw} and $1.37$ from LCSR~\cite{Wu:2006rd}.
\begin{table}[!htbp]
\caption{Ratios of the $D_s^+ \to \phi$ transition form factors at maximum recoil.}\label{tab:ff-Ds-phi}
\renewcommand{\arraystretch}{0.7}
\begin{ruledtabular}
\begin{tabular}{cccccccc}
 Ratio & CCQM    & CQM~\cite{Melikhov:2000yu} & LFQM~\cite{Verma:2011yw} & LCSR~\cite{Wu:2006rd} & HM$\chi$T~\cite{Fajfer:2005ug} & LQCD~\cite{Donald:2013pea} & PDG~\cite{Tanabashi:2018oca} \\
\hline
 $r_2$ 	& $0.99\pm 0.20$	 & 0.73 & 0.86 & $0.53^{+0.10}_{-0.06}$ & 0.52 & $0.74(12)$ & $0.84 \pm 0.11$	\\
							$r_V$	& $1.34\pm 0.27$		 &  1.72 & 1.42 & $1.37^{+0.24}_{-0.21}$ &1.80	& $1.72(21)$ & $1.80 \pm 0.08$	
\end{tabular}
\end{ruledtabular}
\end{table}

The full set of $D_s^+ \to \phi$ form factors was obtained from full LQCD by the HPQCD collaboration~\cite{Donald:2013pea}. The available kinematic range $0\leq q^2 \leq 0.90$ GeV$^2$ in this channel was small enough to be covered entirely in the calculation. Their results for the ratios $r_2$ and $r_V$ are also given in Table~\ref{tab:ff-Ds-phi}, which are in agreement with {\it BABAR} data~\cite{Aubert:2008rs} and with the world average values from PDG.

\begin{figure*}[htbp]
\renewcommand{\arraystretch}{0.3}
\begin{tabular}{cc}
\includegraphics[width=0.45\textwidth]{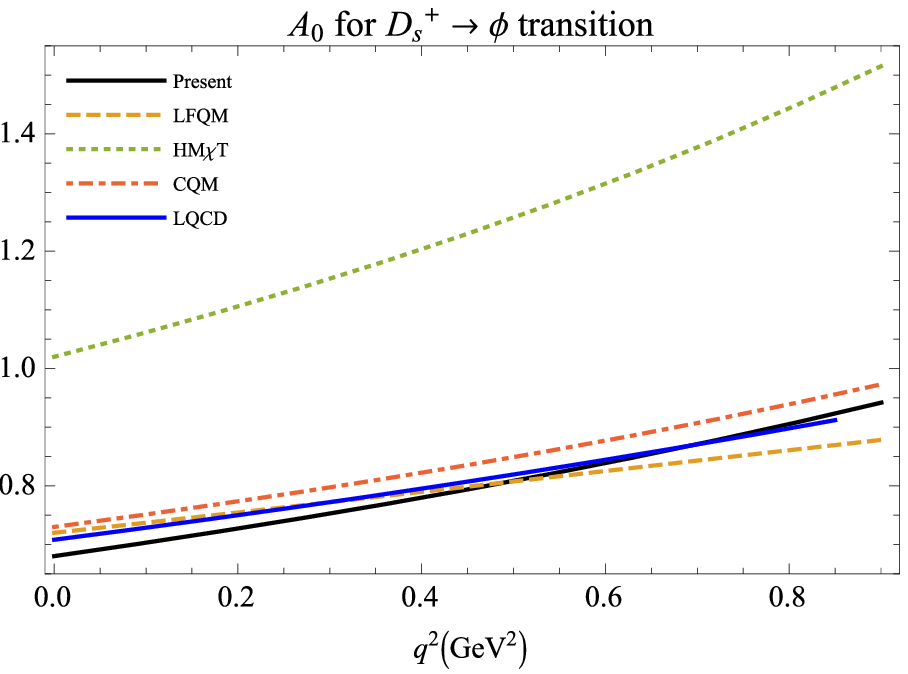} &
\includegraphics[width=0.45\textwidth]{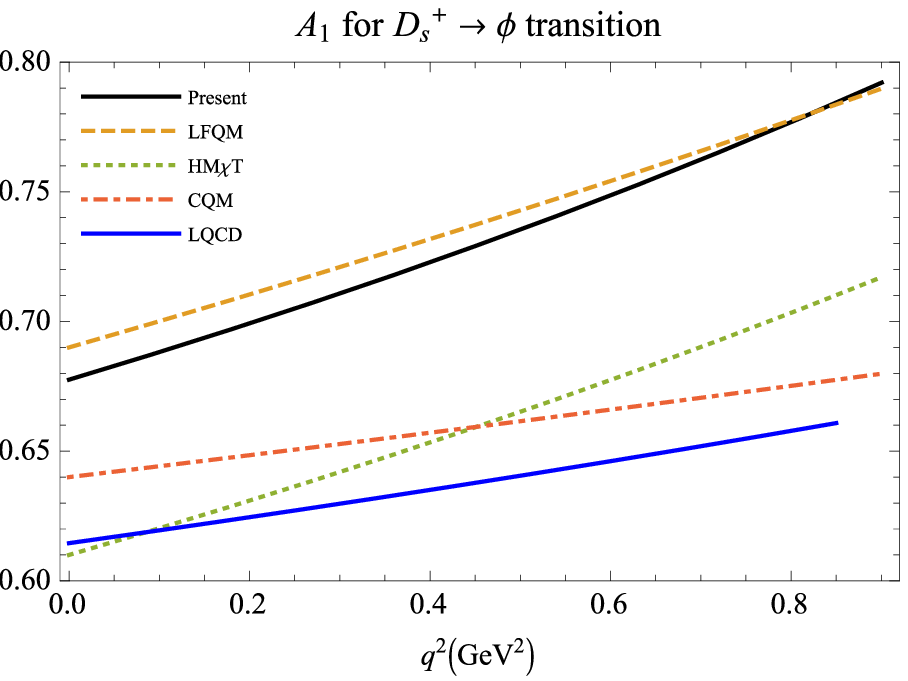} \\
\includegraphics[width=0.45\textwidth]{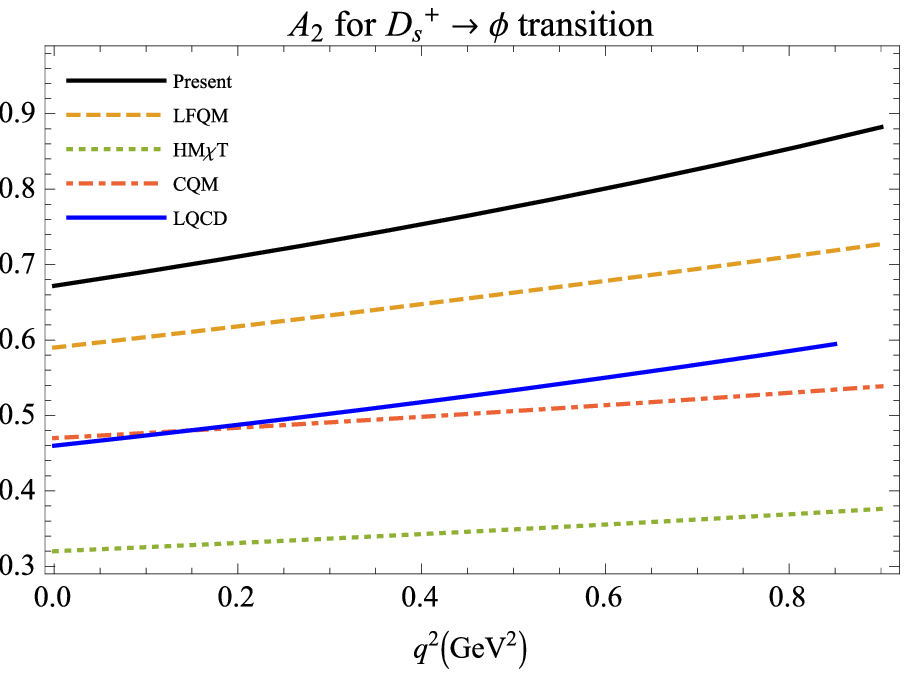} &
\includegraphics[width=0.45\textwidth]{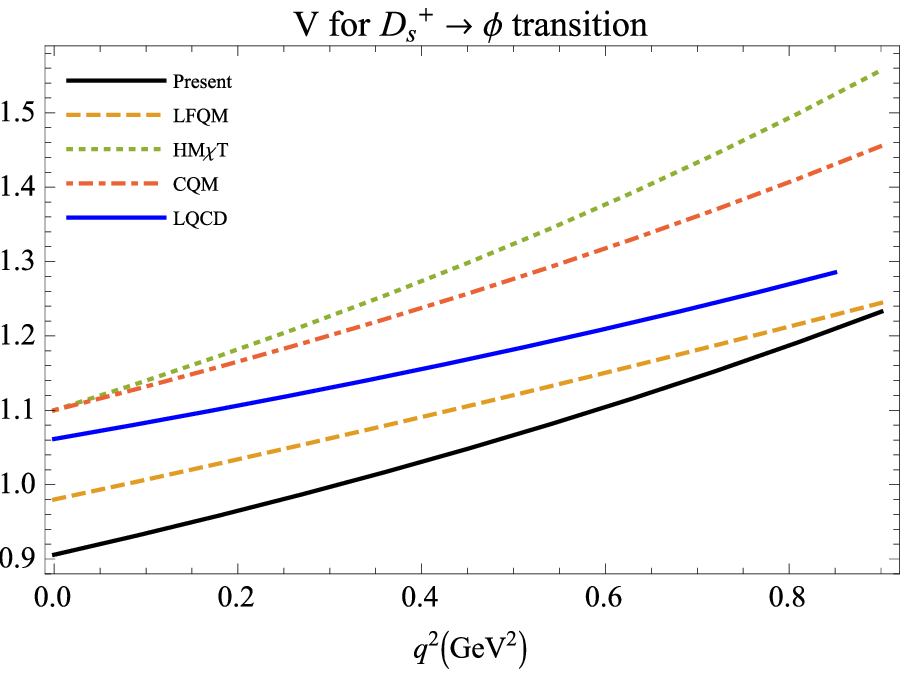}
\end{tabular}
\vspace*{-3mm}
\caption{Form factors for $D_s^+ \to \phi$ in our model, LFQM~\cite{Verma:2011yw}, HM$\chi$T~\cite{Fajfer:2005ug}, CQM~\cite{Melikhov:2000yu}, and LQCD~\cite{Donald:2013pea}.}
\label{fig:Ds-phi}
\end{figure*}

The $q^2$ dependence of the form factors are shown in Fig.~\ref{fig:Ds-phi}. Our form factors are close to the LFQM predictions. The form factor $A_2(q^2)$ varies largely between models. It should be noted that for this decay, the form factors obtained in the CQM~\cite{Melikhov:2000yu} agree well with LQCD calculation~\cite{Donald:2013pea}.

In Table~\ref{tab:Br-Ds-Phi} we present predictions and experimental results for the branching fractions $\mathcal{B}(D_s^+ \to \phi \ell^+ \nu_\ell)$. The decay $D_s^+ \to \phi e^+ \nu_e$ has been measured by {\it BABAR}, CLEO, and BESIII collaborations. The results were averaged by the PDG, giving $\mathcal{B}(D_s^+ \to \phi e^+ \nu_e)=2.39\pm 0.16\,(\%)$~\cite{Tanabashi:2018oca}. The muon mode was measured by BESIII very recently~\cite{Ablikim:2017omq}.
\begin{table*}[!htb]
\caption{Branching fractions for $D_s^+ \to \phi \ell^+ \nu_\ell$ (in $\%$).}
\label{tab:Br-Ds-Phi}
\renewcommand{\arraystretch}{0.7}
\begin{ruledtabular}
\begin{tabular}{lccccccl}
Channel & CCQM &LFQM~\cite{Cheng:2017pcq} & CQM~\cite{Melikhov:2000yu} & LCSR~\cite{Wu:2006rd} & $\chi$UA~\cite{Sekihara:2015iha}  & Experiment & Ref.\\
\hline
$D_s^+ \to \phi e^+ \nu_{e}$			& 3.01	& $3.1 \pm 0.3$	& 2.57 & $2.53^{+0.37}_{-0.40}$ & 2.12 				& $2.26 \pm 0.46$ 				& BESIII~\cite{Ablikim:2017omq}\\	
&	&  &	& & & $2.61 \pm 0.17$ 	& {\it BABAR}~\cite{Aubert:2008rs}\\
&	&	&  & &	& $2.14 \pm 0.19$	& CLEO~\cite{Hietala:2015jqa} \\					
$D_s^+ \to \phi \mu^+ \nu_{\mu}$	&  2.85 &	 $2.9 \pm 0.3$ & 2.57 & $2.40^{+0.35}_{-0.37}$ & 1.94 & $1.94 \pm 0.54$ & BESIII~\cite{Ablikim:2017omq}
\end{tabular}
\end{ruledtabular}
\end{table*}

\subsection{$D^+ \to \eta^{(\prime)} \ell^+ \nu_\ell$ and $D^+_s \to \eta^{(\prime)} \ell^+ \nu_\ell$}
\label{subsec:D-eta}
The semileptonic decays of $D_{(s)}$ into $\eta^{(\prime)}$ are of special interest since they help probe the $\eta-\eta^\prime$ mixing angle and can shed more light on the long-standing question on the gluonic component of these mesons. Many analyses and phenomenological studies have been done to find a definite answer. However, at the moment, this remains an open question since the current experimental and theoretical results are not precise enough to approve or rule out the gluonic contribution. More precise experimental data on $\eta^{(\prime)}$ decays as well as better input from LQCD are therefore highly awaited (see e.g.,~\cite{Anisovich:1997dz, Feldmann:1999uf, DiDonato:2011kr} and references therein).

The $\eta-\eta^\prime$ mixing can be described in two bases. In the singlet-octet basis, these mesons are mixtures of the $|\eta_0\rangle=\frac{1}{\sqrt{3}}|u\bar{u}+d\bar{d}+s\bar{s}\rangle$ and $|\eta_8\rangle=\frac{1}{\sqrt{6}}|u\bar{u}+d\bar{d}-2s\bar{s}\rangle$ components,
\bea
\bigg(\begin{array}{c} |\eta\rangle\\[-1ex]
|\eta^\prime\rangle \end{array}\bigg) =\bigg(\begin{array}{cc} \cos\theta_P & -\sin\theta_P\\[-1ex]
\sin\theta_P & \cos\theta_P \end{array}\bigg)
\bigg(\begin{array}{c} |\eta_8\rangle\\[-1ex]
|\eta_0\rangle \end{array}\bigg).
\ena
While in the quark basis, $\eta$ and $\eta^\prime$ are composed by the nonstrange $|\eta_q\rangle=\frac{1}{\sqrt{2}}|u\bar{u}+d\bar{d}\rangle$ and strange $|\eta_s\rangle=|s\bar{s}\rangle$ states,
\bea
\bigg(\begin{array}{c} |\eta\rangle\\[-1ex]
|\eta^\prime\rangle \end{array}\bigg) =\bigg(\begin{array}{cc} \cos\phi_P & -\sin\phi_P\\[-1ex]
\sin\phi_P & \cos\phi_P \end{array}\bigg)
\bigg(\begin{array}{c} |\eta_q\rangle\\[-1ex]
|\eta_s\rangle \end{array}\bigg).
\ena

The two bases can be transformed between each other. The relation between the mixing angles reads $\theta_P=\phi_P-\arctan\sqrt{2}$. It should be noted that, in general, two mixing angles in each basis are required to properly take into account $SU(3)$ symmetry breaking effects. However, in the quark basis, the two angles are very close, and the use of only one effective mixing angle is adequate~\cite{Feldmann:1998vh}.
Finally, in the quark basis, the gluonic component can be included by using an additional mixing angle $\phi_G$ as follows~\cite{DiDonato:2011kr}:
\bea
|\eta^\prime\rangle &=& \cos\phi_G\left(\sin\phi_P|\eta_q\rangle+\cos\phi_P|\eta_s\rangle\right)+\sin\phi_G|gg\rangle,\\
|\eta\rangle &=& \cos\phi_P|\eta_q\rangle-\sin\phi_P|\eta_s\rangle,
\ena
where the $|gg\rangle$ contribution to the $\eta$ state, which is supposed to be smaller than that to $\eta^\prime$~\cite{DeFazio:2000my}, has been omitted for simplicity.

In our calculation, we assume zero gluonic contribution. We use the quark basis, however, with a different mixing angle $\delta$ as follows:
\bea
\bigg(\begin{array}{c} |\eta\rangle\\[-1ex]
|\eta^\prime\rangle \end{array}\bigg) =-\bigg(\begin{array}{cc} \sin\delta & \cos\delta\\[-1ex]
-\cos\delta & \sin\delta \end{array}\bigg)
\bigg(\begin{array}{c} |\eta_q\rangle\\[-1ex]
|\eta_s\rangle \end{array}\bigg).
\ena
The angle $\delta$ is related to $\phi_P$ and $\theta_P$ by $\delta=\phi_P-\pi/2=\theta_P-\arctan(1/\sqrt{2})$. We take the value $\theta_P=-15.4^\circ$ from~\cite{Feldmann:1998vh}. It is important to note that, in the CCQM, we distinguish between the size parameters of the strange and nonstrange pieces appearing in the quark currents:
\bea
\eta &\rightarrow &-\frac{1}{\sqrt{2}}\sin\delta\, \Phi_{\Lambda_{\eta}^{q\bar q}}(\bar{u}u+\bar{d}d)-\cos\delta\,  \Phi_{\Lambda_{\eta}^{s\bar s}}\bar{s}s,\\
\eta^\prime &\rightarrow &+\frac{1}{\sqrt{2}}\cos\delta\, \Phi_{\Lambda_{\eta^\prime}^{q\bar q}}(\bar{u}u+\bar{d}d)-\sin\delta \, 
\Phi_{\Lambda_{\eta^\prime}^{s\bar s}}\bar{s}s.
\ena
We remind the reader that $\Phi_{\Lambda_X}$ is the vertex function of the state $X$, which contains the corresponding size parameter $\Lambda_X$. This treatment effectively helps take care of $SU(3)$ breaking, and requires four independent size parameters which should be fitted from a number of electromagnetic processes involving $\eta$ and $\eta^\prime$ (see~\cite{Branz:2009cd} for more detail). The results for these parameters are shown in Table~\ref{tab:size_parameter}.

Since we use four independent size parameters, the ``physical" form factors of the transitions $D\to \eta^{(\prime)}$ and $D_s\to \eta^{(\prime)}$ are written as~\footnote{In our recent paper~\cite{Soni:2018adu}, we used the notations $F_{\pm}^{D\to\eta^{(\prime)}}$ and $F_{\pm}^{D_s\to\eta^{(\prime)}}$ for $F_{\pm}^{D\to \eta_q}(\Lambda_{\eta^{(\prime)}}^{q\bar q})$ and $F_{\pm}^{D_s\to \eta_s}(\Lambda_{\eta^{(\prime)}}^{s\bar s})$, respectively, which may cause some misunderstanding. For comparison with experimental results, e.g.,~\cite{Ablikim:2019rjz}, one should multiply the form factors given in~\cite{Soni:2018adu} with the corresponding mixing factors as in Eq.~(\ref{eq:ff-D-eta}).}
\bea
F_{\pm}^{D\to\eta} &=& -\frac{\sin\delta}{\sqrt{2}}\, F_{\pm}^{D\to \eta_q}(\Lambda_{\eta}^{q\bar q}),\,\,\,\,  \qquad
F_{\pm}^{D\to\eta^\prime} = +\frac{\cos\delta}{\sqrt{2}}\, F_{\pm}^{D\to \eta_q}(\Lambda_{\eta^\prime}^{q\bar q}),\nn
F_{\pm}^{D_s\to\eta} &=& -\cos\delta\, F_{\pm}^{D_s\to \eta_s}(\Lambda_{\eta}^{s\bar s}),\qquad
F_{\pm}^{D_s\to\eta^\prime} = -\sin\delta\, F_{\pm}^{D_s\to \eta_s}(\Lambda_{\eta^\prime}^{s\bar s}).
\label{eq:ff-D-eta}
\ena
It should be noted that $\cos\delta>0$, so one has $F^{D_s\to\eta}_+(q^2)<0$ and $F^{D_s\to\eta}_-(q^2)>0$. This change of sign comes from the $SU(3)$ mixing and not from the relevant diagrams. In what follows, we will refer to $F^{D_s\to\eta}_+(q^2)$ as $|F^{D_s\to\eta}_+(q^2)|$ without using the absolute-value notation. Assuming relatively small $SU(3)$ breaking, one would expect that $F_{\pm}^{D\to \eta_q}(\Lambda_{\eta}^{q\bar q})\approx F_{\pm}^{D\to \eta_q}(\Lambda_{\eta^\prime}^{q\bar q})$ and $F_{\pm}^{D_s\to \eta_s}(\Lambda_{\eta}^{s\bar s})\approx F_{\pm}^{D_s\to \eta_s}(\Lambda_{\eta^\prime}^{s\bar s})$. However, in the framework of the CCQM, we observe that this is true only for the case $D_s\to\eta_s$. For $D\to\eta_q$, the form factors $F_{\pm}^{D\to \eta_q}(\Lambda_{\eta}^{q\bar q})$ and $F_{\pm}^{D\to \eta_q}(\Lambda_{\eta^\prime}^{q\bar q})$ differ significantly (see Table~\ref{tab:ff-D-etaq}). This implies that  the $SU(3)$ breaking effects are large in this case.
\begin{table}[!htb]
\caption{Form factors describing the $D_s\to\eta_s$ and $D\to\eta_q$ transitions in the double-pole parametrization Eq.~(\ref{eq:double_pole}).}
\label{tab:ff-D-etaq}
\renewcommand{\arraystretch}{0.7}
\begin{ruledtabular}
\begin{tabular}{lccclccc}
$F$ & $F(0)$ & $a$ & $b$ & $F$ & $F(0)$ & $a$ & $b$\\
\hline
$F_+^{D\to \eta_q}(\Lambda_{\eta}^{q\bar q})$ & 0.67 & 0.93 & 0.12 & $F_-^{D\to \eta_q}(\Lambda_{\eta}^{q\bar q})$ & -0.37 & 1.02 & 0.18\\
$F_+^{D\to \eta_q}(\Lambda_{\eta^\prime}^{q\bar q})$ & 0.76 & 1.23 & 0.23 & $F_-^{D\to \eta_q}(\Lambda_{\eta^\prime}^{q\bar q})$ & -0.064 & 2.29 &1.71\\[1.5ex]
$F_+^{D_s\to \eta_s}(\Lambda_{\eta}^{s\bar s})$ & 0.78 & 0.69 & 0.002 & $F_-^{D_s\to \eta_s}(\Lambda_{\eta}^{s\bar s})$ & -0.42 & 0.74 & 0.008\\
$F_+^{D_s\to \eta_s}(\Lambda_{\eta^\prime}^{s\bar s})$ & 0.73 & 0.88 & 0.018 & $F_-^{D_s\to \eta_s}(\Lambda_{\eta^\prime}^{s\bar s})$ & -0.28 & 0.92 & 0.009
\end{tabular}
\end{ruledtabular}
\end{table}
\begin{table}[!htb]
\caption{Comparison of $F_+(0)$ for $D_{(s)} \to \eta,\eta^\prime$ transitions. In~\cite{Bali:2014pva}, the lattice calculation was done for two different values of pion mass: ${M_\pi = 470\, {\rm MeV}} $ (denoted by a dagger), and ${M_\pi = 370\, {\rm MeV}}$ (denoted by an asterisk).}
\label{tab:ff-D-eta}
\renewcommand{\arraystretch}{0.7}
\begin{ruledtabular}
\begin{tabular}{lcccc}
& $D_s \to \eta$ & $D_s \to \eta'$ & $D \to \eta$ & $D \to \eta'$ \\
 \hline
CCQM	& $0.49\pm 0.07$ & $0.59\pm 0.09$ & $0.36\pm 0.05$	& $0.36\pm 0.05$ \\
LFQM~\cite{Verma:2011yw} 	& 0.48 & 0.59 &  0.39	& 0.32 	 \\
CQM~\cite{Melikhov:2000yu}	& 0.50 & 0.60 & & 		\\
LQCD$^\dagger$\cite{Bali:2014pva} 	& $0.564 (11)$ & $0.437(18)$&	& \\
LQCD$^\ast$\cite{Bali:2014pva} 	&	$0.542(13)$ & $0.404(25)$ & & \\
LCSR~\cite{Offen:2013nma} 	& $0.432\pm 0.033$ & $0.520\pm 0.080$ &	$0.552\pm 0.051$ & $0.458\pm 0.105$ \\
LCSR~\cite{Duplancic:2015zna} & $0.495^{+0.030}_{-0.029}$ &  $0.558^{+0.047}_{-0.045}$ & $0.429^{+0.165}_{-0.141}$ & $0.292^{+0.113}_{-0.104}$\\
BESIII~\cite{Ablikim:2019rjz} & $0.4576(64)$ & $0.490(51)$
\end{tabular}
\end{ruledtabular}
\end{table}

\begin{figure*}[htbp]
\renewcommand{\arraystretch}{0.3}
\begin{tabular}{cc}
\includegraphics[width=0.45\textwidth]{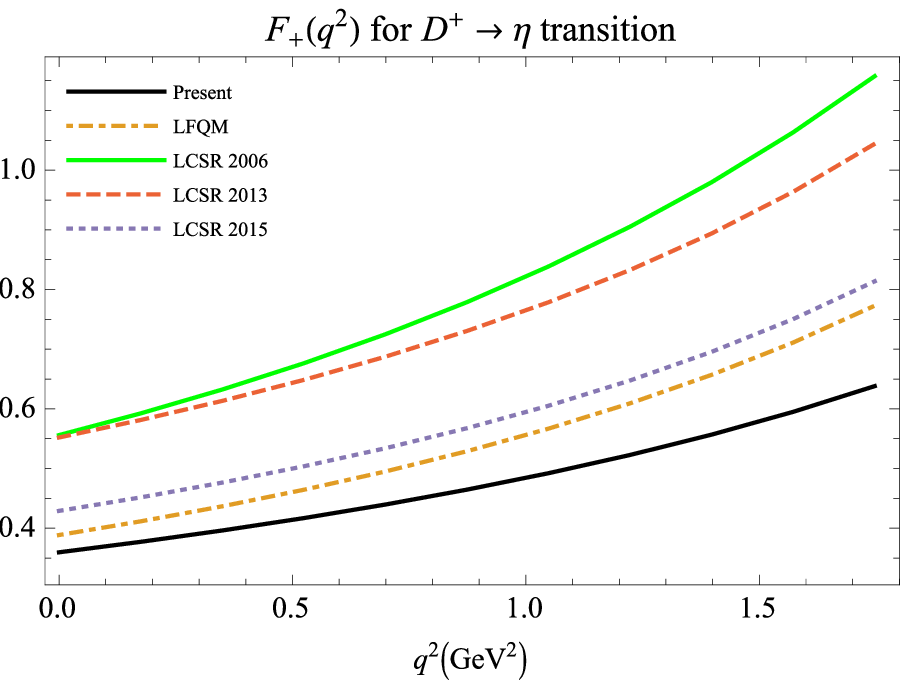}
& \includegraphics[width=0.45\textwidth]{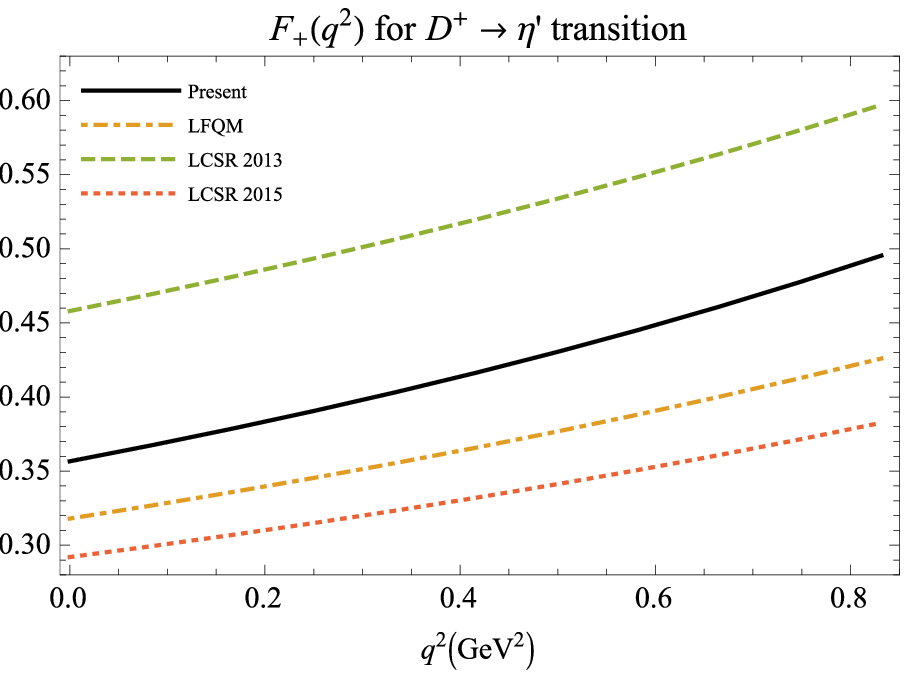}\\
\includegraphics[width=0.45\textwidth]{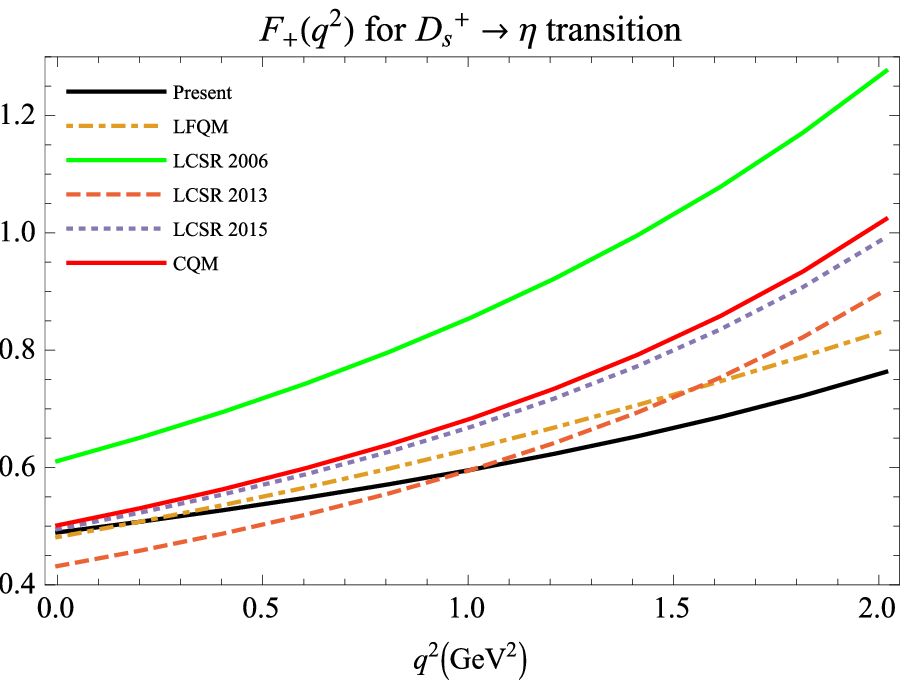}
& \includegraphics[width=0.45\textwidth]{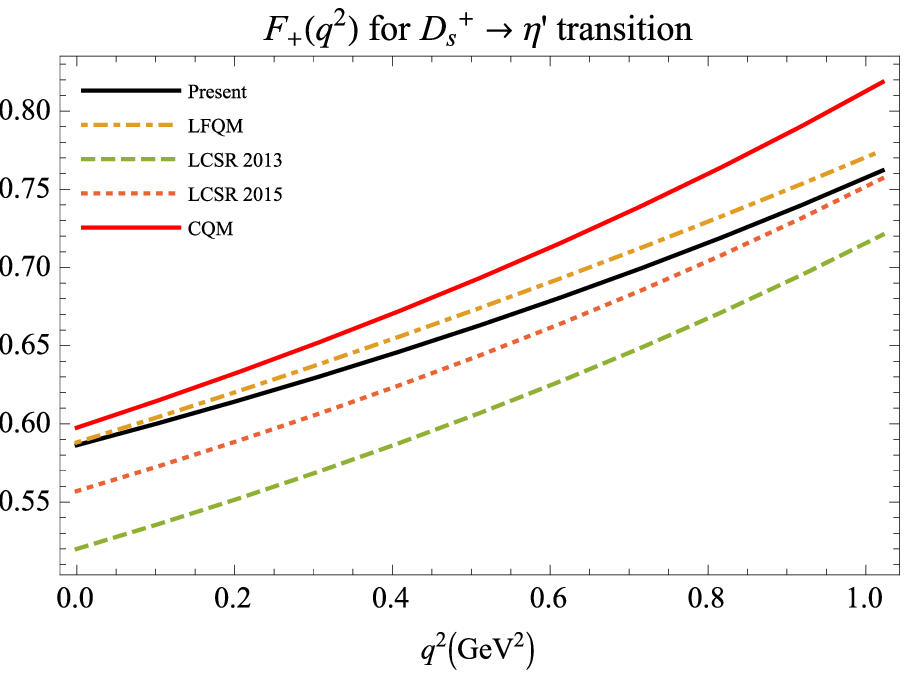}
\end{tabular}
\vspace*{-3mm}
\caption{Form factor $F_+(q^2)$ for $D_{(s)}^+ \to \eta^{(\prime)}$ in our model, LCSR~\cite{Wu:2006rd, Offen:2013nma, Duplancic:2015zna}, LFQM~\cite{Verma:2011yw}, and CQM~\cite{Melikhov:2000yu}.}
\label{fig:D_Eta}
\end{figure*}

In Table~\ref{tab:ff-D-eta} we summarize the results for $F_+(0)$. It is seen that for the $D_s\to \eta^{(\prime)}$ channels, current theoretical predictions are quite consistent between each other, and agree reasonably well with the first experimental data obtained very recently by BESIII~\cite{Ablikim:2019rjz}. However, the lattice calculation~\cite{Bali:2014pva} shows the tendency $F^{D_s\to \eta}_+(0) > F^{D_s\to \eta^\prime}_+(0)$, which is opposite to what is observed in other approaches and from BESIII result. It is interesting to note that the predictions in different quark models (including ours) for $F^{D_s\to \eta^{(\prime)}}_+(0)$ are identical. Regarding the transitions $D\to\eta^{(\prime)}$, no experimental data is available so far. Our predictions show that $F^{D\to \eta}_+(0)\approx F^{D\to \eta^\prime}_+(0)$ while LFQM and LCSR calculations observed a mild tendency that $F^{D\to \eta}_+(0)> F^{D\to \eta^\prime}_+(0)$. This would serve as a good test for the model predictions, given that measurements of $F^{D\to \eta^{(\prime)}}_+(0)$ will soon be available. In Fig.~\ref{fig:D_Eta}, we plot the $q^2$ dependence of the form factors $F_+(q^2)$ obtained in various approaches.

Theoretical predictions and experimental data for the branching fractions of the decays $D_{(s)}^+\to (\eta,\eta^{\prime})\ell^+\nu_\ell$ are summarized in Table~\ref{tab:Br-D-eta}. It is seen that our results agree very well with CLEO and BESIII measurements. Regarding the ratios of branching fractions, we got $\mathcal B(D_s^+ \to \eta' e^+ \nu_e)/\mathcal B(D_s^+ \to \eta e^+ \nu_e) = 0.37$ which coincides with the result $0.36\pm 0.14$ obtained by CLEO~\cite{Yelton:2009aa} and the more recent one $0.40 \pm 0.14$ by BESIII~\cite{Ablikim:2016rqq}. Our result $\mathcal B(D^+ \to \eta' e^+ \nu_e)/\mathcal B(D^+ \to \eta e^+ \nu_e) = 0.21$ agrees with the values $0.19\pm 0.05$ and $0.18\pm 0.05$ extracted from experimental data by CLEO~\cite{Yelton:2010js} and BESIII~\cite{Ablikim:2018lfp}, respectively.  Our prediction $\mathcal B(D_s^+ \to \eta' \mu^+ \nu_\mu)/\mathcal B(D_s^+ \to \eta \mu^+ \nu_\mu) = 0.36$ is in good agreement with the recent result $0.44\pm 0.23$ of BESIII~\cite{Ablikim:2017omq}. Finally, we predict $\mathcal B(D^+ \to \eta' \mu^+ \nu_\mu)/\mathcal B(D^+ \to \eta \mu^+ \nu_\mu) = 0.21$ which will be tested by future experiments.
\begin{table*}[!htb]
\caption{Branching fractions for $D^+\to \eta^{(\prime)}\ell^+\nu_\ell$ (in $10^{-4}$) and $D^+_s\to \eta^{(\prime)}\ell^+\nu_\ell$ (in \%).}
\label{tab:Br-D-eta}
\renewcommand{\arraystretch}{0.7}
\begin{ruledtabular}
\begin{tabular}{lcllllll}
Channel & CCQM 	&  LFQM~\cite{Cheng:2017pcq} & LCSR~\cite{Wu:2006rd} & LCSR~\cite{Duplancic:2015zna} & LCSR~\cite{Offen:2013nma} & Experiment & Ref.\\
\hline
$D^+ \to \eta e^+ \nu_e$ & 9.38 & $12.0 \pm 1.0$	& $8.60^{+1.60}_{-1.50}$ & $14.2 \pm 11.0$ & $24.5\pm 5.26$ & $10.74 \pm 0.96$ & BESIII~\cite{Ablikim:2018lfp}\\	
& & & & & & $11.4 \pm 1.0$ & CLEO~\cite{Yelton:2010js}\\
$D^+ \to \eta \mu^+ \nu_\mu$ & 9.12 & $12.0 \pm 1.0$ & $8.40^{+1.60}_{-1.40}$\\
$D^+ \to \eta' e^+ \nu_e$ & 2.00 & $1.80 \pm 0.20$ & {} & $1.52 \pm 1.17$ & $3.86 \pm 1.77$ & $1.91 \pm 0.53$	& BESIII~\cite{Ablikim:2018lfp}\\
& & & & & & $2.16 \pm 0.53$	& CLEO~\cite{Yelton:2010js}\\
$D^+ \to \eta' \mu^+ \nu_\mu$ & 1.90 & 	$1.70 \pm 0.20$\\ \\
$D_s^+ \to \eta e^+ \nu_e$ 	& 2.24 	& $2.26 \pm 0.21$ & $1.27^{+0.26}_{-0.20}$ & $2.40 \pm 0.28$ & $2.00 \pm 0.32$ &  $2.29 \pm 0.19$	& PDG~\cite{Tanabashi:2018oca} \\	
& & & & & & $2.32\pm 0.09$ & BESIII~\cite{Ablikim:2019rjz}\\
$D_s^+ \to \eta \mu^+ \nu_\mu$ & 2.18 & $2.22\pm 0.20$ & $1.25^{+0.25}_{-0.20}$ &  &  & $2.42 \pm 0.47$ & BESIII~\cite{Ablikim:2017omq} \\
$D_s^+ \to \eta' e^+ \nu_e$	& 0.83 & $0.89\pm 0.09$ & {} & $0.79 \pm 0.14$ & $0.75 \pm 0.23$ & $0.74 \pm 0.14$	& PDG~\cite{Tanabashi:2018oca}\\
& & & & & & $0.82\pm 0.08$ & BESIII~\cite{Ablikim:2019rjz} \\
$D_s^+ \to \eta' \mu^+ \nu_\mu$ & 0.79	& $0.85\pm 0.08$ & & &  & $1.06 \pm 0.54$	& BESIII~\cite{Ablikim:2017omq}
\end{tabular}
\end{ruledtabular}
\end{table*}

\subsection{$D^+_{(s)} \to D^0 e^+ \nu_e$}
\label{subsec:D-D0}
The decays $D^+_{(s)} \to D^0 e^+ \nu_e$  are of particular interest because they are induced by the light quark decay $d(s)\to u$, while the heavy $c$ quark acts as the spectator. The available $q^2$ range in these decays are very small, $0\leq q^2\leq(m_{D_{(s)}}-m_{D_0})^2\approx 10^{-5} (10^{-2})$ GeV$^2$, and the only possible final charged lepton is the positron. This phase-space suppression makes the decays very rare in the SM, and the observation of $D^+ \to D^0 e^+ \nu_e$ is far beyond the current experimental capacity. However, the decay $D^+_s \to D^0 e^+ \nu_e$ can be reached in the near future by Belle~II. Recently, the first experimental constraint on the branching fraction $\mathcal{B}(D^+ \to D^0 e^+\nu_e)$ was obtained by BESIII~\cite{Ablikim:2017tdj}. However, the upper limit is still far above the SM predictions.

From the theoretical point of view, the small phase space helps to reduce significantly the theoretical errors, and the branching fractions can be calculated with high precision. This provides a precise test of the SM in future experiments and deepens our understanding of the light quark transitions in the background field of the heavy spectator. The first theoretical calculation of these rare decays was done with the help of flavor $SU(3)$ symmetry in the light quark sector~\cite{Li:2007kgb}. By assuming $SU(3)$ symmetry, the form factors are normalized to unity at maximum recoil, $F^{DD_{0(s)}}_+(0)=1$. Their $q^2$ dependence is then obtained by using the simple pole extrapolation. Finally, the uncertainties are estimated based on $SU(3)$ breaking effects due to the strange quark mass. A more detailed study later was done using a similar approach with an extension to the semileptonic and nonleptonic decays of heavy baryons~\cite{Faller:2015oma}. 

In the CCQM, we obtain $F_+^{D D_0}(q^2)\approx 0.91$ and $F_+^{DD_s}(q^2)\approx (0.92-0.93)$ for the whole $q^2$ range. In Table~\ref{tab:rare_branching}, we summarize the results for the semileptonic decays $D_{(s)}^+ \to D^0 e^+\nu_e$. The branching fractions obtained in the CCQM are comparable with other theoretical calculations mentioned above. 

\begin{table*}[!htb]
\caption{Branching fractions for $D_{(s)}^+ \to D^0 e^+ \nu_e$, with $|V_{us}|=0.2243$ and $|V_{ud}|=0.9742$~\cite{Tanabashi:2018oca}.}
\label{tab:rare_branching}
\renewcommand{\arraystretch}{0.75}
\begin{ruledtabular}
\begin{tabular}{cclccl}
Channel & CCQM & Other & Ref. & Experiment & Ref.\\
\hline
$D^+ \to D^0 e^+ \nu_e$ 		& $2.23 \times 10^{-13}$ 	& $2.78 \times 10^{-13}$ & \cite{Li:2007kgb} & $<1.0\times 10^{-4}$ 		& BESIII~\cite{Ablikim:2017tdj}\\
											&										& $2.71 \times 10^{-13}$				& \cite{Faller:2015oma}\\
$D_s^+ \to D^0 e^+ \nu_e$ 	& $2.52 \times 10^{-8}$		& $(2.97 \pm 0.03) \times 10^{-8}$	&  \cite{Li:2007kgb} & & \\
											&										& $3.34 \times 10^{-8}$					& \cite{Faller:2015oma}
\end{tabular}
\end{ruledtabular}
\end{table*}

\subsection{Polarization observables}
\label{subsec:pol}
In Sec.~\ref{sec:4fold}, starting with the full angular distribution, we have defined a number of physical observables that can be studied experimentally. These observables include the forward-backward asymmetry $\mathcal{A}_{FB}^\ell(q^2)$, the longitudinal $P_L^\ell(q^2)$ and transverse $P_T^\ell(q^2)$ polarizations of the final charged lepton, the lepton-side $C_F^\ell(q^2)$ and hadron-side $C_F^h(q^2)$ convexity parameters, the polarization fractions of the final vector meson $F_{L,T}(q^2)$, and the trigonometric moments $W_{T,I,A}(q^2)$. These observables provide a more detailed picture of the physics in semileptonic decays of hadrons, and help study the lepton-mass effects. At $B$-factories, polarization observables have been measured extensively, especially in light of several LFU violation signals accumulated recently in $B$ meson decays (for a recent review, see e.g.~\cite{Bifani:2018zmi}).

Due to currently limited statistics, it is difficult to measure the $q^2$ dependence of these observables. Instead, they are usually averaged over the whole $q^2$ range. It should be noted that, in order to calculate the $q^2$ average, one has to multiply both the numerator and denominator of e.g. Eq.~(\ref{eq:fbAsym}) by the $q^{2}$-dependent piece $C(q^2)=|{\bf p_2}|(q^2-m_\ell^2)^2/q^2$ of the phase-space factor, and then integrate the two separately. For example, the mean forward-backward asymmetry is calculated as follows:
\be
\langle \mathcal{A}_{FB}\rangle = \frac34 
\frac{\int dq^{2} C(q^{2})\big({\cal H}_P-4\delta_\ell{\cal H}_{SL}\big)}
{\int dq^{2} C(q^{2}){\cal H}_{\rm tot}}.
\label{eq:FBint}
\en
\begin{table}[htbp]
\caption{Mean values for forward-backward asymmetry, longitudinal and transverse polarizations, and lepton-side convexity parameter.}
\renewcommand{\arraystretch}{0.7}
\begin{ruledtabular}
\begin{tabular}{lcrcccccc}
Channel &$\left\langle\mathcal{A}_{FB}^e\right\rangle$ & $\left\langle\mathcal{A}_{FB}^\mu \right\rangle$ & $\left\langle P_L^e \right\rangle$ & $\left\langle P_L^\mu \right\rangle$ & $10^2\left\langle P_T^e \right\rangle$ & $\left\langle P_T^\mu \right\rangle$ & $\left\langle C_F^e \right\rangle$ & $\left\langle C_F^\mu \right\rangle$\\\hline
$D\to \pi\ell^+\nu_\ell$ & $-4.1\times 10^{-6}$ & $-0.04$ & $1.00$ & $0.88$ & $-0.22$ & $-0.36$ & $-1.5$ & $-1.37$\\
$D\to K\ell^+\nu_\ell$ & $-6.4\times 10^{-6}$ & $-0.06$ & $1.00$ & $0.83$ & $-0.28$ & $-0.43$ & $-1.5$ & $-1.32$\\
$D\to\eta\ell^+\nu_\ell$ & $-6.4\times 10^{-6}$ & $-0.06$ & $1.00$ & $0.83$ & $-0.28$ & $-0.44$ & $-1.5$ & $-1.32$\\
$D\to\eta^\prime\ell^+\nu_\ell$ & $-13.0\times 10^{-6}$ & $-0.10$ & $1.00$ & $0.70$ & $-0.42$ & $-0.59$ & $-1.5$ & $-1.19$\\
$D\to\rho\ell^+\nu_\ell$ & $-0.21$ & $-0.24$ & $1.00$ & $0.92$ & $-0.09$ & $-0.13$ & $-0.44$ & $-0.36$\\
$D\to\omega\ell^+\nu_\ell$ & $-0.21$ & $-0.24$ & $1.00$ & $0.92$ & $-0.09$ & $-0.12$ & $-0.43$ & $-0.35$\\
$D\to K^\ast\ell^+\nu_\ell$ & $-0.18$ & $-0.21$ & $1.00$ & $0.91$ & $-0.11$ & $-0.15$ & $-0.47$ & $-0.37$\\
$D_s\to K\ell^+\nu_\ell$ & $-5.0\times 10^{-6}$ & $-0.05$ & $1.00$ & $0.86$ & $-0.24$ & $-0.39$ & $-1.5$ & $-1.35$\\
$D_s\to \eta\ell^+\nu_\ell$ & $-6.0\times 10^{-6}$ & $-0.06$ & $1.00$ & $0.84$ & $-0.27$ & $-0.42$ & $-1.5$ & $-1.33$\\
$D_s\to \eta^\prime\ell^+\nu_\ell$ & $-11.2\times 10^{-6}$ & $-0.09$ & $1.00$ & $0.75$ & $-0.38$ & $-0.54$ & $-1.5$ & $-1.23$\\
$D_s\to \phi\ell^+\nu_\ell$ & $-0.18$ & $-0.21$ & $1.00$ & $0.91$ & $-0.11$ & $-0.14$ & $-0.43$ & $-0.34$\\
$D_s\to K^{\ast}\ell^+\nu_\ell$ & $-0.22$ & $-0.25$ & $1.00$ & $0.92$ & $-0.09$ & $-0.11$ & $-0.40$ & $-0.33$
\end{tabular}
\end{ruledtabular}
\label{tab:pol}
\end{table}

In Table~\ref{tab:pol} we summarize our predictions for the $q^2$-average of the polarization observables.~\footnote{We take this chance to consider a correction in our recent paper~\cite{Soni:2018adu} where formulae Eqs.~(7-9) were obtained for the $(\ell^-\bar{\nu}_\ell)$ final-state configuration, and not $(\ell^+\nu_\ell)$. As a result, our predictions for the polarization observables in Table~X of~\cite{Soni:2018adu} were actually made for the $(\ell^-\bar{\nu}_\ell)$ final state, and not $(\ell^+\nu_\ell)$ as they should have been.}
One sees that the lepton-mass effect in $\langle\mathcal{A}_{FB}^\ell\rangle$ is small for the $D_{(s)}\to V$ transitions: the absolute values of the asymmetry for the muon modes are only about $15\%$ larger than those for the corresponding positron ones. For the $D_{(s)}\to P$ transitions, one has $\langle\mathcal{A}_{FB}^\mu\rangle/\langle\mathcal{A}_{FB}^e\rangle\sim 10^4$, simply because the forward-backward asymmetry in this case is proportional to the lepton mass squared. In the zero lepton mass limit, leptons are purely longitudinally polarized. The longitudinal polarization of the positron is very close to the zero-lepton-mass value of unity. It is essentially the muon mass that brings in a difference of about $10\%$--$30\%$ between $\langle P_L^\mu \rangle$ and $\langle P_L^e \rangle$. The lepton-mass effect in the transverse polarization is much more significant than that in the longitudinal one due to the fact that $P_T^\ell$ is directly proportional to $m_\ell$. It is seen explicitly in Table~\ref{tab:pol} that $\langle P_T^\mu \rangle/\langle P_T^e \rangle\sim 10^2$. Regarding the lepton-side convexity parameter, the lepton-mass effect yields a difference of about $10\%$--$20\%$ between $\left\langle C_F^\mu\right\rangle$ and $\left\langle C_F^e\right\rangle$. For $D_{(s)}\to P$, $\left\langle C_F^e\right\rangle$ are all very close to $-1.5$, which is the exact value in the zero lepton mass limit. 

In Table~\ref{tab:W} we present our predictions for the longitudinal polarization fraction of the final vector meson and the trigonometric moments. The lepton-mass effect in these observables is rather small, about $2\%$--$13\%$. For the positron modes, one has $\left\langle F_L\right\rangle>0.5$, which indicates that the final vector mesons are polarized slightly more longitudinally than transversely. However, this tendency reduces, and can even be reversed in the muon modes, as one can see $\left\langle F_L\right\rangle=0.49$ for $D_s\to K^{\ast}\mu^+\nu_\mu$. It is important to note that this should be confirmed by experiments, since such small difference (0.49 vs. 0.50) lies within the theoretical uncertainty of the CCQM. For $D_{(s)}\to P$, one simply has $W_T(q^2)=W_I(q^2)=W_A(q^2)=0$, and $F_L(q^2)=1$.
\begin{table}[htbp]
\caption{Longitudinal polarization fraction of final vector mesons and trigonometric moments [see Eq.~(\ref{eq:W})] for positron and muon (in parentheses) modes.}
\renewcommand{\arraystretch}{0.7}
\begin{ruledtabular}
\begin{tabular}{lcccc}
Channel & $\left\langle F_L\right\rangle$& $\left\langle W_T\right\rangle$ & $\left\langle W_I\right\rangle$& $\left\langle W_A\right\rangle$\\
\hline
$D\to\rho\ell^+\nu_\ell$ & 0.53 (0.51)&$-0.091$ ($-0.089$) & 0.054 (0.051) & $-0.067$ ($-0.061$)\\
$D\to\omega\ell^+\nu_\ell$ & 0.52 (0.50) & $-0.093$ ($-0.091$) & 0.054 (0.051) & $-0.066$ ($-0.060$)\\
$D\to K^\ast\ell^+\nu_\ell$ & 0.54 (0.52) & $-0.097$ ($-0.094$) & 0.055 (0.051) & $-0.056$ ($-0.049$))\\
$D_s\to \phi\ell^+\nu_\ell$ & 0.53 (0.50) & $-0.101$ ($-0.098$) & 0.055 (0.052) & $-0.055$ ($-0.048$)\\
$D_s\to K^{\ast}\ell^+\nu_\ell$ & 0.51 (0.49) & $-0.094$ ($-0.092$) & 0.054 (0.051) & $-0.068$ ($-0.062$)
\end{tabular}
\end{ruledtabular}
\label{tab:W}
\end{table}

The fraction $F_L(q^2)$ has been studied widely, both theoretically and experimentally, in beauty decays. In charm decays, the equivalent observable $\Gamma_L/\Gamma_T$ has traditionally been used more often in the literature, where $\Gamma_{L(T)}$ denotes the partial decay rate of the vector meson in the final state with longitudinal (transverse) polarization. The relation between the two observables reads
\be
\frac{\Gamma_L}{\Gamma_T}=\frac{\left\langle F_L\right\rangle}{\left\langle F_T\right\rangle}=\frac{\left\langle F_L\right\rangle}{1-\left\langle F_L\right\rangle}.
\en 
In Table~\ref{tab:GLT} we summarize theoretical predictions for $\Gamma_L/\Gamma_T$. The ratio has been measured in the decays $D\to K^\ast\ell^+\nu_\ell$ and $D_s\to \phi\ell^+\nu_\ell$, with the average values of $1.13\pm 0.08$ and $0.72\pm 0.18$, respectively~\cite{Tanabashi:2018oca}. The later is quite small, compared to the predictions of $1.11$ in our model and $1.13$ in the LFQM~\cite{Wang:2008ci}.
\begin{table}[htbp]
\caption{Ratio $\Gamma_L/\Gamma_T$ of a final vector meson for positron and muon (bold text) modes.}
\renewcommand{\arraystretch}{0.7}
\begin{ruledtabular}
\begin{tabular}{lccl|lccl}
Channel & CCQM & Other & Ref. &Channel & CCQM & Other & Ref.\\
\hline
$D\to\rho\ell^+\nu_\ell$ & 1.13 ({\bf 1.04})& 1.16 & CQM~\cite{Melikhov:2000yu}  &  $D\to K^\ast\ell^+\nu_\ell$ & 1.18 ({\bf 1.07}) & 1.28 & CQM~\cite{Melikhov:2000yu}\\
& & $1.17(9)$ & LCSR~\cite{Wang:2002zba} & &  & $1.15(10)$ & LCSR~\cite{Wang:2002zba}\\
& & $0.86(6)$ & QCDSR~\cite{Ball:1991bs} & & & $1.2(3)$ & LQCD~\cite{Allton:1994ui}\\
$D_s\to \phi\ell^+\nu_\ell$ & 1.11 ({\bf 1.01}) & 1.13  & LFQM~\cite{Wang:2008ci} & $D_s\to K^{\ast}\ell^+\nu_\ell$ & 1.04 ({\bf 0.97}) & 1.09 & LFQM~\cite{Wang:2008ci}\\
$D\to\omega\ell^+\nu_\ell$ & 1.10 ({\bf 1.02}) & &  & & & 1.21 & CQM~\cite{Melikhov:2000yu}
\end{tabular}
\end{ruledtabular}
\label{tab:GLT}
\end{table}

\section{Summary and Conclusion}
\label{sec:conclusion}
We provided a detailed theoretical description of charm semileptonic decays with strong emphasis on phenomenological applications. Starting with the decay matrix elements, we defined the hadronic form factors and discussed their parametrizations commonly used in the literature. The full angular decay distribution was obtained with the help of the helicity technique in a pedagogical manner. Systematically, a large set of physical observables were introduced based on the angular distribution, which provides a detailed framework for the study of semileptonic decays. All formulae were derived in their full form, taking into account the lepton mass.

The review included recent experimental data on the $D$ and $D_s$ semileptonic decays and the corresponding theoretical predictions from various studies. However, we focused mainly on our predictions obtained in the covariant confining quark model, which allowed us to calculate all nescessary hadronic form factors in the full kinematic regions. A thorough comparison of the form factor normalization and shape was provided between the CCQM and other theoretical approaches, as well as experimental data whenever possible. In general, our predictions for the form factor $F_+(0)$ and the ratios $r_{2,V}$ agree reasonably well with experimental data. However, the prediction $r_V^{D_s\phi}=1.37\pm 0.27$ is quite small compared to the world average value $1.80\pm 0.08$ given by the PDG~\cite{Tanabashi:2018oca}. By systematically comparing our form factors with those from other approaches, we observed that: (i) our form factors are less steep near $q^2_{\rm max}$; (ii) the CCQM tends to predict higher values for $A_1(q^2)$ and lower values for $V(q^2)$ in comparision with other theoretical studies, which results in smaller values for the ratio $r_V$; (iii) both the normalization and shape of our form factors are very close to those in the covariant LFQM.

Our results for the decay branching fractions and their ratios are in good agreement with other theoretical approaches and with recent experimental data given by {\it BABAR}, CLEO, and BESIII collaborations. In particular, the ratios of branching fractions are in full agreement with experimental data. Regarding the branching fraction, for all decays, our results agree with experimental data within 35\% theoretical uncertainty, and in many cases, within 15\%. The only exception is the branching fraction $\mathcal{B}(D_s^+ \to K^0 e^ + \nu_e)=2.0$, which is twice as small as CLEO result of $3.9\pm 0.9$~\cite{Hietala:2015jqa}, and disagrees with the recent BESIII value $3.25\pm 0.41$~\cite{Ablikim:2018upe} by about 60\%. Finally, we discussed the lepton mass effect and provided the first ever theoretical predictions for a large set of polarization observables, including, in particular, the forward-backward asymmetries, the lepton longitudinal and transverse polarizations, and the final vector meson polarizations, which are important for future experiments.

\begin{acknowledgments}
P.~S. acknowledges support from Istituto Nazionale di Fisica Nucleare, I.S. QFT\_\,HEP. J.~N.~P. acknowledges financial support from University Grants Commission of India under Major Research Project F.No.42-775/2013(SR). M.~A.~I., J.~G.~K., and C.~T.~T. acknowledge support from Heisenberg-Landau Grant for their collaboration. M.~A.~I. acknowledges financial support from PRISMA Cluster of Excellence at Mainz University. N.~R.~S. thanks Bogoliubov Laboratory of Theoretical Physics (JINR) for warm hospitality during Helmholtz-DIAS International Summer School ``Quantum Field Theory at the Limits: from Strong Field to Heavy Quarks'' where this work was initiated. C.~T.~T. thanks Nguyet~Le  for her permanent support. The authors thank Hai-Bo~Li for the invitation to write the review.
\end{acknowledgments}

\end{document}